\documentclass[pra,
twocolumn,
english,
superscriptaddress,
floatfix,
longbibliography]{revtex4-2}

\usepackage{physics}
\usepackage{dcolumn}
\usepackage{dsfont}
\usepackage{comment}
\usepackage{mathtools}
\usepackage{xurl}
\usepackage[export]{adjustbox}
\usepackage{hyperref}
\usepackage{breakurl}
\hypersetup{%
     colorlinks   = true,
     breaklinks   = true,
     linkcolor    = {blue},
     anchorcolor  = {blue},
     filecolor    = {blue},
     citecolor    = {blue},
     urlcolor     = {blue},
}
\usepackage{float}
\usepackage{algorithm}
\usepackage{algcompatible}
\usepackage{layouts}
\usepackage{caption}
\usepackage{subcaption}
\usepackage{graphicx}
\usepackage{mathtools}

\usepackage{todonotes}
\usepackage{tikz}

\usetikzlibrary{shapes,arrows}    
\usetikzlibrary{positioning}
\tikzstyle{block} = [rectangle, draw, fill=blue!20, dashed,
    text width=8em, text centered, rounded corners, minimum height=2em]
\tikzstyle{line} = [draw, -latex']
\tikzstyle{block2} = [rectangle, draw, fill=blue!20, dashed,
    text width=13em, text centered, rounded corners, minimum height=2em]
\tikzstyle{block3} = [rectangle, draw, fill=blue!20, dashed,
    text width=10em, text centered, rounded corners, minimum height=2em]
\tikzstyle{blockSmall} = [rectangle, draw, fill=blue!20, dashed,
    text width=5em, text centered, rounded corners, minimum height=2em]

\tikzstyle{blockGreen} = [rectangle, draw, line width=0.3mm, fill=green!20, dotted, text width=8em, text centered, rounded corners, minimum height=2em]
\tikzstyle{blockSmallGreen} = [rectangle, draw, line width=0.3mm, fill=green!20, dotted,
    text width=5em, text centered, rounded corners, minimum height=2em]
\tikzstyle{blockSmallGreenLong} = [rectangle, draw,line width=0.3mm, fill=green!20,  dotted,
    text width=10em, text centered, rounded corners, minimum height=2em]

\tikzstyle{blockSmallRed} = [rectangle, draw, line width=0.1mm, fill=red!20, rounded corners, 
    text width=5em, text centered, minimum height=2em]
 \tikzstyle{blockRed} = [rectangle, draw, line width=0.1mm, fill=red!20, rounded corners, 
   text width=8em, text centered, minimum height=2em]  
 \tikzstyle{blockRedSmall} = [rectangle, draw, line width=0.1mm, fill=red!20, rounded corners,
    text width=5em, text centered, minimum height=2em]  

\begin{document}
\title{\color{black}A Hybrid Quantum-Classical Heuristic to solve large-scale Integer Linear Programs}
\author{Marika Svensson}
\email[e-mail: ]{marika.svensson@jeppesen.com}
\affiliation{Jeppesen, 411 03 Gothenburg, Sweden}
\affiliation{Department of Computer Science and Engineering, Chalmers University of Technology, 412 96 Gothenburg, Sweden}
\author{Martin Andersson}
\affiliation{Jeppesen, 411 03 Gothenburg, Sweden}
\author{Mattias Gr{\"o}nkvist}
\affiliation{Jeppesen, 411 03 Gothenburg, Sweden}
\author{Pontus Vikst{\aa}l}
\affiliation{Wallenberg Centre for Quantum Technology, Department of Microtechnology and Nanoscience, Chalmers University of Technology, 412 96 Gothenburg, Sweden}
\author{Devdatt  Dubhashi}
\affiliation{Department of Computer Science and Engineering, Chalmers University of Technology, 412 96 Gothenburg, Sweden}
\author{Giulia Ferrini}
\affiliation{Wallenberg Centre for Quantum Technology, Department of Microtechnology and Nanoscience, Chalmers University of Technology, 412 96 Gothenburg, Sweden}
\author{G{\"o}ran Johansson}
\affiliation{Wallenberg Centre for Quantum Technology, Department of Microtechnology and Nanoscience, Chalmers University of Technology, 412 96 Gothenburg, Sweden}
\date{\today}
\color{black}
\begin{abstract}
%
%
%
 %
%
We present a method that integrates any quantum algorithm capable of finding solutions to integer linear programs into the Branch-and-Price algorithm, which is regularly used to solve
large-scale integer linear programs with a specific structure. The role of the quantum algorithm is to find integer
solutions to subproblems appearing in  Branch-and-Price. Obtaining optimal or near-optimal integer solutions to these
subproblems can increase the quality of solutions and reduce the depth and branching factor of the Branch-and-Price algorithm and hence reduce the overall running time.
We investigate the viability of the approach by considering the Tail Assignment problem and the Quantum Approximate Optimization Algorithm (QAOA). 
Here, the master problem is the optimization problem Set Partitioning or its decision version Exact Cover and can be expressed as finding the ground state of an Ising spin glass Hamiltonian.
%
For Exact Cover, our numerical results indicate that the required algorithm depth decreases with the number of feasible solutions for a given success probability of finding a feasible solution.
For Set Partitioning, on the other hand, we find that for a given success probability of finding the optimal solution, the required algorithm depth can increase with the number of feasible solutions if the Hamiltonian is balanced poorly, which in the worst case is exponential in the problem size. %
We therefore address the importance of properly balancing the objective and constraint parts of the Hamiltonian. 
We empirically find that the approach is viable with QAOA if polynomial algorithm depth can be realized on quantum devices.
\color{black}
\end{abstract}

\maketitle
\section{Introduction \label{sec:Introduction}}
Large-scale Integer Linear Programs (ILPs) appear in the real world frequently as they model problems such as planning, scheduling and resource allocation. These problems are characterized by their large size, a linear cost function, affine inequality and/or equality constraints, as well as variables required to be integers.  

Airline planning problems such as Crew Rostering, Crew Pairing~\cite{QUESNEL20201040,DEVECI201854} and Tail Assignment~\cite{Gronkvist,Gronkvist2005} fall into this category. These problems are made more difficult by very complex rules and regulations imposed by aviation authorities, airlines and unions~\cite{QUESNEL20201040}. 
\color{black}
These rules can even be hard to express in mathematical optimization models and the models can furthermore have objective functions that are nonlinear in some optimization formulations~\cite{PAQS1,Branch-and-Price}.\color{black}

\color{black}One way to address these difficulties is to formulate the optimization problem with a very large number of variables and \color{black} to separate the problem into a generation problem and a selection problem. 
\color{black} With this formulation, standard approaches such  as Branch-and-Bound or Branch-and-Cut~\cite{alma991020577069705251} can not be used directly to solve these problems due to their large size, where even enumerating the legal decision variables can require exponential time and space, see Sec. \ref{sec:colgen} for a more detailed explanation\color{black}. Instead, by starting with an empty set of variables,
the generation problem is responsible for generating new variables (aircraft routes in the Tail Assignment problem) to the selection problem (an ILP for the Tail Assignment problem). The task of the selection problem is to find the subset of the generated variables that in the most cost-effective way satisfy all the constraints in the ILP (in the most basic Tail Assignment problem, this corresponds to having each flight in the schedule covered by exactly one aircraft).
This process is generalized in the Branch-and-Price algorithm~\cite{Branch-and-Price} which combines Branch-and-Bound~\cite{Land60anautomatic} and Column Generation~\cite{DanzigColGen,LubbeckeDesrosiers} and has generally been successful for large-scale ILPs with this type of structure. The benefit of separating the problem is that the complex rules only affect the generation problem, whereas the selection problem is often a pure Set Cover or Set Partitioning problem. 

%
In the Column Generation algorithm, the generation and selection problems are solved iteratively \color{black} until optimal conditions hold\color{black}. In this context, the selection problem is called the Restricted Master Problem (RMP) and the generation problem is called the Pricing Problem (PP). The RMP, which only contains a subset of the decision variables of the original problem, is solved as a Linear Program (LP). 
%
Column Generation is generally insufficient to solve the original ILP since the solution is most likely fractional. To remedy this, Column Generation is combined with Branch-and-Bound for finding the integer solution. For readers unfamiliar with Branch-and-Price, details are given in Appx.~\ref{sec:BandBfixing}.
%
 %

%
With the results for factoring with Shor's algorithm~\cite{Shor_1997} and unstructured database search with Grover's algorithm~\cite{GroversAlg}, providing subexponential and quadratic speed-up, respectively, it is natural to ask if quantum algorithms also can provide speed-up for ILPs \color{black} even though superpolynomial  speed-up for these problems is not expected. \color{black} 
 The adiabatic quantum algorithm~\cite{Farhi_2001} and quantum annealing~\cite{Kadowaki_1998} have subsequently been proposed. 
 Other quantum algorithms for combinatorial optimization problems~\cite{zahedinejad2017combinatorial, Montanaro_2020} such as Grover's adaptive search algorithm~\cite{gilliam2020grover} have also been proposed. In recent years, much interest has been given to the Quantum Approximate Optimization Algorithm (QAOA)~\cite{Farhi2014} \color{black} for solving combinatorial optimization problems, as it may be a suitable algorithm to run on near-term gate-based quantum computers and to demonstrate quantum advantage or quantum supremacy~\cite{farhi2019quantum}.\color{black}
%
%
 
Experiments performed in~\cite{GoogleQsupremacy}  \color{black}have reported to demonstrate quantum supremacy for a problem that is not related to optimization\color{black}. Such devices can be classified as Noisy Intermediate-Scale Quantum (NISQ) computers, where qubits are controlled imperfectly and quantum error correction is generally not considered~\cite{Preskill_2018}. Moreover, QAOA was demonstrated in~\cite{arute2020quantum} for the Sherrington-Kirkpatrick model and MaxCut, where experiments agree well with simulations. 
Such results further motivate investigating QAOA for ILPs \color{black}and distinctly large-scale ILPs. \color{black} 
 
 Here we address the open question of whether quantum algorithms can provide any advantage for large-scale ILPs, \color{black} where we stress that these  problems can require exponential time and space even to {\em generate} the full ILP or the continuous relaxation counterpart. The large number of decision variables therefore in practice rules out a direct application of any quantum algorithm capable of solving an ILP, as well as standard classical algorithms for ILPs and the continuous relaxation\color{black}. We propose here \color{black} instead \color{black} a method that incorporates any quantum algorithm capable of finding an optimal or near-optimal solution to ILPs with Branch-and-Price by utilizing the quantum algorithm to solve RMP instances. The method can reduce the time to solution, improve solution quality and is \color{black}importantly \color{black} favorable to NISQ computers.
%
We investigate the method numerically by considering QAOA and the real-world problem Tail Assignment that generalizes Set Partitioning and its decision version Exact Cover, which are NP-hard and NP-complete problems~\cite{GJ}. 
%
 The results have been obtained by simulating ideal QAOA circuits applied to instances with one or more feasible solutions, extracted from a heuristic Branch-and-Price algorithm~\cite{Gronkvist}. The numerical results expand on~\cite{vikstl2019applying}, where QAOA was applied to instances with a single feasible solution \color{black}and mapped as an Exact Cover problem (the decision version of the optimization problem Set Partitioning)\color{black}, also extracted from Tail Assignment. 

The paper is organized as follows. In Sec.~\ref{sec:TailAssignment} we introduce the Tail Assignment problem. We present the method for integrating a quantum algorithm with Branch-and-Price in Sec.~\ref{sec:integQAOA}. In Sec.~\ref{sec:qaoa} we review QAOA and the chosen mapping of Exact Cover and Set Partitioning to an Ising spin glass Hamiltonian. In Sec.~\ref{sec:DataInstances} the extracted RMP instances are presented. We present and motivate the chosen optimization strategy for studying larger algorithm depths in Sec.~\ref{sec:OptInterpolation}. Results are given in Sec.~\ref{sec:results} first for Exact Cover and second for Set Partitioning. Last, in Sec.~\ref{sec:Conclusion} we summarize the findings and discuss interesting open questions that are beyond the scope of this work. 
\section{Tail Assignment - An example of a Real-World large-scale Integer Linear Program \label{sec:TailAssignment}}
Airlines regularly face several large NP-hard planning problems such as Fleet Assignment, Crew Pairing, Crew Rostering and Tail Assignment in the planning process~\cite{Gronkvist,inbook}. 
For Tail Assignment, the task is to determine, given a set of flights and a set of aircraft, what flights are operated by which individual aircraft and what order under the constraint that each flight is flown exactly once such that some objective is optimized. Operational constraints such as minimum connection times, airport curfews, maintenance, and preassigned activities must also be respected, and can be considered part of the input to Tail Assignment. A set of flights operated by an aircraft is referred to as a route, where the operational constraints distinguish legal routes from illegal routes. This means that a solution consists of a set of legal routes that cover all flights exactly once in the most cost-effective way. 
As an example, an airline can encounter problems with one thousand flights per day with hundreds of aircraft, where the aircraft are of ten different types~\cite{Gronkvist}. \color{black} In the worst case, this means that the number of possible routes to determine if they are legal or illegal would be $2^{|F|}$, where $F$ is the set of flights. By considering restrictions such as the arrival time must be less than the departure time of two flights following each other in a route  the combinatorial explosion can be decreased. However, typically the number of legal routes will be very large and too large to solve without separating the problem into a selection problem and a generation problem.\color{black}

Tail assignment can thus be classified as a large-scale ILP, where we refer the readers to~\cite{NW88} and~\cite{alma991020577069705251} for a comprehensive view of established algorithms for solving ILPs \color{black} and to~\cite{Branch-and-Price, LubbeckeDesrosiers, clec.SPRINGERLINK978038725486920050101, clc.d0f3e45d.6401.4506.92d0.e61126e72d9c20020101} for large-scale ILPs\color{black}. The classical algorithm we consider here used to find optimal or near-optimal solutions to Tail Assignment in~\cite{Gronkvist} is a heuristic Branch-and-Price. The heuristic Branch-and-Price can be understood as the Branch-and-Price algorithm where the branching step is replaced with a fixing step that is better suited for Tail Assignment by diving into a branch of the full search tree. 

For consistency, we give the details of the algorithms Branch-and-Bound, Column Generation, Branch-and-Price and the heuristic Branch-and-Price in Appx.~\ref{sec:BandBfixing}. 
\subsection{The Set Partitioning problem and the Exact Cover problem\label{sec:SPEC}}
We define a simple path-based model of Tail Assignment as a Set Partitioning problem
\begin{align}
\text{minimize} \  & \sum_{r \in R} c_r x_r, \label{eq:sp1}\\
\text{subject to} \ & \sum_{r \in R} a_{fr}x_r = 1 \ \forall f \in F, \label{eq:sp2}\\
       & x_r \in \{0, 1\} \ \forall r \in R, \label{eq:sp3}
\end{align}
where $F$ is the set of flights and $R$ is the set of legal aircraft routes. 
In the linear objective function, Eq.~\eqref{eq:sp1}, $c_r \in \mathds{Z}$ corresponds to the cost of using route $r$. The entries $a_{fr} \in \{0, 1\}$ are elements of a constraint matrix $A$ indicating if flight $f$ is part of route $r$. A column in the constraint matrix is therefore a route. Furthermore, Eq.~\eqref{eq:sp2} enforces the requirement that the set of routes in a solution should contain flight $f$ exactly once. Finally, the decision variables $x_r\ \forall r \in R$ indicate which routes are used. 

The Tail Assignment problem can, in practice, also be described by the decision problem Exact Cover, for cases where the objective is to find any feasible solution and not the optimal solution necessarily. The Exact Cover problem can be modeled as an ILP where the objective function in Eq.~\eqref{eq:sp1} is ignored and set to 0 for any assignment of the decision variables. 

We now define the set $S_{\text{feasible}}$ to be the set of feasible solutions to the Set Partitioning problem and the Exact Cover problem as %
\begin{equation}
  S_{\text{feasible}} = \left\{\vec{x}\in \{0,1\}^{|R|}: \sum_{r\in R} a_{fr}x_r = 1\ \forall f \in F \right\}. \label{eq:Sfeas}
\end{equation}
If we consider a linear system of equations modulus 2 
\begin{equation}
  A\vec{x}=\vec{b} \text{ mod } 2
  \label{BoolSysEq}
\end{equation}
where the matrix $A$ is of dimension $|F|\times|R|$, $\vec{x}$ is a column vector with $|R|$ unknown variables and $\vec{b}$ is a column vector with $|F|$ entries. The elements of $A$, $\vec{b}$ and $\vec{x}$ are either 0 or 1, respectively. The system of equations has 
\begin{equation}
  2^{|R| - \text{rank}(A)}
    \label{eq:booleanNbrSols}
\end{equation}
number of solutions as long as the linear system of equations in Eq.~\eqref{BoolSysEq} has at least one solution~\cite{Montanari}. 
For Set Partitioning $2^{|R| - \mathrm{rank}(A)}$ constitutes an upper bound on the number of feasible solutions $|S_{\text{feasible}}|$~\cite{Seliverstov}, since any feasible solution to Set Partitioning is also a solution modulus 2 to the system of equations in Eq.~\eqref{BoolSysEq} where all entries in $\vec{b}$ is set to one. It is therefore possible that the number of feasible solutions is significantly smaller than the upper bound. 
Furthermore, as the counting version of Exact Cover and Set Partitioning is \#P-complete~\cite{10.1016/j.ipl.2008.10.009}, obtaining the actual number of feasible solutions for typical instances for Tail Assignment becomes intractable.

We have investigated the number of feasible solutions for generated RMP instances of Tail Assignment with CPLEX~\cite{cplex2009v12}. We find that the number of feasible solutions for two sets of generated instances  can be larger than $5\cdot 10^6$ for problems with 700-800 decision variables. We can therefore not rule out that the number of feasible solutions can be very large in practice, and the consequence is to investigate if a large feasible set is a limiting factor in the performance for QAOA. 
\section{Integrating a Quantum Algorithm with Branch-and-Price \label{sec:integQAOA}}
\begin{figure}[ht] 
 \centering
\begin{tikzpicture}[node distance = 1.4cm, auto]
\node[shape=circle,draw=black!0!white] (c2d) at (4.2,-4.2)  {$No$};
\node[shape=circle,draw=black!0!white] (c2d) at (3.4,-5)  {$Yes$};

\node[shape=circle,draw=black!0!white] (c2d) at (1.4,-4)   {$Yes$};
\node[shape=circle,draw=black!0!white] (c2d) at (0.4,-5.1)   {$No$}; 

\node[shape=circle,draw=black!0!white] (c2d) at (3.4,-10)  {$Yes$};
\node[shape=circle,draw=black!0!white] (c2d) at (4.7,-9.2)  {$No$};

\node[shape=circle,draw=black!0!white] (c2d) at (-1.7,-8)  {$No$};
\node[shape=circle,draw=black!0!white] (c2d) at (0.4,-9)   {$Yes$};

\node[shape=circle,draw=black!0!white] (c2d) at (1.25,-10.5)  {$Yes$};
\node[shape=circle,draw=black!0!white] (c2d) at (3.4,-11.3)   {$No$};
 \node [blockGreen]           (input)      {Find initial solution};
 \node [blockGreen, below of=input,node distance=1.3cm] (LP)      {Solve LP relaxed Restricted Master Problem (RMP)};
 \node [blockSmallGreen, below of=LP, node distance=1.8cm]     (ColGen)    {Solve Pricing Problem (PP)};
 \node [blockGreen, below of=ColGen, node distance=1.4cm] (ImprovingCols)  {$\exists$ $r$ s.t. $\bar{c}_r < 0$? };
 \node [blockGreen, below of=ImprovingCols, node distance=1.7cm] (ColGenSol) {Solution to LP relaxed MP, $\vec{x}_{\text{MP}}^*$, found}; 
 
 \node [block,   below of=ColGenSol, node distance=2cm]  (IntSol) {$\vec{x}_{\text{MP}}^*$ feasible to original problem?};
 \node [blockSmall,   left of=ColGen, node distance=2cm]    (Fix)   {Branch};
 \node [blockGreen, right of=ColGen, node distance=3cm] (exp2)  {Add improving columns};
 \node [blockSmall, below of=IntSol, node distance=2.6cm]        (sol)   {Exit};
 \node [blockSmallRed, below of=exp2, node distance=1.3cm]     (QAOAcheck){Is RMP promising?};
\node [blockRed, below of=QAOAcheck, node distance=1.2cm]   (QAOApre)   {Preprocess to reduce RMP};
\node [blockRed, below of=QAOApre, node distance=1.3cm, line width=0.4mm]   (QAOA)   {Solve RMP w. quantum algorithm};

\node [blockRed, below of=QAOA, node distance=1.3cm]   (QAOApost)   {Postprocess solution};

\node [blockRed, below of=QAOApost, node distance=1.2cm]        (solQAOA) {Is solution $x_{\text{RMP}}^*$ feasible?};
\node [blockRed, below of=solQAOA, node distance=1.35cm]      (doneQAOA) {$\vec{c}_{\text{RMP}}^T\vec{x}_{\text{RMP}}^*\leq C$?};
\node [blockRed,below of=doneQAOA, node distance=1.27cm](upperBound) {Store upper bound $\vec{c}_{\text{RMP}}^T \vec{x}_{\text{RMP}}^*$ };

 \path [line] (input) -- (LP);
 
 \path [line] (IntSol)     -| (Fix) ;
 \path [line] (Fix)      |- (LP) ;
 \path [line] (exp2)      -- (QAOAcheck);
 \path [line] (QAOAcheck)   -- (QAOApre);
  \path [line] (QAOApre)   -- (QAOA);
 \path [line] (LP)       -- (ColGen);
 \path [line] (ColGen)     -- (ImprovingCols);
 \path [line] (ImprovingCols) -- (ColGenSol);
 \path [line] (ColGenSol)   -- (IntSol);
 \path [line] (IntSol)     --(sol);
 \path [line] (QAOA)      -- (QAOApost);
  \path [line] (QAOApost)   -- (solQAOA);
 \path [line] (solQAOA)    -- (doneQAOA);
 \path [line] (doneQAOA)    -- (upperBound);
 \path [line] (doneQAOA)    -- (sol);
 
 \draw [-] (solQAOA) to (5,-9.4); 
 \draw [line] (5,-9.4) |- (LP);

 \draw [-] (upperBound) to (5.1,-12); 
 \draw [line] (5.1,-12) |- (LP);
 
 \draw [-] (QAOAcheck) to (4.9,-4.4); 
 \draw [line] (4.9,-4.4) |- (LP);
 
 \draw [-] (ImprovingCols) to (1.8,-4.5);
 \draw [line] (1.8,-4.5) to (1.8,-3.55);
 
\end{tikzpicture}
\caption{High level depiction of the Branch-and-Price algorithm integrated with a quantum algorithm capable of finding solutions to ILPs. The variable $\bar{c}_r$ is the reduced cost of route $r$, $\vec{c}_{RMP}$ is the cost vector of an RMP instance, $\vec{x}_{RMP}^*$ is the solution provided by a quantum algorithm with postprocessing and the constant $C$ is a threshold for the accepted quality of a solution} 
\label{HeuristicBranchAndBound}
\end{figure}
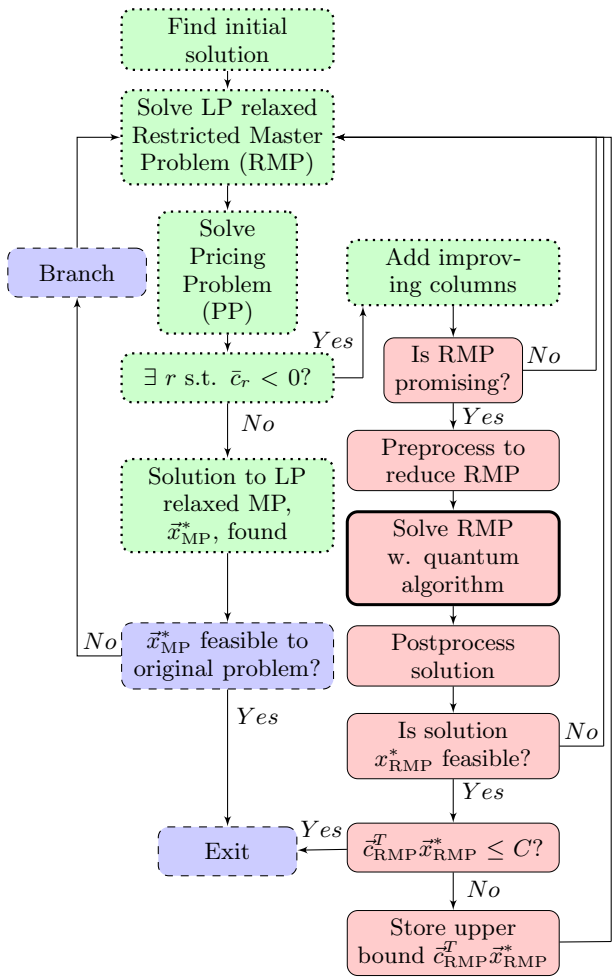

In this section we present the method where the Branch-and-Price algorithm is augmented by integrating any quantum algorithm capable of finding optimal or near-optimal integer solutions to RMP instances. The integrated Branch-and-Price algorithm is depicted in Fig.~\ref{HeuristicBranchAndBound} where Branch-and-Price is distinguished with green and blue colored boxes, and dotted and dashed borders. 
 The green boxes with dotted borders highlight the Column Generation algorithm, and the blue boxes with dashed borders are distinctive for the Branch-and-Bound algorithm.
 The red boxes with solid borders give the integration of a quantum algorithm. 
 
 The integrated method utilizes a quantum algorithm for each Column Generation iteration if the RMP is deemed promising. We remind the reader that since routes are generated dynamically by the Column Generation algorithm each iteration corresponds to a new ILP instance, which means that each iteration provides a possibility to find a new integer solution to the problem via a quantum (or classical) algorithm.
For example, we might want to avoid using a quantum algorithm in the beginning of the Column Generation process as it will, in general, be more likely to find good integer solutions in later iterations. 
However, determining how often to use a quantum algorithm will be a trade-off that depends on if the RMP instance is expected to contain integer solutions with reasonable quality, the run-time of the algorithm for practical instances, the quality of solutions the quantum algorithm can find and its potential to be used in parallel with the Branch-and-Price algorithm. 
Additionally, prior to utilizing a quantum algorithm classical preprocessing techniques are applied to the RMP instance and the  output of  the quantum algorithm is used as input to classical postprocessing techniques. We note that the method is similar to those explored in~\cite{Danna2005} and shares similarities to the use of a quantum device for scheduling problems in~\cite{Tran2016AHQ}. However, our proposed method is the first to our knowledge that considers the hybrid classical and quantum approach for large-scale ILPs \color{black} and is inspired by the integration of classical IP solvers for 0-1 integer programs into a generation and selection approach for large-scale ILPs in  \cite{PAQS1}\color{black}.
 
 The addition of a quantum algorithm can improve  the classical algorithm in several ways. 
Firstly, the quantum algorithm can provide a set of optimal or near-optimal integer solutions to RMP instances, which means that the quantum algorithm can be used as a primal heuristic in the Column Generation algorithm. This technique is sometimes referred to as the restricted master heuristic~\cite{ColGenPrimalHeuristics}. 
In the restricted master heuristic, a subset, which is a fixed number of columns and variables, is chosen from the RMP and the resulting problem is solved as a static Integer Program (IP). 
However, we do not wish to restrict the number of variables and columns to solve as a static IP. Instead, we propose to use the whole RMP instance unless we are required to leave out variables due to limitations in the size of a quantum device.
Such heuristics can improve solution quality as  observed in~\cite{NUNEZARES20161002} by simply obtaining optimal or near-optimal solutions to RMP instances. Furthermore, as primal heuristics have been shown to be very important for solving mixed integer programs, heuristics that leverage a quantum algorithm seems to be a natural step for Branch-and-Price. 
Moreover, by finding a set of integer solutions, some flexibility is introduced as it is possible to compare the quality of several solutions with respect to more parameters than each solution's cost. This is mainly an advantage for a real-world problem, where buffers occurring in solutions can improve sensitivity to disruptions.
%
%

Secondly, the quantum algorithm can provide tighter upper bounds in the branching step, which can be utilized in pruning decisions directly without sacrificing optimality. When we have access to tighter upper bounds, these bounds are compared to the lower bounds found in the Column Generation algorithm. If the lower bound is greater or equal to the upper bound, we can discard the subproblem as we can prune by bound. If we do not have access to these tighter upper bounds, more subproblems are created and explored. This means that the upper bounds can reduce the search tree's size, which leads to a reduced running time of the algorithm. The upper bounds can also reduce the number of iterations required in the Column Generation algorithm as noted in~\cite{Danna2005} by computing the Lagrangian lower bound, where the stopping criteria is given when the Lagrangian lower bound is greater than the best known upper bound. 

%
We can also consider introducing heuristic pruning rules that can reduce the running time of Branch-and-Price. 
We remark that finding a good solution fast can be preferable to finding the optimal solution for real-world problems. Heuristic pruning rules guided by optimal or near-optimal solutions to RMP instances can therefore be beneficial. 
However, as even optimal integer solutions to RMP instances do not guarantee an optimal solution to the subproblem in Branch-and-Price, the pruning decisions will be heuristic and do not guarantee an optimal solution. 
By introducing heuristic pruning rules, the goal is thus to obtain high quality solutions faster. 
The pruning decision can be determined by comparing the solution quality for different RMP instances by monitoring the iterative change in the objective and the LP lower bound gap.
%
If the Branch-and-Price is based on variable fixing decisions, the solutions from a quantum algorithm can indicate if certain variables can be chosen to be fixed. The procedure of fixing a variable is such that if a variable $x_i$ is set to 1 for a majority of the obtained solutions, the variable can be fixed to 1 and the Branch-and-Bound algorithm dives into this particular branch of the search tree. 
Further techniques as in RQAOA in~\cite{Bravyi_2020} can also be utilized where it is possible to find relations between two decision variables $z_i = \sigma_{ij} z_j$ where $\sigma_{ij} = \text{sign}(\matrixelement*{\vec{\gamma}^*, \vec{\beta}^*}{\hat{\sigma}_{i}^z\hat{\sigma}_{j}^z}{\vec{\gamma}^*, \vec{\beta}^*})$ and $(i,j)$ is an edge in the graph $G = (V, E)$ such that $(i,j) = \text{argmax}_{(i', j') \in E }\{
|\matrixelement*{\vec{\gamma}^*, \vec{\beta}^*}{\hat{\sigma}_{i'}^z\hat{\sigma}_{j'}^z}{ \vec{\gamma}^*, \vec{\beta}^*}| \}$ of an Ising model that encodes an ILP.
Such heuristic pruning rules would be similar to the ones of diving heuristics (which can be greedy, random or based on rounding strategies) or local branching heuristics~\cite{PrimHeurLocBranch}. 

Finally, the quantum algorithm can reduce the running time if it finds some integer solution below a given threshold or sufficiently close to the lower bound of the original problem as the algorithm, in that case, stops even though the search tree of Branch-and-Price is not explored fully.

Since the method is heuristic, the running time can best be evaluated by executing it on real problems and quantum devices, which is currently intractable due to the current size of quantum computers. We expect that as quantum hardware matures, such experiments will be of interest. 
We can, on the other hand, note that the general Branch-and-Bound algorithm has worst-case running time $\mathcal{O}(Mb^d)$ where $b$ is branching factor, $d$ is the search depth and $M$ is the upper bound on the running time to explore a subproblem fully. If we can obtain optimal or near-optimal integer solutions to subproblems, the number of nodes we can prune is larger and thus reduces the algorithm's running time. 

Furthermore, as the augmented algorithm is valid for any Branch-and-Price algorithm applied to problems with  Master Problems (MPs) possible to solve by some quantum algorithm, the method can be suitable for a large class of ILPs. In particular, this framework can be employed for airline planning problems such as Tail Assignment, Crew Pairing and Crew Rostering 
but also other large-scale ILPs such as vehicle routing problems~\cite{BaBVehicleRoutingProblems}. 

Whilst this approach prohibits applying a quantum algorithm to the Tail Assignment problem and other large-scale ILPs  directly, 
it reduces the number of required  decision variables and qubits. 
In particular, the MP for Tail Assignment, which is a Set Partitioning problem, is mapped such that the number of decision variables corresponds exactly to the required number of qubits (this is true also for other MPs that are 0-1 variable LPs  with equality constraints). 
We could map the Tail Assignment problem directly to an Ising model using an arc-based formulation (see in~\cite{Gronkvist} Eq. (4.1)-(4.7)), but this  would require $10^{7}$ qubits for a problem with $10^3$ flights and 10 aircraft prior preprocessing. For typical RMP instances we instead expect to require around $10^{3}-10^{4}$ decision variables for the path-based formulation in Eq.~\eqref{eq:sp1}-\eqref{eq:sp3}. The proposed method is thus much more suitable for NISQ computers.
The arc-based formulation has an additional disadvantage beyond the resource requirement of qubits for problems as Tail Assignment, which are the recursive maintenance requirements. These are non-trivial to map to an Ising model, and removing the constraints would likely result in infeasible solutions.

%
%
Moreover, we expect that if RMP instances can be solved approximately with sufficiently shallow circuit depth, the circuits can be realized on NISQ~\cite{Leymann_2020} computers. 
 In contrast to QAOA, Shor's algorithm has been given estimations where implementation requires 
 in the worst case one billion physical qubits~\cite{MoscaShor}
 and more recently $20\cdot10^6$ qubits in~\cite{gidney2019factor}. 

As mentioned earlier in this section, we propose a preprocessing step using classical integer programming techniques~\cite{10.2307/253992,presolve} in order to reduce the
number of variables and constraints of the problem prior to utilizing a quantum computer. Reducing the number of
variables (required qubits) and constraints (problem graph connectivity) is important for the limited
NISQ computers to be able to address real-world problems. The level of sophistication can range from very basic to
very advanced techniques and the level of sophistication used will be a trade-off between the computational time
of the preprocessor and the size and performance of the quantum computer.
%
%
We also consider classical postprocessing of the output from a quantum algorithm, where additional local searches can be done and we can combine good RMP solutions to obtain improved solutions with standard or specialized classical solvers.
Infeasible solutions can additionally be attempted to be corrected to feasible solutions by heuristic classical algorithms. 
%

We stress that the benefit in separating the original problem with the Branch-and-Price algorithm is that the master problem often is a pure Set Partitioning or Set Cover problem without any additional side constraints. The PP, on the other hand, is often a resource constrained shortest path problem that considers the complex rules. Thus, the method is not based on being more suitable for NISQ devices but is based on known successful methods for solving complex large-scale ILPs.
%
Furthermore, by simplifying a real-world problem to a pure Set Cover or Set Partitioning problem we also avoid tackling an ILP with potentially many complicated side constraints with quantum algorithms. This also means that the intricate task of balancing multiple constraint penalties is simplified. 

If the method is favorable for large-scale ILPs depends on how complicated the constraints are and the resource requirements of various formulations. The method proposed here can be expected to provide constant speed-up and improve the quality of the solutions. However, it is unclear if the method can provide polynomial speed-up as the addition of a quantum algorithm provides no guarantee for a speed-up and is tied to the Column Generation algorithm, which limits the possible speed-up we can expect.
If it is possible to use an alternate formulation that is not required to be separated into a generation problem and a selection problem, it might be beneficial to map the problem directly to an Ising spin glass Hamiltonian. 
However, as we have pointed out, this often requires significantly more decision variables and qubits to be applicable to real problems and will be more challenging for NISQ devices.
\section{The Quantum Approximate Optimization Algorithm \label{sec:qaoa}}
Farhi, Goldstone and Gutmann presented in~\cite{Farhi2014} QAOA, which is a hybrid classical and quantum variational algorithm capable of finding approximate solutions to combinatorial optimization problems. The algorithm is inspired by the adiabatic quantum algorithm but is designed for gate-based quantum computers. Furthermore, evidence that a classical computer can not simulate a QAOA circuit without exponential overhead was presented later in~\cite{farhi2019quantum}.
The algorithm consists of a Trotterized approximation to the time evolution which alternates the operators $e^{-i\gamma_k \hat{H}_f}$ and $e^{-i\beta_k\hat{H}_M}$ for $k=1, 2, \dots, p$, where $p$ is the depth of the algorithm. 
An ideal QAOA circuit applied to the initial state $\ket{+}=\frac{1}{\sqrt{2^n}}\sum_{i=0}^{2^{n}-1}\ket{i}$ gives the QAOA state for depth $p$
\begin{align*}
 |\vec{\gamma}, \vec{\beta} \rangle &= e^{-i\beta_p \hat{H}_M} e^{-i\gamma_p \hat{H}_f} \dots e^{-i\beta_1 \hat{H}_M} e^{-i\gamma_1 \hat{H}_f} \ket{+}
\end{align*}
where $\hat{H}_M= \sum_{i=1}^n \hat{\sigma}_{i}^x$ is the mixing Hamiltonian and $\hat{H}_f = \sum_{\vec{x} \in \{0, 1\}^n} f(\vec{x}) \ket{\vec{x}}\bra{\vec{x}}$ is a diagonal cost Hamiltonian with respect to the computational basis. The cost Hamiltonian encodes an objective function $f(\vec{x})$ which represents a combinatorial optimization problem. With optimal angles $\vec{\gamma}^*$ and $\vec{\beta}^*$ and sufficiently large algorithm depth, the QAOA state should have a large proportion in states that are close to the ground state and equal to the ground state. 
By repeating the process of constructing the state and performing measurements in the computational basis, a solution that is equal or close to the ground state of the cost Hamiltonian can be found. 

The sum of the angles $\sum_{k=1}^p |\gamma_k| + |\beta_k|$, also referred to as variational time parameters, is proportional to the total running time to execute the quantum circuit, as the implementation of the gates  associated to the cost Hamiltonian is graph and hardware architecture dependent. If we let the algorithm depth go to infinity and restrict the angles to be small, the algorithm becomes exact~\cite{Farhi2014}. 

For an ILP problem $\hat{H}_f$
will consist of one partial Hamiltonian that corresponds to the objective function and another that corresponds to constraints, not unlike common penalty methods~\cite{NoceWrig06}. 
If $f(\vec{x})$ represents a minimization problem the optimal angles $\vec{\gamma}^*$ and $\vec{\beta}^*$ can be found by solving the classical optimization problem %
\begin{align}
 \text{argmin } & 
 \matrixelement*{\vec{\gamma}, \vec{\beta}}
 {\hat{H}_f}
 {\vec{\gamma}, \vec{\beta}}, \label{eq:classicalOpt1}
 \\
 \text{subject to } & \gamma_i \in [0, 2\pi] \ \forall i=1,..,p,\label{eq:classicalOpt2a}
 \\
 & \beta_i \in [0, \pi] \ \forall i=1,..,p \label{eq:classicalOpt2}
\end{align}
as $\matrixelement*{\vec{\gamma}, \vec{\beta}}{\hat{H}_f}{\vec{\gamma},\vec{\beta}}=f(\vec{x}^*)$ if $|\vec{\gamma}, \vec{\beta}\rangle = \ket{\vec{x}^*}$ where $\vec{x}^*$ is the optimal solution to the problem $f(\vec{x})$ represents. The function in Eq.~\eqref{eq:classicalOpt1} is the expectation value function and can be referred to as the energy landscape. The domain in Eq.~\eqref{eq:classicalOpt2a} and~\eqref{eq:classicalOpt2} holds for Hamiltonian $\hat{H}_f$ with integer eigenvalues~\cite{vikstl2019applying}.

As far as we know, instances extracted from the real-world problem Tail Assignment has previously only been studied for QAOA in the context of Exact Cover in~\cite{vikstl2019applying} where success probabilities close to unity for instances up to 25 qubits with one feasible solution could be obtained for $p\leq 20$ for ideal QAOA circuits. 
Recently, the vehicle routing problem was also studied up to 20 qubits~\cite{utkarsh2020solving} where a clear dependency was established between the problem size and the performance of QAOA. On the other hand, real-world problems have been studied for quantum annealing, such as for flight gate assignment in~\cite{stollenwerk2018flight}, where the authors address the issue of bin packing the cost vector of the objective function. However, the complication of degenerate problem instances have not been discussed to a large extent in the context of QAOA, nor has much focus been given to how suitable weights are found to balance the constraints and the objective part of the Hamiltonian  $\hat{H}_f$. In Sec.~\ref{sec:results}, we focus on the effect of choosing suitable weights on the required algorithm depth given a success probability and if having a large feasible space is a limiting factor for the performance. 
\subsection{Mapping Set Partitioning and Exact Cover \label{sec:mappingArticle}}
It is possible to map the Set Partitioning and Exact Cover problem to the Ising spin glass Hamiltonian with an underlying graph $G=(V, E)$ with  nodes  given by the set $V$ and the edges  given by the set $E$, where the Hamiltonian is $\hat{H}=\sum_{i=1}^{|V|}h_i \hat{\sigma}_i^z + \sum_{(i,j) \in E} J_{ij}\hat{\sigma}_i^z\hat{\sigma}_j^z $  as presented in~\cite{Lucas2014}. 
In this case, the Hamiltonian has at most two spin interaction terms $\hat{\sigma}_i^z\otimes\hat{\sigma}_j^z$ and is 2-local~\cite{Gharibian_2015}, albeit this does not correspond to a geometric locality with respect to hardware architecture. 


By introducing a quadratic penalty on the constraints in Eq.~\eqref{eq:sp2} a nonlinear integer optimization problem is obtained. The quadratic penalty results in a Hamiltonian which has two parts (when ignoring a constant energy shift), a Hamiltonian which is related to the objective function and a Hamiltonian related to the constraints. These parts are weighted with constants $\mu_1$ and $\mu_2$ accordingly
\begin{align*}
\hat{H}^{\text{Set Partitioning}}&= \sum_{r\in R} \text{\Large[}\mu_1\cdot h_r^{\text{Objective}} \\
& + \mu_2\cdot h_r^{\text{Exact Cover}}\text{\Large]}\hat{\sigma}_r^z 
\\
 & + \mu_2\cdot\sum_{r'>r} J_{rr'}^{\text{Exact Cover}}\hat{\sigma}_r^z \hat{\sigma}_{r'}^z,
\end{align*}
where 
\begin{align*}
 & h^{\text{Objective}}_r = \frac{c_r}{2},
 \\
 & h^{\text{Exact Cover}}_r = \sum_{f \in F} a_{fr}\left(\sum_{r' \in R}\frac{a_{fr'}}{2} -1 \right)
\end{align*}
and 
\begin{align*}
 & J^{\text{Exact Cover}}_{rr'} = \sum_{f \in F}\frac{a_{fr} a_{fr'}}{2}.
\end{align*}
We observe that the terms $h^{\text{Objective}}_r$ are given by the objective function in Eq.~\eqref{eq:sp1} and therefore indicate the cost of an assignment of the decision variables $\vec{x}\in \{0, 1\}^{|R|}$. The terms $h^{\text{Exact Cover}}_r$ and $J^{\text{Exact Cover}}_{rr'}$ are due to the constraints in Eq.~\eqref{eq:sp2}, where $J^{\text{Exact Cover}}_{rr'}$ gives a penalty for each overlapping flight in route $r$ and $r'$ and with the terms $h^{\text{Exact Cover}}_r$ gives a penalty if the combination of routes in an assignment does not cover all flights.

The problem graph $G=(V, E)$ is given by the coefficients $h^{\text{Exact Cover}}_r, h^{\text{Objective}}_r$ and $J^{\text{Exact Cover}}_{rr'}$ in the Hamiltonian where the graph itself can be thought also as a conflict graph of the variables. Finally, the detailed mapping of Exact Cover to an Ising spin glass model was presented in~\cite{vikstl2019applying} and further expanded for the mapping of the Set Partitioning problem in Appx.~\ref{sec:mapping}. Mappings for other minimization problems common for large-scale ILPs such as Set Cover can also be found in~\cite{Lucas2014}.
\section{Problem Instances \label{sec:DataInstances}} 
The instances~\cite{datainstances} have been extracted from the real-world problem Tail Assignment by finding a set of different integer solutions when executing the heuristic Branch-and-Price algorithm. The different solutions are found by permuting the cost of routes randomly during the execution of the algorithm. From this set, 35 instances have been constructed with varying number of routes and number of feasible solutions by combining complete and partial solutions. 

Typically, the instances have very large costs and can be as large as $10^6$, making the energy landscape numerically hard to search. 
The objective function has therefore been further simplified to study qualitative differences in the performance of RMP instances for QAOA. 
The costs have been simplified such that the smallest cost $c_r^{\text{min}}$ is set to 1, larger costs have been modified such that each cost $c_r$ has a unique value and that the optimal solution is unique. 
For real instances this is not a proposed methodology, as it can disturb the order of the solutions with respect to quality significantly. An option for real instances is to either increase the weight for the penalty of the constraints, which results in a numerically challenging energy landscape to optimize or we can disturb the costs such that they are easier to handle but preserves the objective function with some accuracy. 

We can modify the costs by subtracting all costs with a constant and dividing all costs with another constant, finally the costs are rounded to integers. There is a limit to how much we can disturb the costs such that the order of solutions with respect to cost is not changed significantly. One should choose to divide by a constant that separates the costs $c_r$ by at least a constant integer, which results in a better preservation of the objective function compared to choosing a larger constant to divide the costs by. Here we have assumed a simple objective function to study the performance of QAOA. 

The instances are identified by the number of decision variables $|R|$ and the number of feasible solutions $|S_{\text{feasible}}|$. The number of decision variables are 6, 8, 10, 12, 14 and 20. The number of feasible solutions vary from 1 to $|R|/2$. We denote a problem graph associated to an instance $G_{r=|R|}^{s=|S_{\text{feasible}}|}$ which gives the set of graphs as
\begin{equation*}
 \{ G_{r}^{s} \}_{r=6-20}^{s=[r/2]}.
\end{equation*}
Additionally, in~\cite{vikstl2019applying} it was observed that the average node degree of the problem graphs affects the performance of QAOA, in that obtaining near unity success probability require greater algorithm depth as the average node degree, $\langle d_G(v)\rangle$, of the problem graphs increases. 
 \begin{figure}[h!]
 \centering
 \begin{subfigure}[b]{0.37\textwidth}
 \centering
 \includegraphics[trim=10 20 10 20, clip,width=\textwidth]{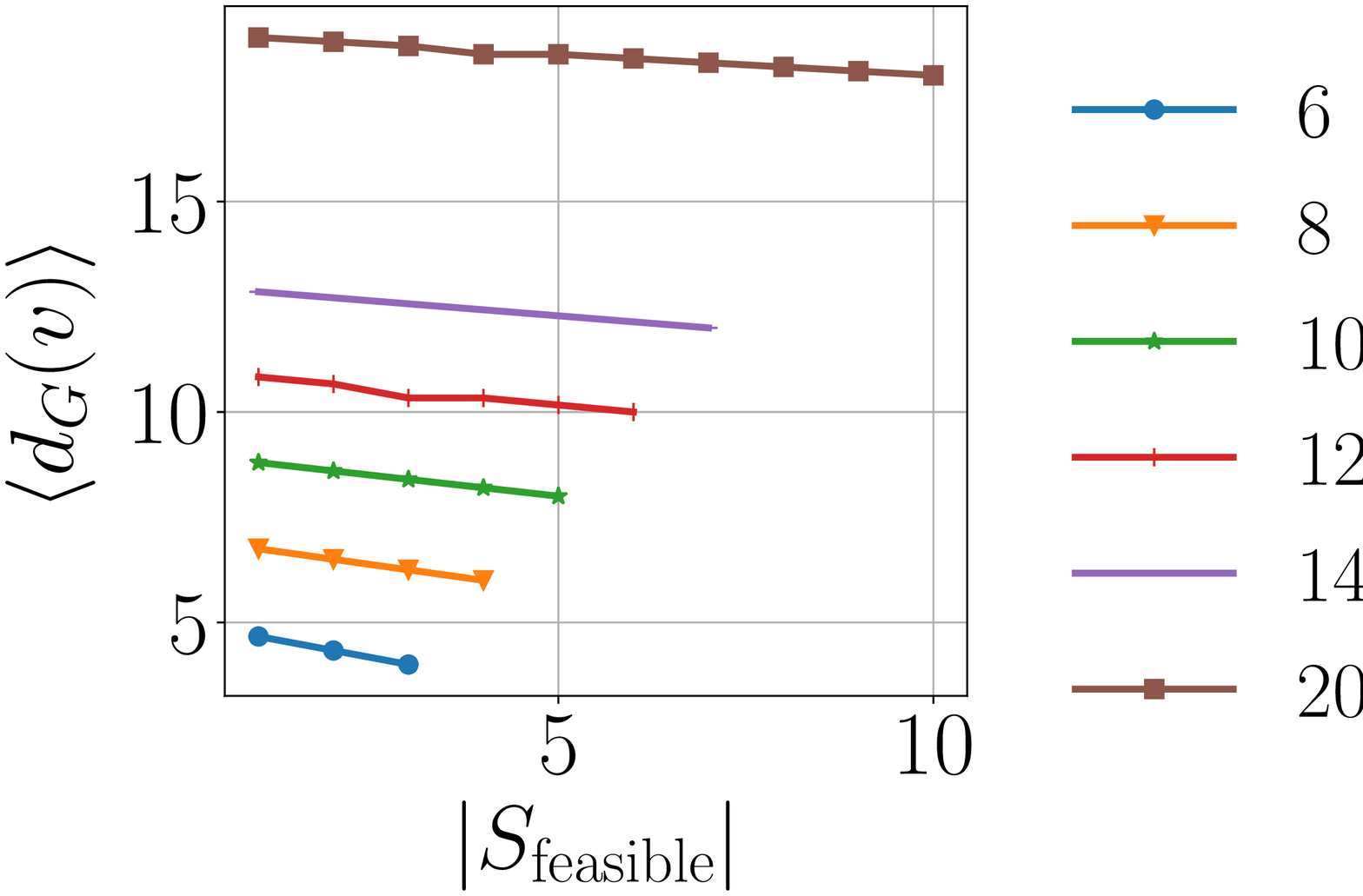}
 \caption{Average node degree of problem graphs for generated instances}
 \label{fig:node_degree}
 \end{subfigure}
\begin{subfigure}[b]{0.1\textwidth}
         \centering
         \includegraphics[trim=40 40 40 40, clip,width=\textwidth]{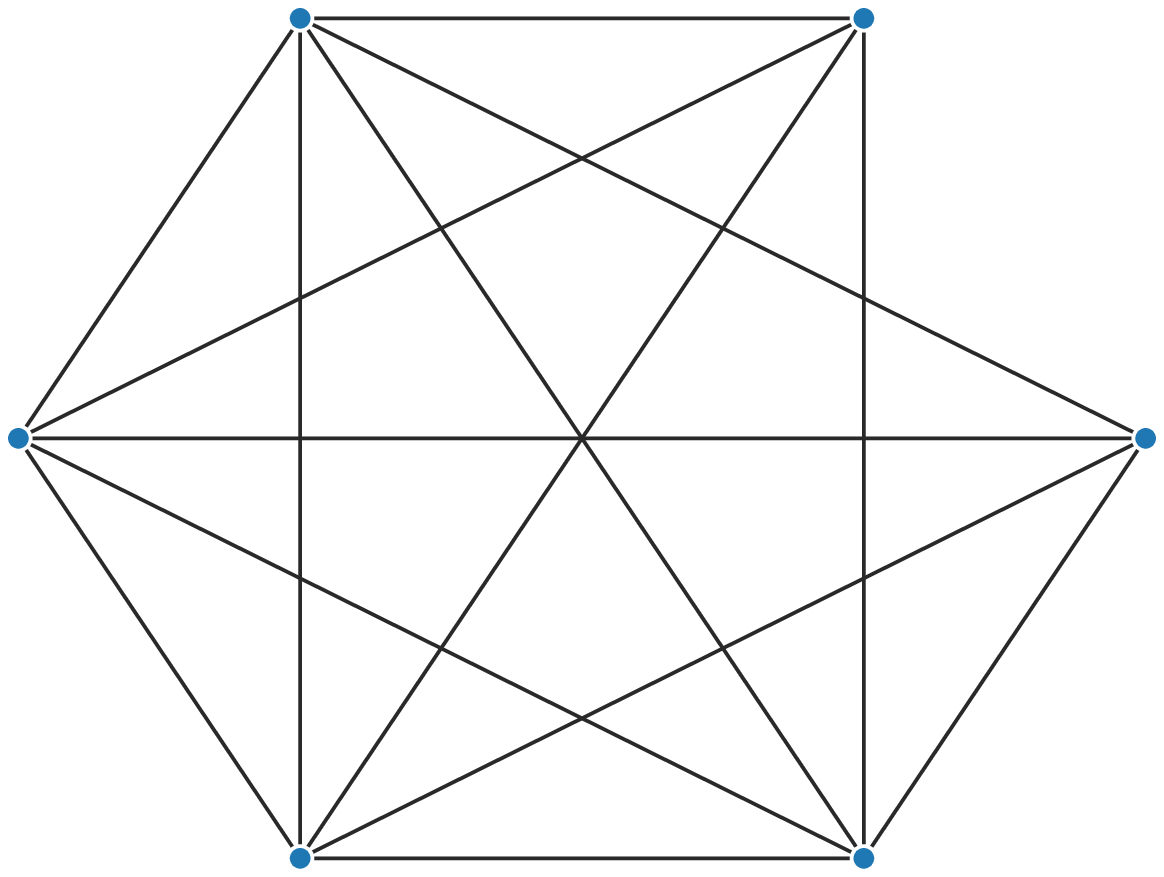}
          \includegraphics[trim=40 40 40 40, clip,width=\textwidth]{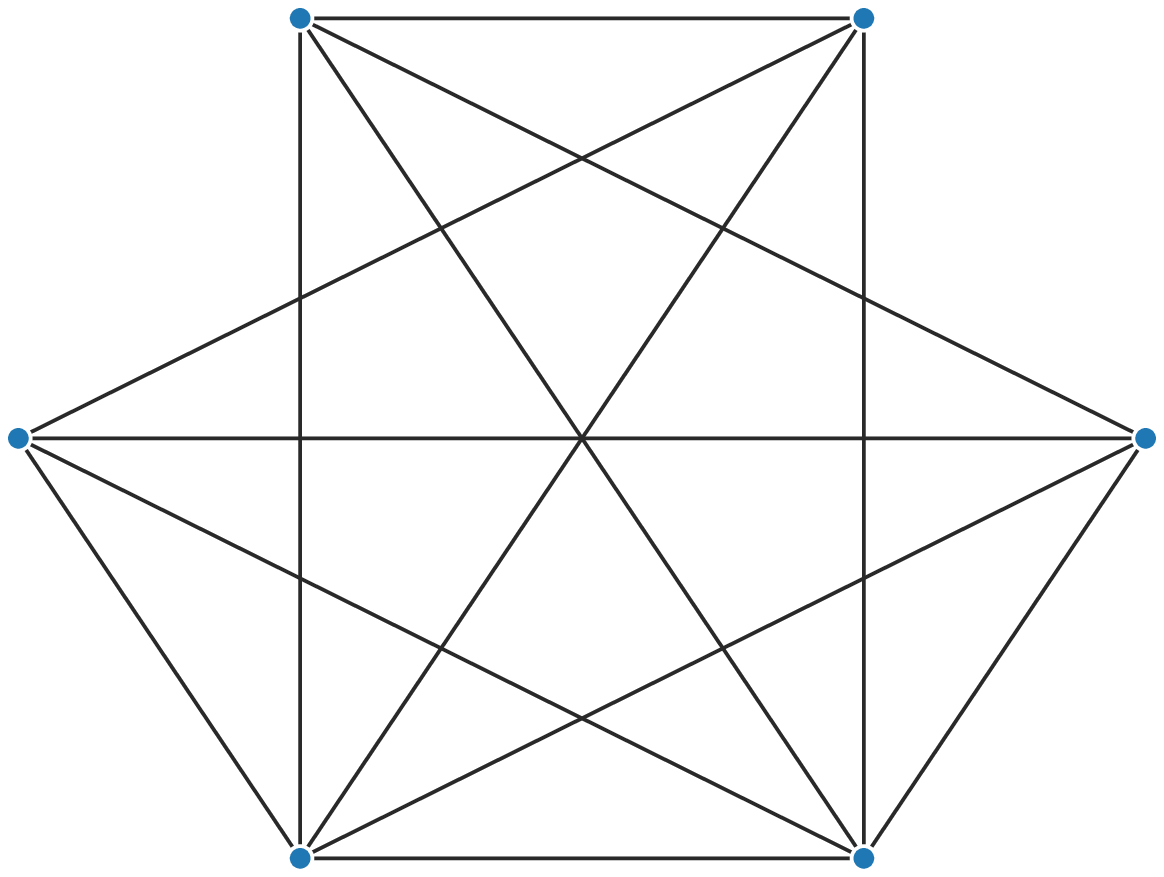}
         \includegraphics[trim=40 -10 40 40, clip,width=\textwidth]{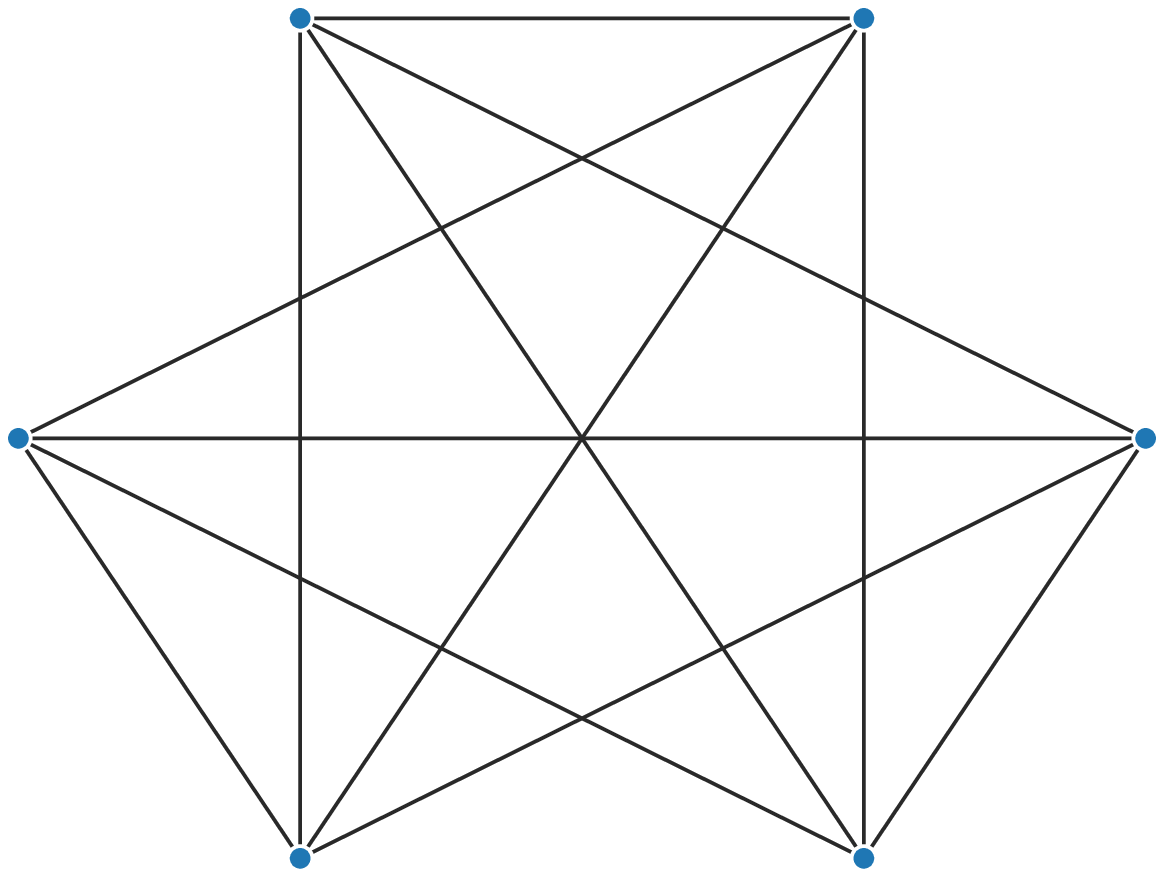}
 \caption{Problem graphs}
 \label{fig:graphs}
 \end{subfigure}
 \caption{Average node degree of the problem graphs  are shown in Fig.~\ref{fig:node_degree}. Problem graphs $\{ G_{r}^{s} \}_{r=6}^{s=1,2,3}$ are depicted in Fig.~\ref{fig:graphs}}
 \label{fig:problem_graph_densities}
\end{figure}
The effect of the average node degree was found to dominate over the problem size such that for a given success probability, the required algorithm depth was greater for instances with 15 qubits compared to instances with 25 qubits.
We have extracted the average node degree of each problem graph, depicted in Fig.~\ref{fig:problem_graph_densities}. It can be noted that the average node degree increases with the problem size and decreases as the number of feasible solutions increases. We further noted that the problem graphs are close to being complete graphs, i.e., each node's degree is $|R|-1$ or $|R|-2$. It is thus expected that such instances are hard for QAOA to solve with respect to problem size.
\section{Optimization Strategy\label{sec:OptInterpolation}}
Finding the solution to the optimization problem in Eq.~\eqref{eq:classicalOpt1}-\eqref{eq:classicalOpt2} is NP-hard~\cite{rieffel2019ansatze,bittel2021training} in itself. Furthermore, each query of the function in Eq.~\eqref{eq:classicalOpt1} requires either executing the QAOA circuit on a quantum device or a simulation on a classical computer. As we are currently prohibited from executing QAOA for the problem instances with sufficient algorithm depths on a quantum device the remaining option is to simulate the algorithm with a classical computer. Moreover, since simulating the quantum circuits is exponential in the number of qubits, the consequence is that a function evaluation is computationally expensive. Furthermore, in order to study the performance of QAOA more accurately, we wish to study intermediate to large algorithm depths, which makes the simulations even more expensive as the dimension of the expectation value function in Eq.~\eqref{eq:classicalOpt1} is 2 times the algorithm depth. 

Compared to problems as MaxCut with uniform weights set to 1 or versions thereof~\cite{Farhi2014,willsch2019,Wang_2018} the Set Partitioning problem and Exact Cover problem have coefficients in the Hamiltonian $h_i$ and $J_{ij}$ that are governed by the constraint matrix and objective function that grow with the chosen weights. These coefficients are thus not constrained to 0,1 or -1 and can be large. 
The difference in coefficients results in complicated energy landscapes, that oscillate rapidly, to optimize with multiple local minima. Moreover, we can see this from the closed form expression of the energy landscape for $p=1$ for an Ising spin glass Hamiltonian, associated to a graph $G=(V, E)$ with edge weights $J_{ij}$ and node weights $h_i$, which is given by 
 \begin{align}
   \bra{\gamma \beta}\hat{H} \ket{\gamma \beta} =& 
   \sum_{i=1}^n h_i \text{sin}(2\beta)\text{sin}(2\gamma h_i)
   \smashoperator{\prod_{j:(i, j)\in E}} \text{cos}(2\gamma J_{ij}) \nonumber
   \\ 
  +\smashoperator{\sum_{(i, j)\in E}}\frac{J_{ij}}{2} & \text{\Huge(}\text{sin}^2(2\beta)
  \smashoperator{\prod_{\substack{(i,k)\in E\\ (j,k)\notin E}}}
    \text{cos}(2\gamma J_{ik})
  \smashoperator{\prod_{\substack{(j,k)\in E \\ (i,k) \notin E}}}
    \text{cos}(2\gamma J_{jk}) 
   \nonumber \\ 
   \times  \text{\Huge[}
   &
   \text{cos}(2\gamma(h_i-h_j)) 
   \smashoperator{ \prod_{\substack{(i,k)\in E \\(j,k)\in E}}} 
   \text{cos}(2\gamma (J_{ik} - J_{jk}))
   \nonumber \\ 
   -&\text{cos}(2\gamma(h_i+h_j))
  \smashoperator{ \prod_{\substack{(j,k)\in E \\(i,k) \in E}}}
  \text{cos}(2\gamma (J_{ik} + J_{jk}))
   \text{\Huge]}
   \nonumber \\ 
   +\text{sin}&(4\beta)\text{sin}(2\gamma J_{ij})\text{\Huge[} 
   \text{cos}(2\gamma h_i) \smashoperator{\prod_{k\neq j:(i,k)\in E}}\text{cos}(2\gamma J_{ik}) \nonumber
   \\ 
   &\ \ \ \ \ \ +\text{cos}(2\gamma h_j) \smashoperator{\prod_{l\neq i:(j,l)\in E}} \text{cos}(2\gamma J_{jl}) 
   \text{\Huge])}\label{eq:expvalend},
 \end{align}
 as presented in~\cite{ozaeta2020expectation}. We derive the expression for consistency in Appx.~\ref{sec:expvalproof}. 
 
The complicated energy landscape underlies our motivation to focus on
obtaining good locally optimal angles via the interpolation strategy presented by Zhou, in~\cite{Zhou_2020}, in order to study the success probability for QAOA with intermediate to large algorithm depth $p$. 
The first step in the interpolation algorithm is to perform global optimization for algorithm depth $k=1$ and for algorithm depth $k>1$ locally optimal angles $(\vec{\gamma}^{L^*}, \vec{\beta}^{L^*})$ angles are found by providing a good starting point $(\vec{\gamma}^{L}, \vec{\beta}^{L})$ to a local search algorithm. The starting point for local search is determined by interpolating previously found locally optimal angles. The algorithm iterates for $k=2,\dots, p$. The following definition gives the interpolation in each step
\begin{align*}
  & \eta_{k+1, i}^L = 
  \begin{cases}
  \eta_{k, 1}^{L^*} & \text{ if } i=1
    \\
    \frac{i-1}{k}\eta_{k, i-1}^{L^*} + \frac{k-i+1}{k}\eta_{k, i}^{L^*} & \text{ if } i=2,\dots,k
    \\
    \eta_{k, k}^{L^*} & \text{ if } i = k+1
 \end{cases}
\end{align*}
where $\eta$ is $\gamma$ or $\beta$. The index $i$ denotes the $i$:th element of locally optimal angles found for algorithm depth $k$ and index $k$ denotes the best found angles of algorithm depth $k$. The distinction between $L$ and $L^*$ is the separation of the starting point and angles found after a local search. In our case, the global optimization was performed with python's differential evolution routine. The local optimization was performed with L-BFGS-B, which is also a standard solver in python.
\section{Numerical Results for Restricted Master Problem Instances\label{sec:results}}
We present the numerical results obtained for ideal QAOA circuits where the variational parameters have been obtained via the interpolation strategy first for Exact Cover in Sec.~\ref{sec:resultsEC} and second for Set Partitioning in Sec.~\ref{sec:resultsSP}. 
\subsection{Solving the Exact Cover problem\label{sec:resultsEC}}
\begin{figure}[h!]
         \includegraphics[trim=10 15 20 19,clip,height=3.3cm]{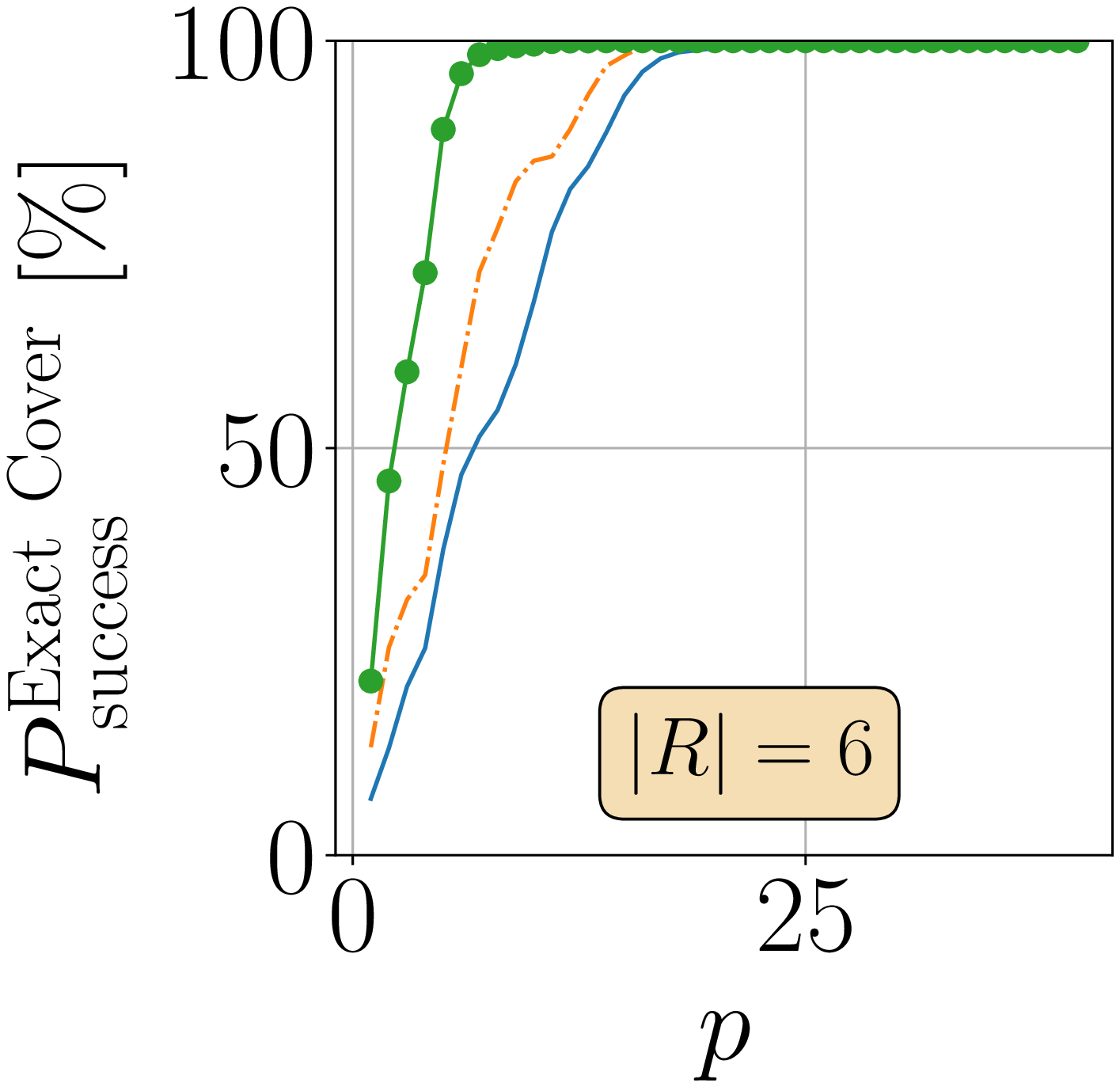}
         \includegraphics[trim=125 15 20 19,clip,height=3.3cm]{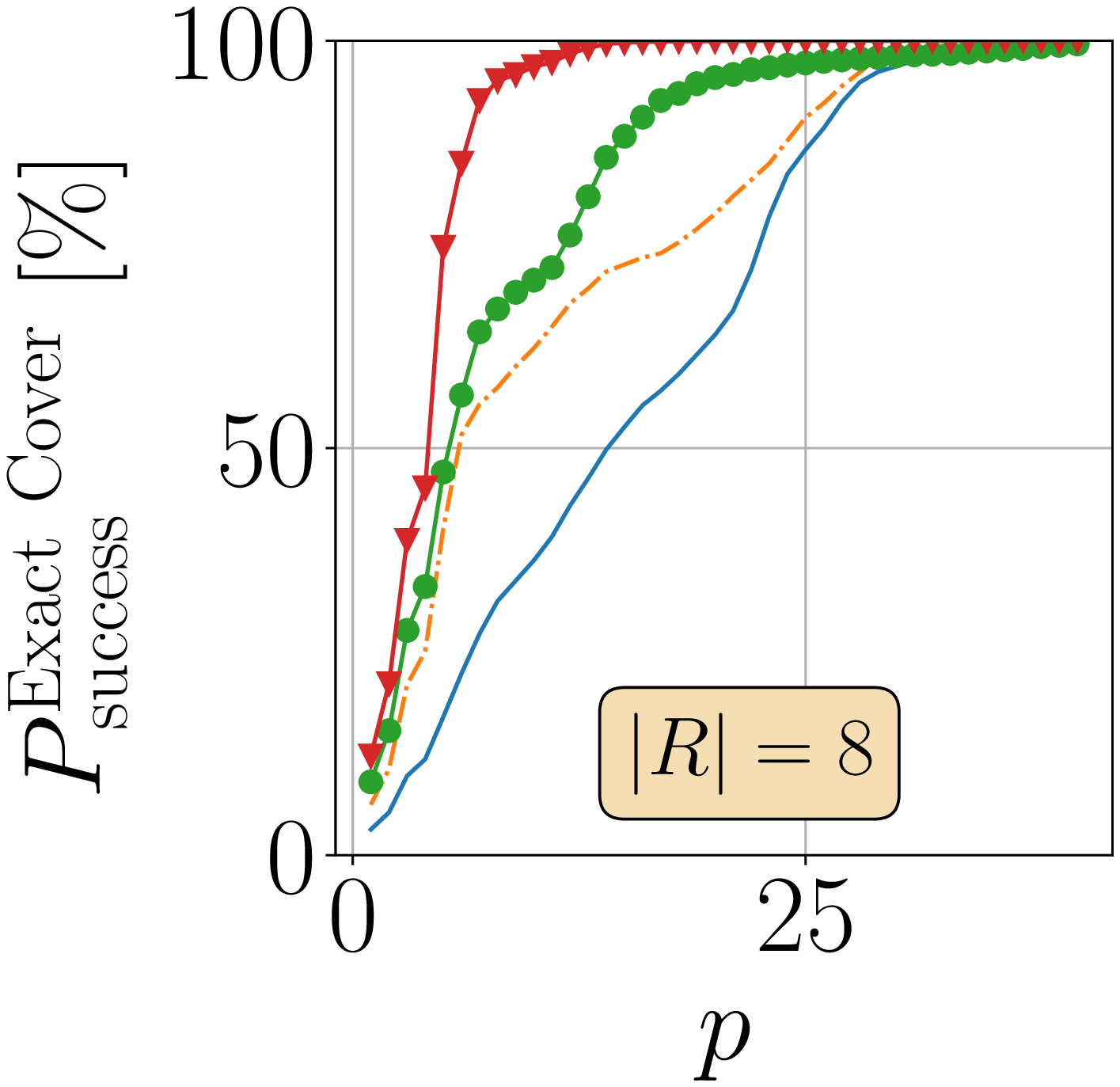}
         \includegraphics[trim=125 15 20 19,clip,height=3.3cm]{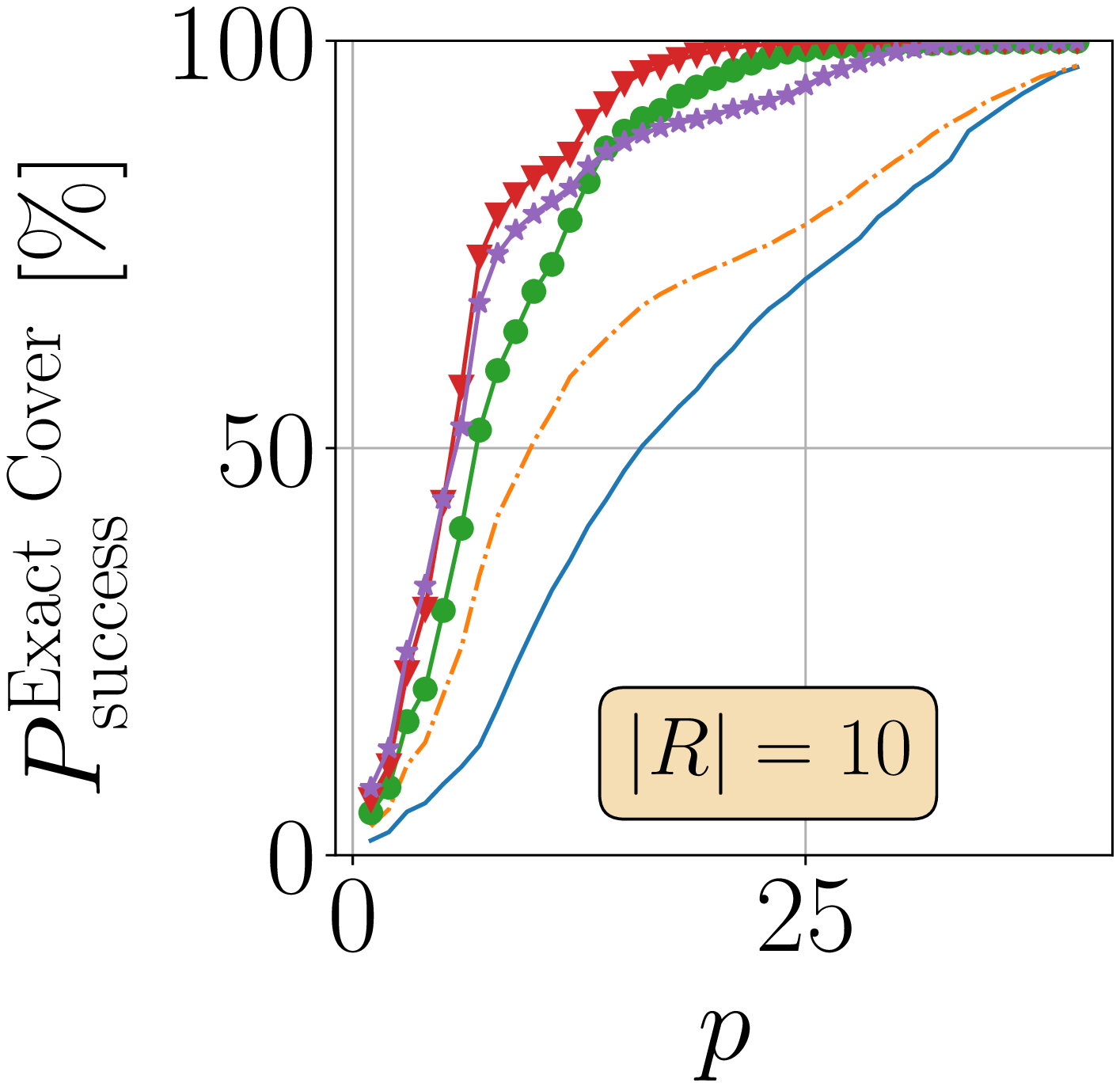}
         \includegraphics[trim=10 15 20 19,clip,height=3.3cm]{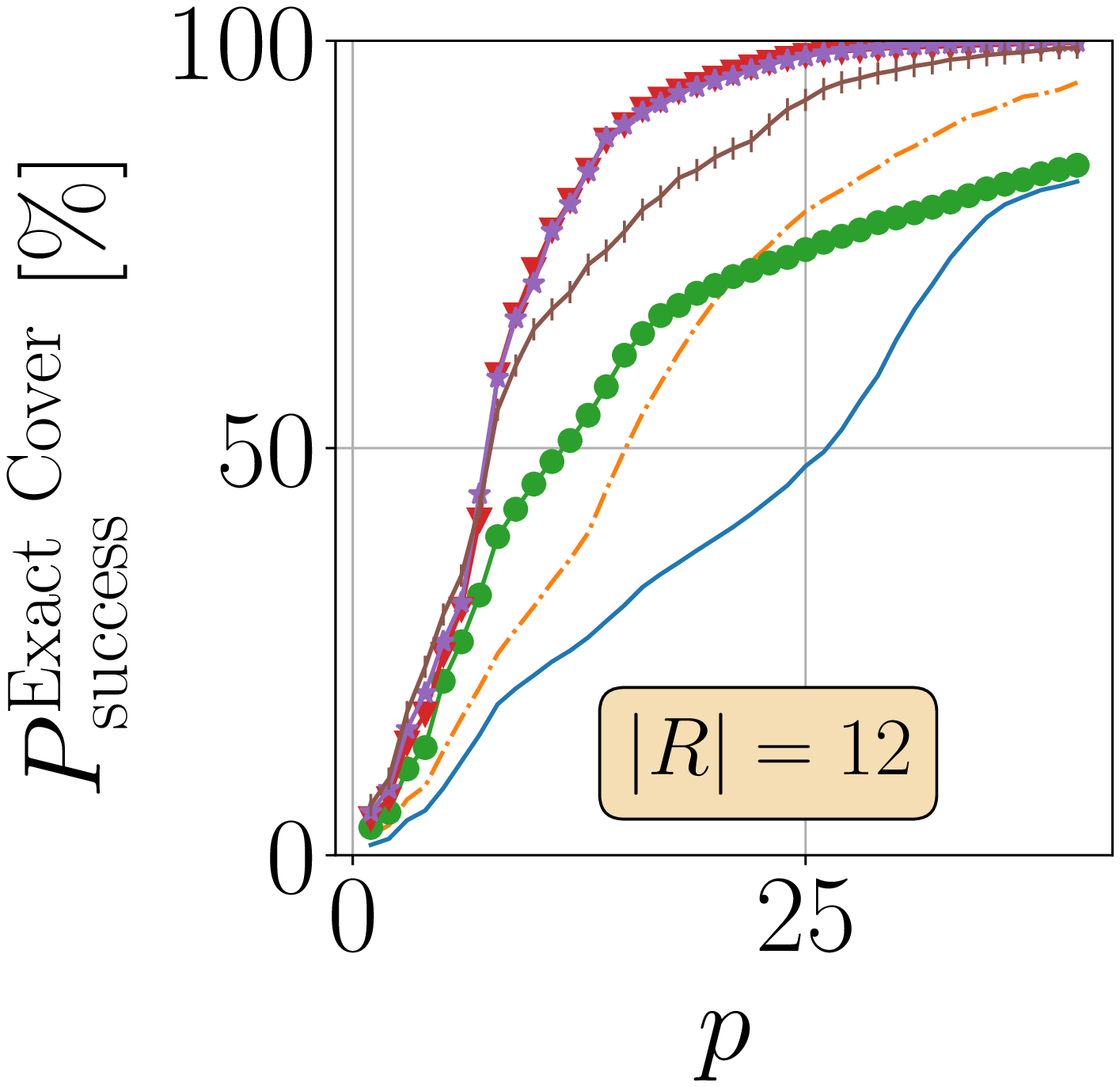}
         \includegraphics[trim=125 15 20 19,clip,height=3.3cm]{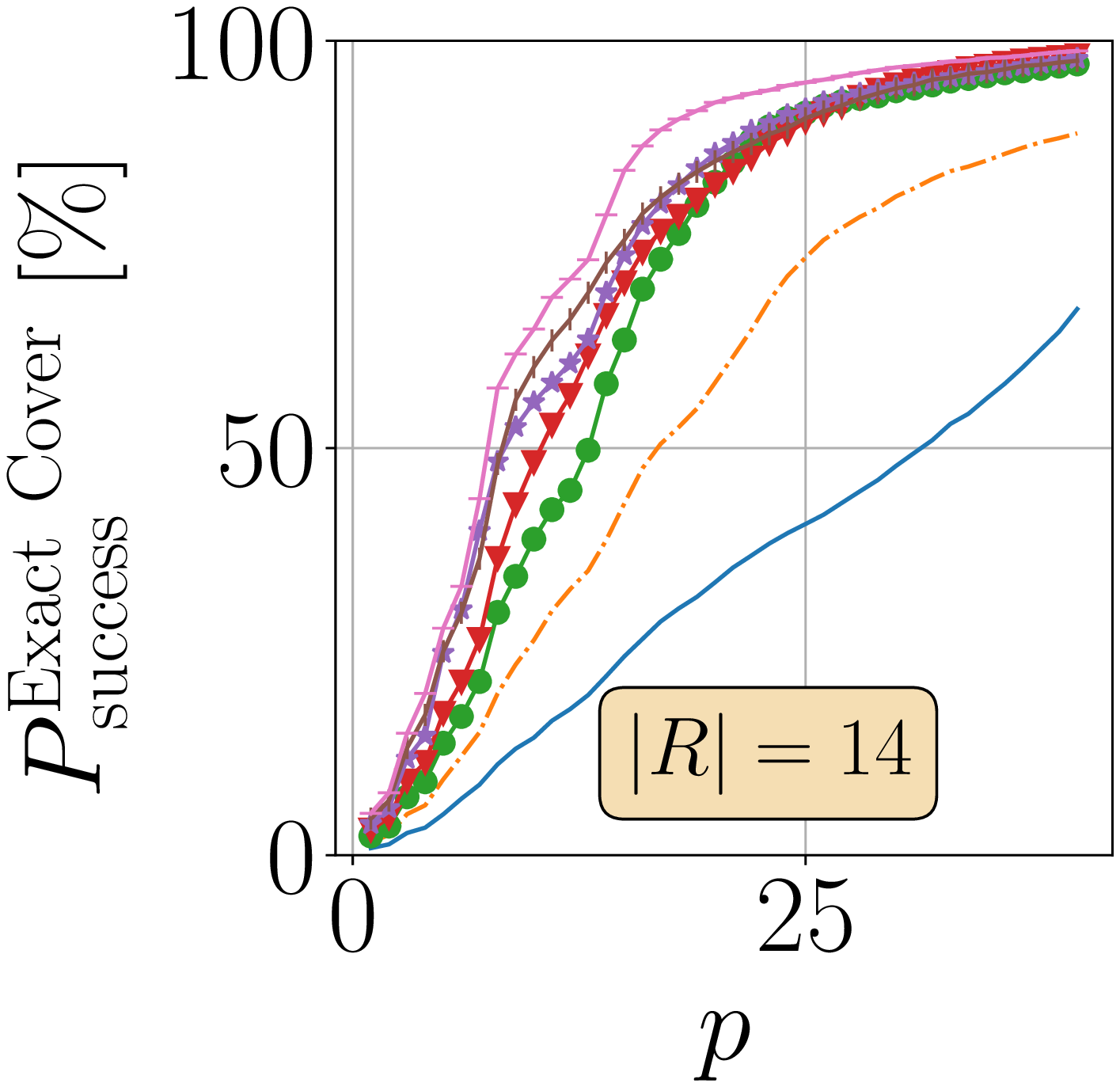}
         \includegraphics[trim=125 15 17 19,clip,height=3.3cm]{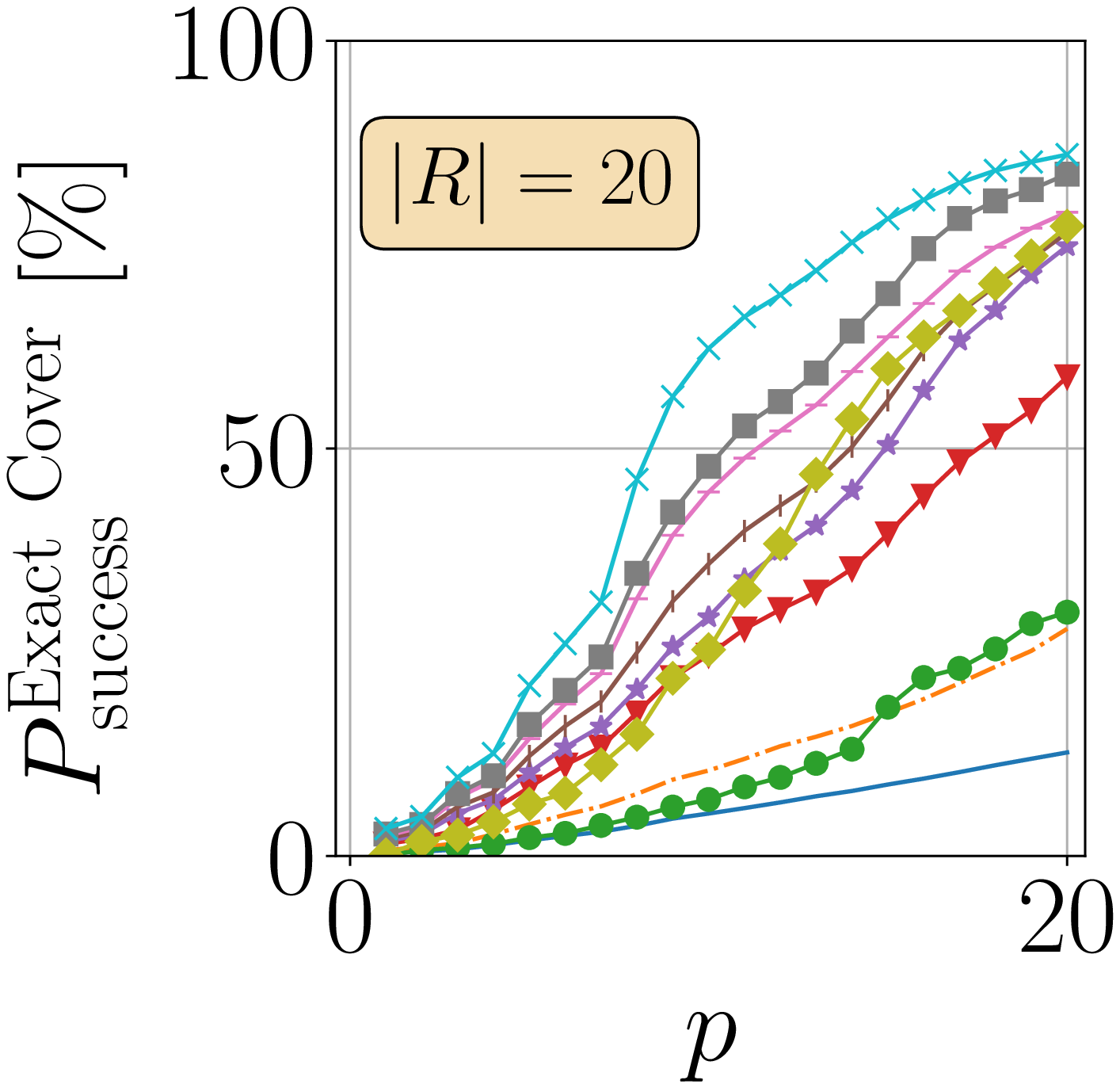}
         \includegraphics[trim=-100 80 100 0 ,clip,width=0.5\textwidth]{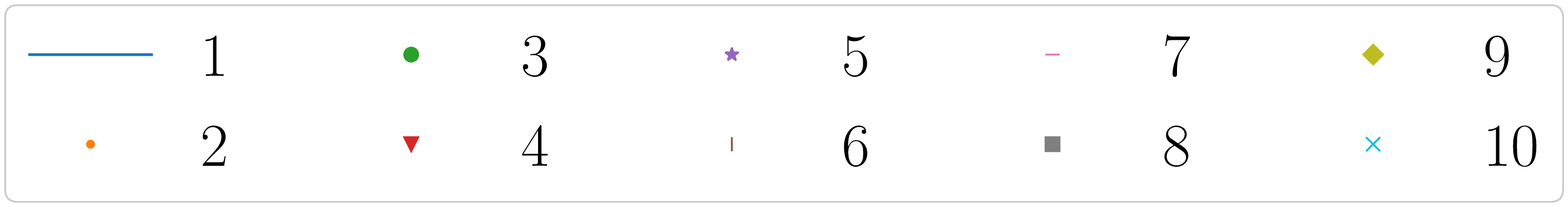}
         \caption{ Success probabilities for Exact Cover. The colors and markers indicate  $|S_{feasible}|$}
     \label{fig:EC_interpolation6_20}
\end{figure}

For Exact Cover,  we only require to obtain a feasible solution  $\vec{x}_i \in S_{\text{feasible}}$. For such purpose, the most natural choice of mapping is by ignoring the objective part of the Hamiltonian, i.e., the cost Hamiltonian is expressed as 
\begin{equation*}
    \hat{H}_f = \hat{H}^{\text{Exact Cover}}.
\end{equation*}
Furthermore, it is straightforward to define the success probability as the probability of obtaining any of the feasible solutions
\begin{align*}
       P_{\text{success}}^{\text{Exact Cover}} =   
       \sum_{\vec{x}_i \in S_{\text{feasible}}} 
       |\langle\vec{x}_i|\vec{\gamma}^{L^*}, \vec{\beta}^{L^*}\rangle|^2.  
\end{align*}
The success probabilities for QAOA applied to the Exact Cover instances are plotted in Fig.~\ref{fig:EC_interpolation6_20}. 
We remark that the required algorithm depth decreases for a given success probability as the number of feasible solutions increases in general, albeit increases when the problem size increases. 
These results correspond to results found in~\cite{vikstl2019applying}, wherein  Fig.~\ref{fig:problem_graph_densities} we presented the average node degree of the instances, that decreases with the number of feasible solutions whilst increasing more significantly as the problem size increases. 

As the most challenging  cases are those where the number of feasible solutions is small, we observe that obtaining a success probability above 50\%  can require an algorithm depth that is more than $|R|$ by at least a constant, where $|R|$ is the number of decision variables and qubits of the instance. Therefore, it is unknown how well QAOA can perform for instances with $10^3-10^4$ decision variables when executed on a NISQ device as decoherence is a limiting factor currently.
\subsection{Solving the Set Partitioning problem\label{sec:resultsSP}}
When we consider applying QAOA to the Set Partitioning problem, two additional aspects are of interest. The first aspect is  how one should choose good weights that balance the objective part of the Hamiltonian and the Exact Cover (constraints given by a quadratic penalty) part of the Hamiltonian. The second aspect is a consequence of the first, namely how the chosen weights affect the required algorithm depth for a given success probability.
The total cost Hamiltonian is a combination of the two partial Hamiltonians accordingly
\begin{equation*}
  \hat{H}_f = \hat{H}^{\text{Set Partitioning}} = \mu_1 \hat{H}^{\text{Objective}} + \mu_2\hat{H}^{\text{Exact Cover}}.
\end{equation*}
We have chosen the weight $\mu_1 \in \{\mathds{Z}^+\cup\{0\}\}$ depending on a factor $f$ 
\begin{align*}
  & \mu_1 =
  \begin{cases}
  0 &\text{ if } f=\infty \\
  1 &\text{ otherwise} 
  \end{cases},
\end{align*}
and $\mu_2 \in \mathds{Z}^+$  depending on the largest eigenvalues of the partial objective and Exact Cover Hamiltonians, and factor $f$ 
\begin{align*}
  & \mu_2 = 
  \begin{cases}
  1 & \text{ if } f=\infty
  \\
  \left\lfloor f\cdot \frac{\lambda_{\text{Objective}}^{\text{max}}}{\lambda_{\text{Exact Cover}}^{\text{max}}} \right \rceil 
   & \text{ otherwise} 
  \end{cases}.
\end{align*}
By choosing the weights to be integers, the domain is preserved in the optimization problem defined in Eq.~\eqref{eq:classicalOpt1}-\eqref{eq:classicalOpt2}.
Thus, $f$$=$$\infty$ corresponds to the mapping where $\hat{H}^{\text{Set Partitioning}} = \hat{H}^{\text{Exact Cover}}$. We then define the success probability as the probability of finding the optimal solution
\begin{equation*}
  P_{\text{success}}^{\text{Set Partitioning}} = |\langle\vec{x}^*| \vec{\gamma}^{L^*}, \vec{\beta}^{L^*}\rangle|^2
\end{equation*}
where $\vec{x}^*$ is the solution to the Set Partitioning problem, i.e the binary vector that corresponds the minimal value of Eq.~\eqref{eq:sp1} such that $\vec{x}^*\ \in S_{\text{feasible}}$.

\begin{figure}[h!]
   \raggedright
   \includegraphics[trim=10 15 20 20,clip,height=3.3cm]{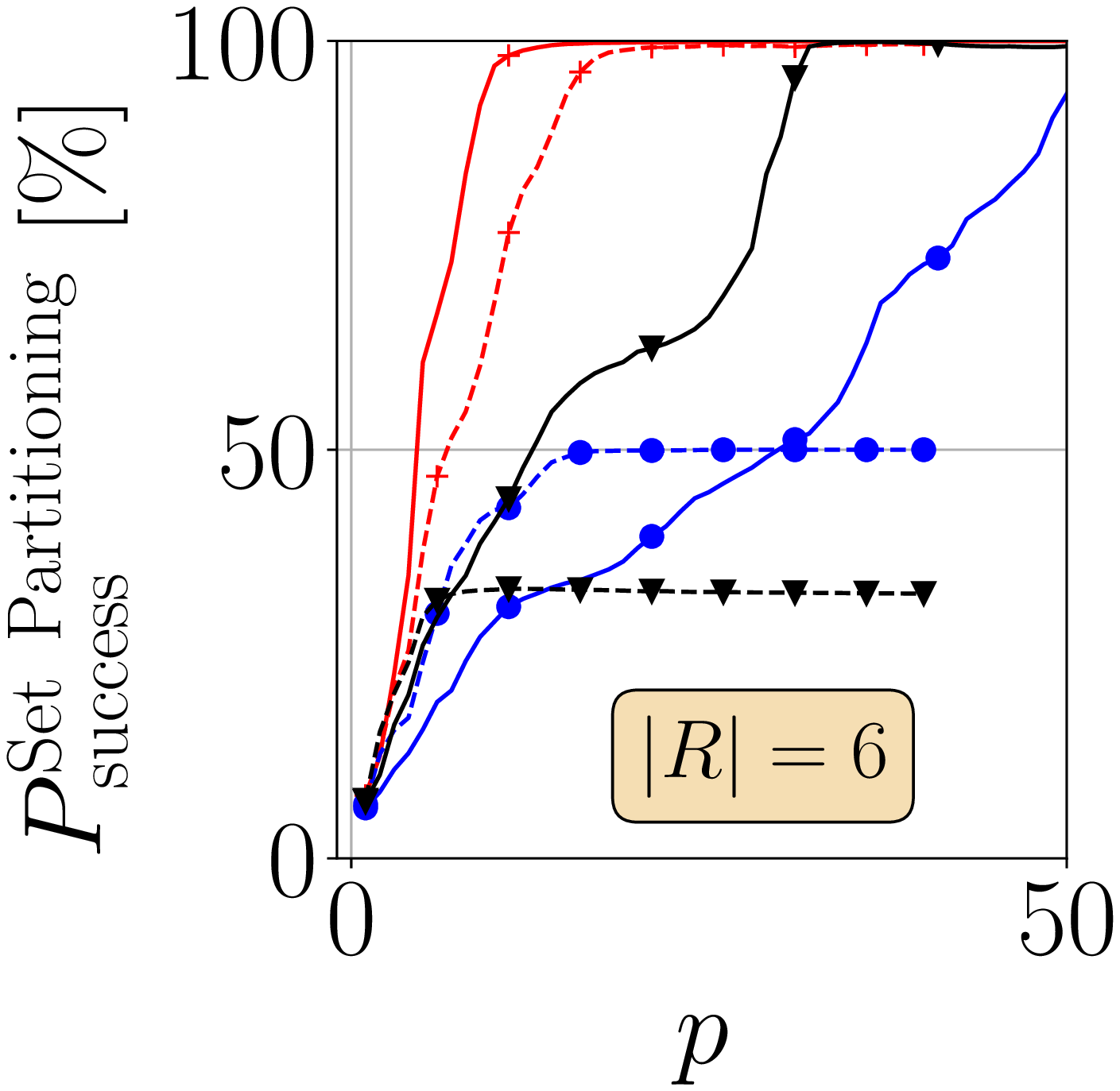}
   \includegraphics[trim=125 15 20 20,clip,height=3.3cm]{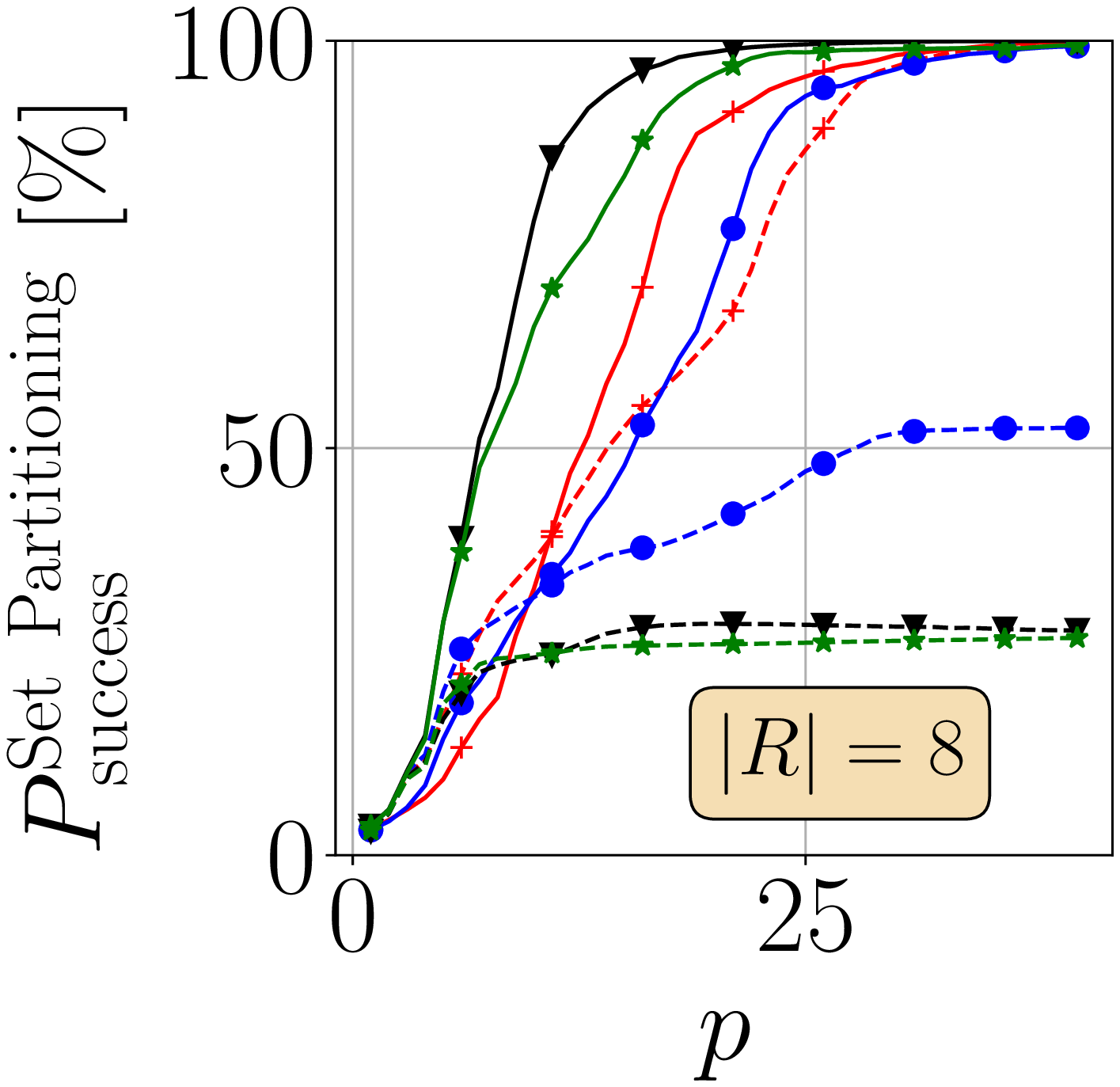}
   \includegraphics[trim=125 15 20 20, clip,height=3.3cm]{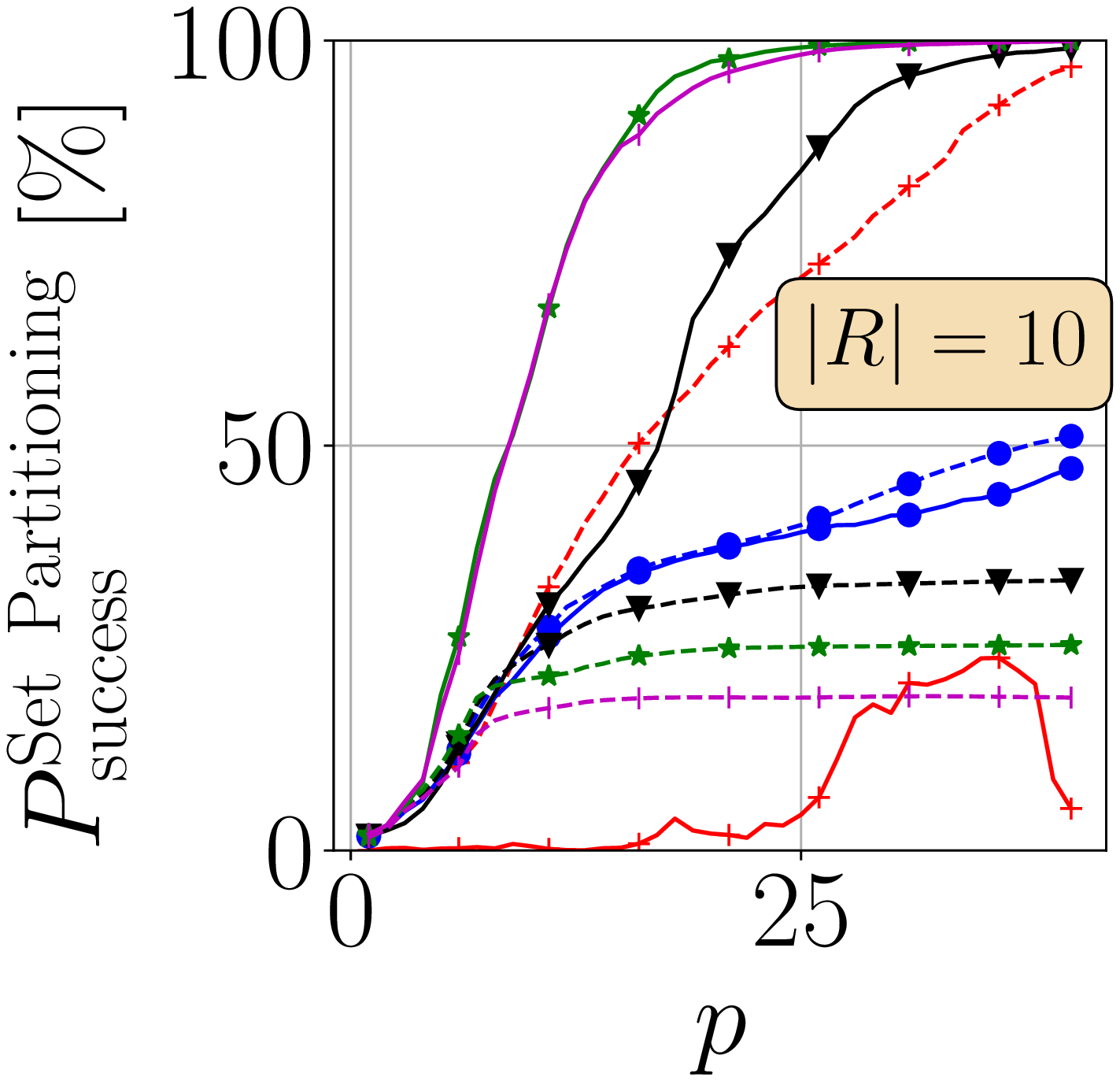}
   \\
   \raggedright
   \includegraphics[trim=10 15 20 20, clip,height=3.3cm]{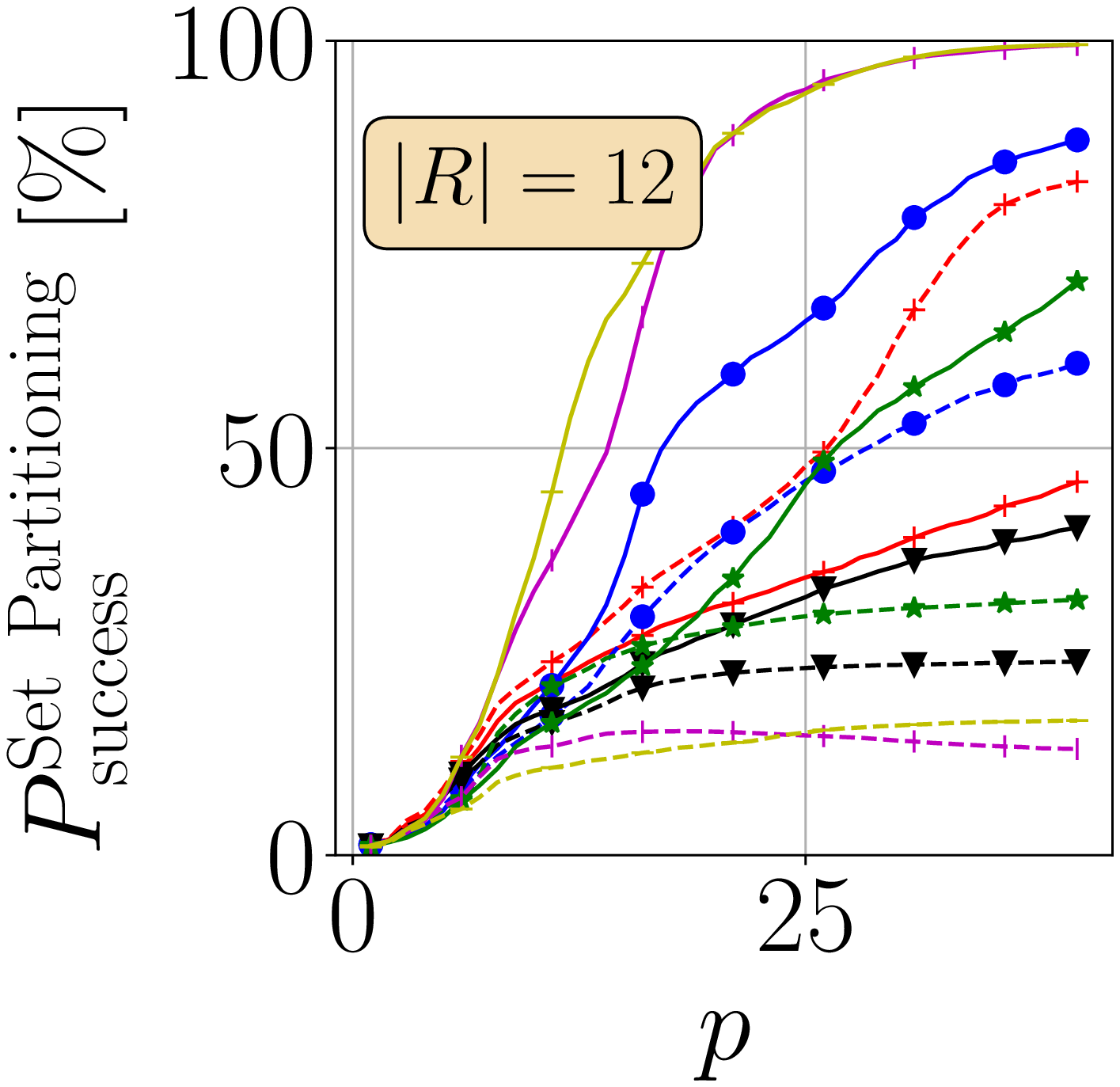}
   \includegraphics[trim=125 15 20 20,clip,height=3.3cm]{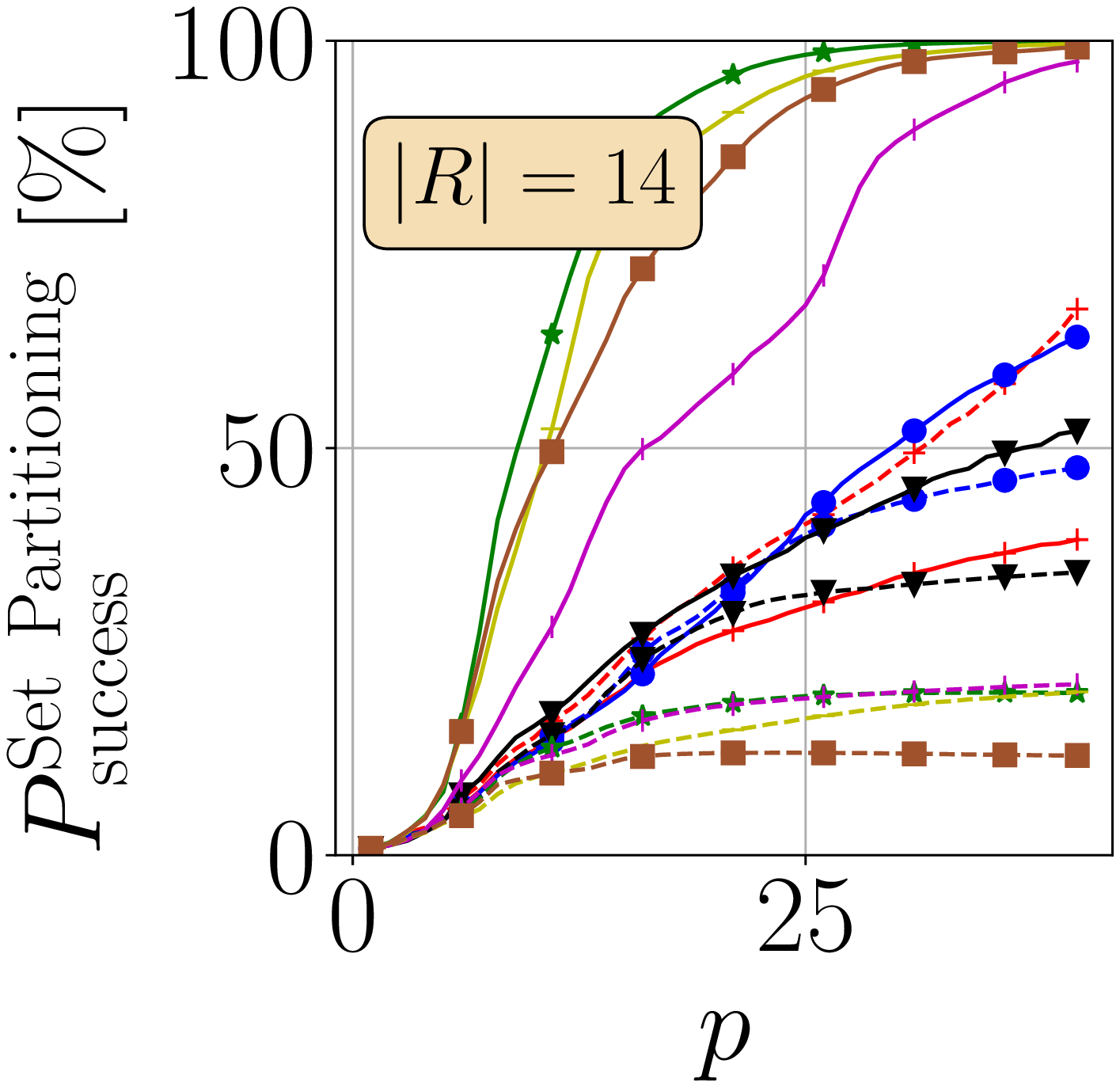}
   \includegraphics[trim=-70 30 5000 30, height=3cm]{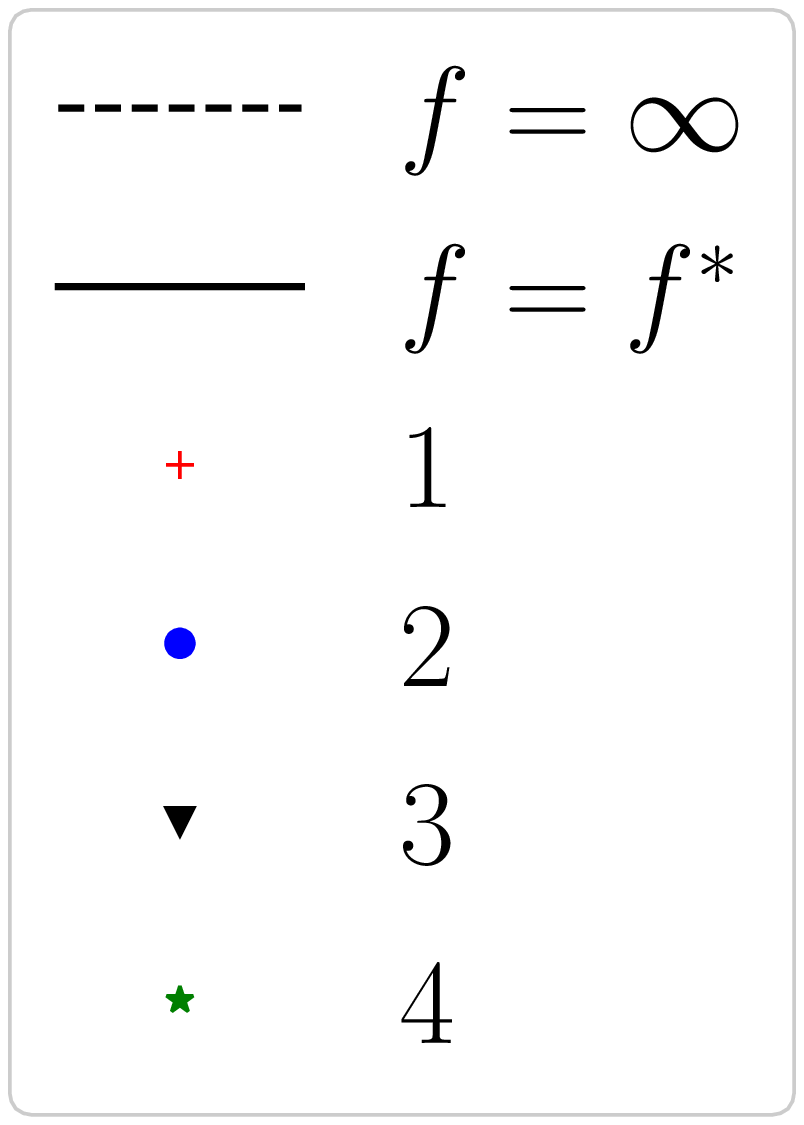}
   \\
   \raggedright
   \includegraphics[trim=0 15 20 20 ,clip,height=3.3cm]{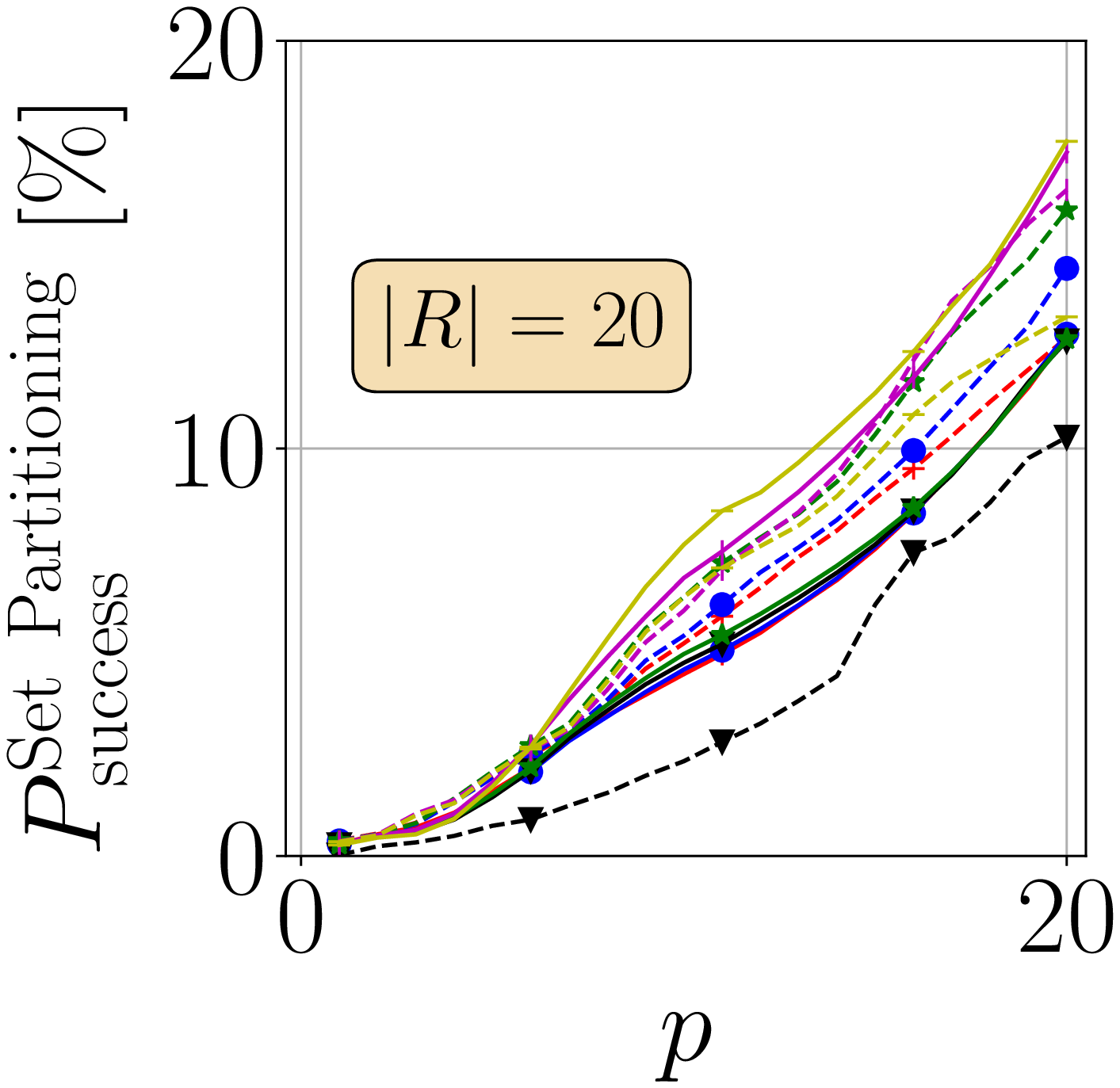} 
   \includegraphics[trim=60 15 20 20, clip,height=3.3cm]{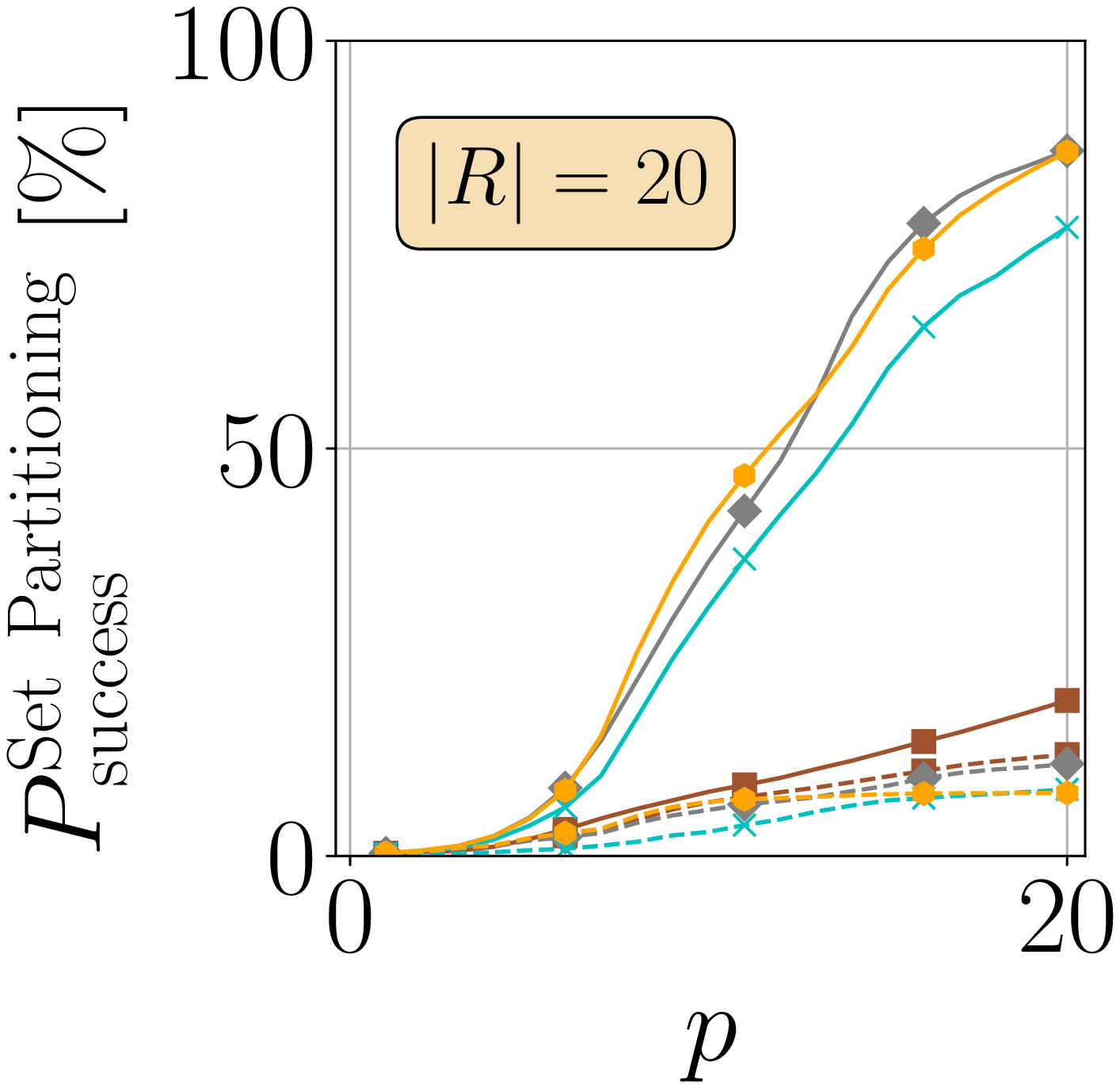}
   \includegraphics[trim=-30 30 400 30 ,height=3cm]{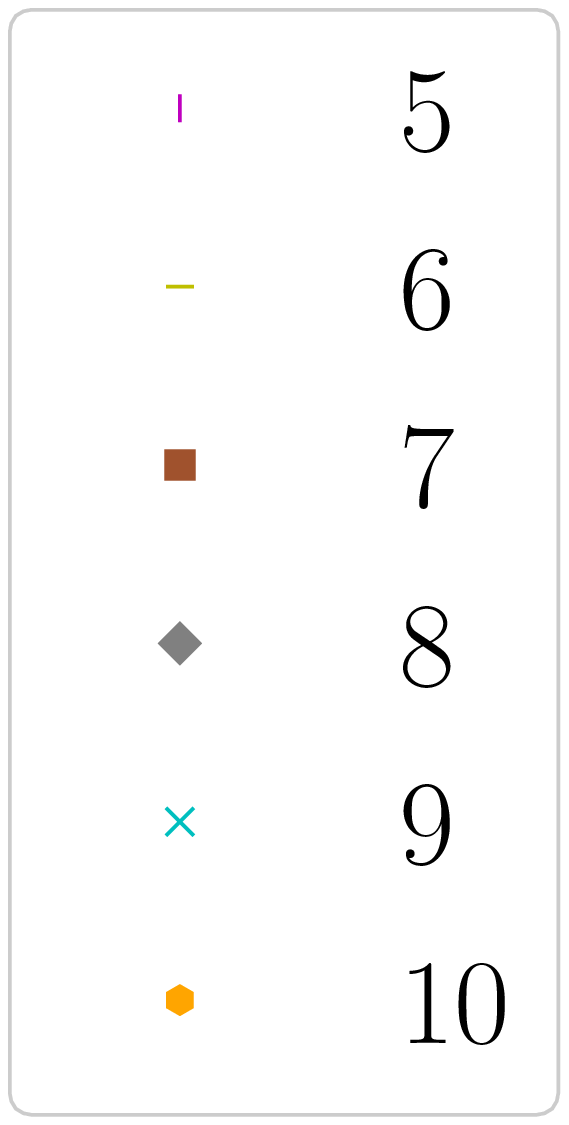}
   \caption{Success probability for solving the Set Partitioning problem depending on the choice of weights $\mu_1$ and $\mu_2$, the dashed lines correspond to $f=\infty$ and solid lines correspond to the best found factors $f=f^*$. The colors and markers of the lines indicate the value of $|S_{\text{feasible}}|$}
\label{fig:SP_summary}
\end{figure}

The success probabilities of Set Partitioning are plotted in Fig.~\ref{fig:SP_summary} for ideal QAOA circuits. Dashed lines distinguish the lines for factor $f=\infty$ and the best found factors $f^*$ are distinguished by the solid lines.
Furthermore, success probabilities are tabulated for additional factors for a given algorithm depth in Appx. \ref{sec:SP_success_prob}, where the factors have been chosen to construct cost Hamiltonians with the constraint   that the ground state corresponds to the optimal solution $\vec{x}^*$. 

 It is clear from the results of the numerical simulations in Fig.~\ref{fig:SP_summary} and Table~\ref{tab:success_prob} 
 that the success probability of solving Set Partitioning can be increased (and thus reducing the required algorithm depth) with a suitable choice of weights $\mu_1$ and $\mu_2$ for 22 instances of the 29 instances with more than one feasible solution.
 We also observe that a good choice of weights for instances with a single feasible solution corresponds to $f=\infty$ for all problem sizes.
%
 We observe that the success probability can decrease with the number of feasible solutions to $P_{\text{success}}^{\text{Set Partitioning}} \approx \frac{1}{|S_{\text{feasible}}|}$ if the weights are chosen poorly, 
which in the worst case is exponential in the problem size. To avoid requiring a considerable algorithm depth, finding good weights is thus required to solve the optimization problem with NISQ devices.

\begin{figure}[h!]
 \includegraphics[trim=0 30 0 0 30, clip,height=3.8cm]
 {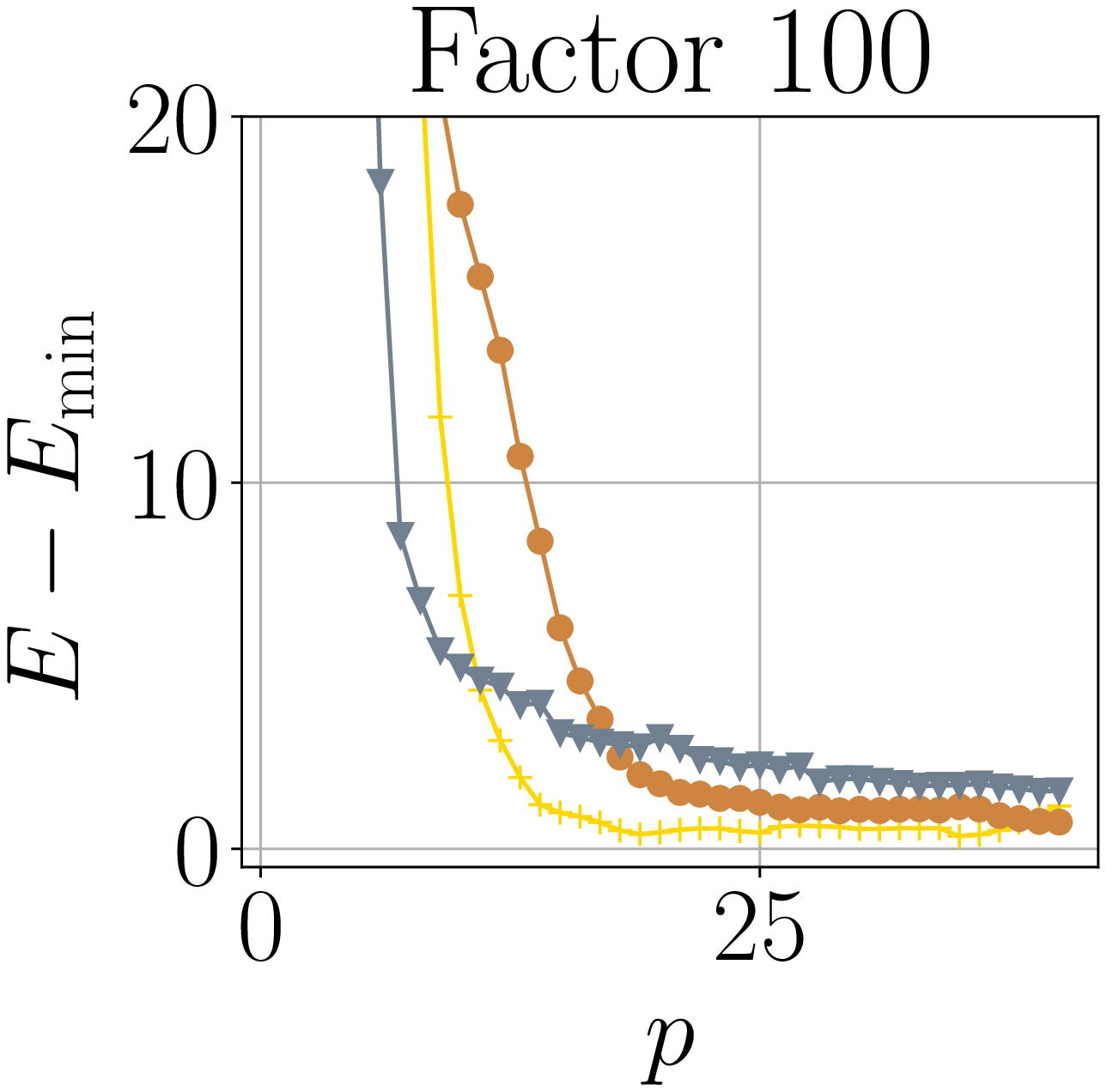}
 \includegraphics[trim=105 30 0 30, clip,height=3.5cm]
 {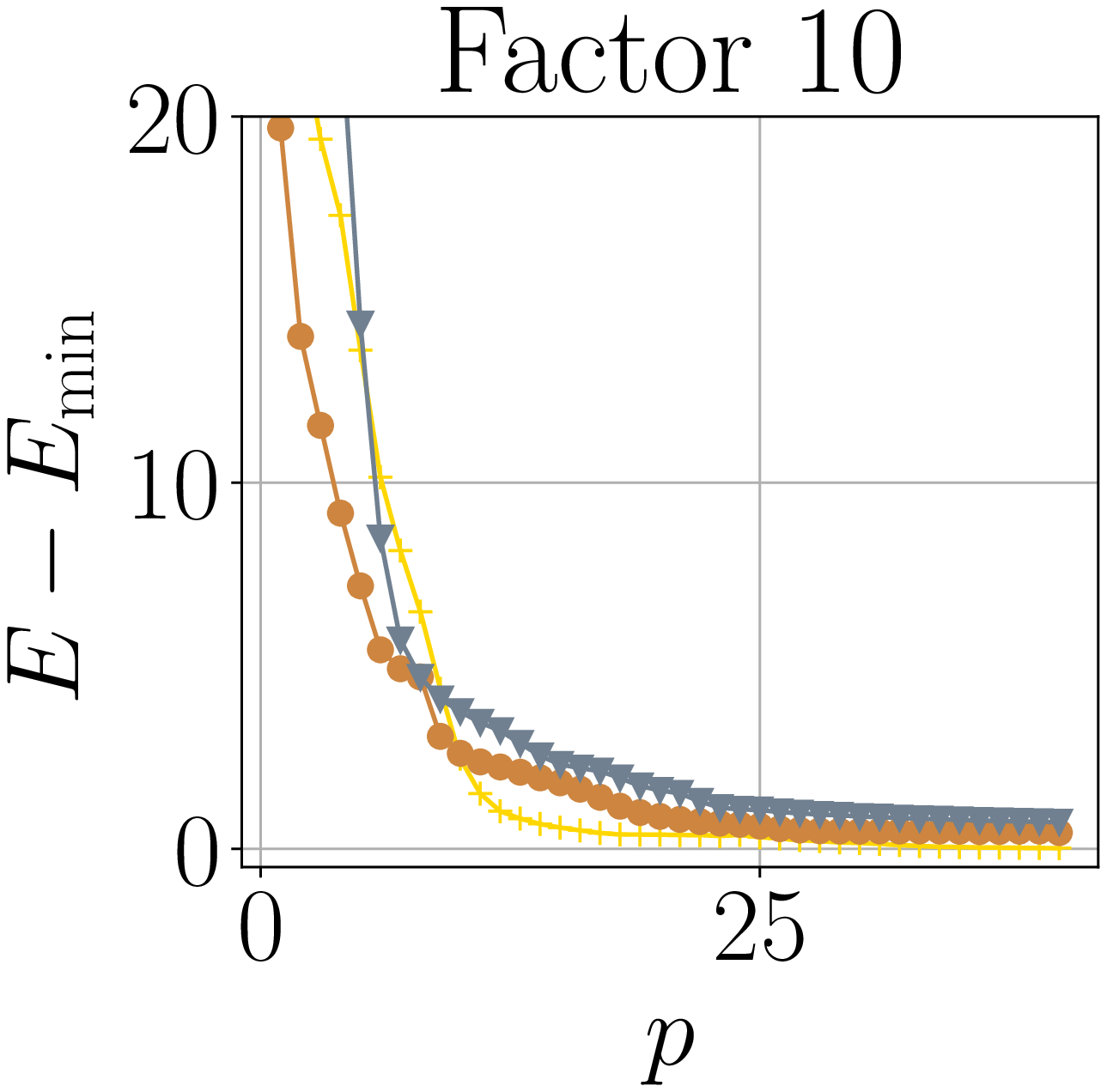}
 \includegraphics[trim=0 30 0 30, clip,height=3.5cm]
 {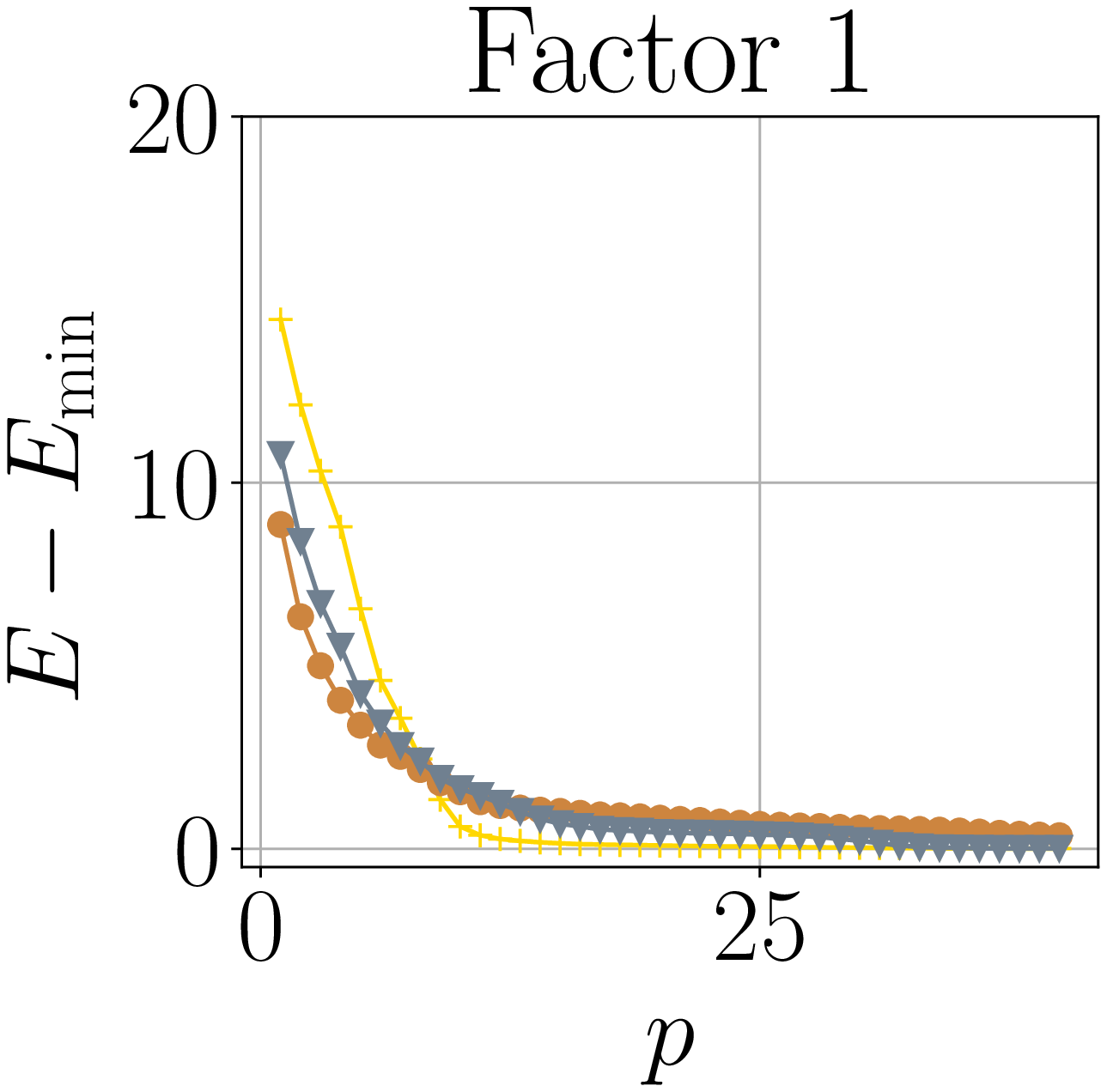}
 \includegraphics[trim=105 30 0 30, clip,height=3.5cm]
 {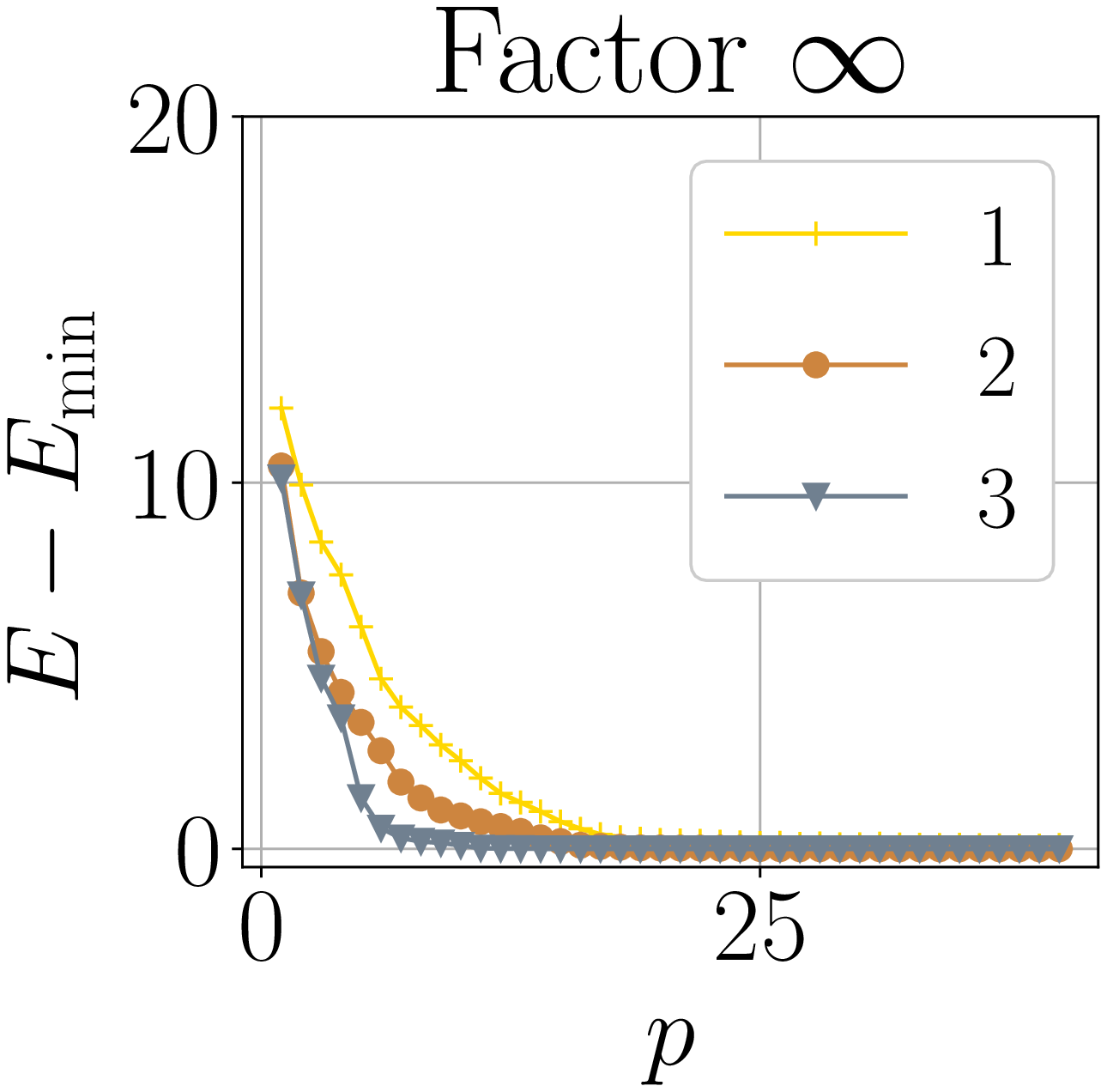}

\caption{Regret of the expectation value function for instances corresponding to problem graphs $G_{6}^{1, 2, 3}$ for factors 1, 10, 100 and $\infty$. The regret is defined as the difference between 
$E=\matrixelement*{\vec{\gamma}^{L^*},\vec{\beta}^{L^*}}{\hat{H}_f}{\vec{\gamma}^{L^*}, \vec{\beta}^{L^*}}$
where $(\vec{\gamma}^{L^*},\ \vec{\beta}^{L^*})$ are the locally optimal angles found by the interpolation strategy
and $E_{\text{min}}=\bra{\vec{x}^*}\hat{H}_f\ket{\vec{x}^*}$}
\label{fig:expval6}
\end{figure}

Moreover, the regret (the difference between the minimum expectation value found during the optimization procedure and the optimal solution) of the expectation value function is depicted in Fig.~\ref{fig:expval6} for instances with 6 routes with varying weights. 
\color{black}
We observe for $f=\infty$ that the regret is reduced to near zero, whilst  failing to increase the success probability significantly above $\frac{1}{|S_{\text{feasible}}|}$. For factors 10 and 100, the regret is  greater compared to the best found factor for a given algorithm depth\color{black}. The difference in regret corresponds to decreased required algorithm depth for the best found factor compared to factors 10 and 100 to achieve near unity success probability for Set Partitioning, see Table~\ref{tab:success_prob}. 

\begin{figure}[h!]
   \includegraphics[trim=-60 300 40 0, clip, width=0.23\textwidth]{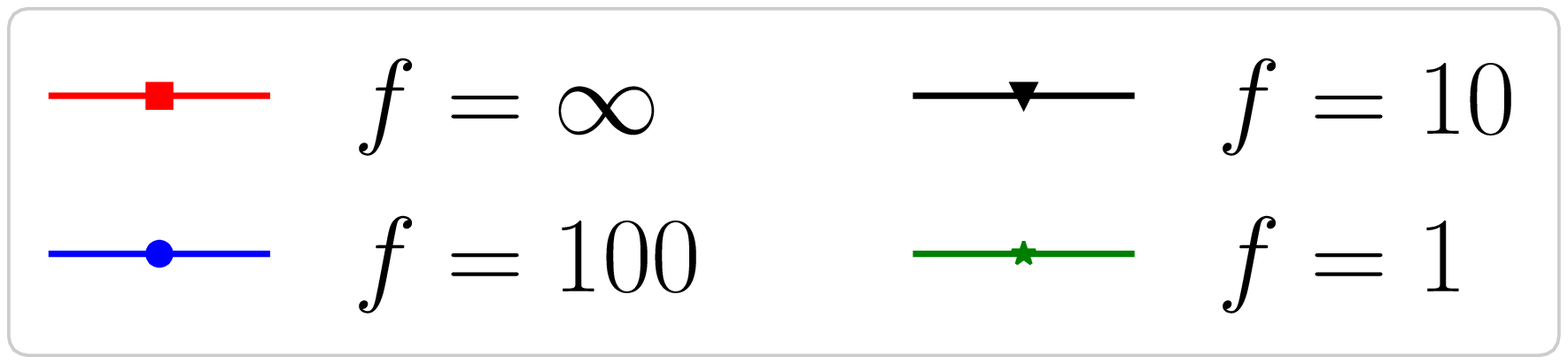}
   \includegraphics[trim=-60 320 60 0, clip, width=0.23\textwidth]{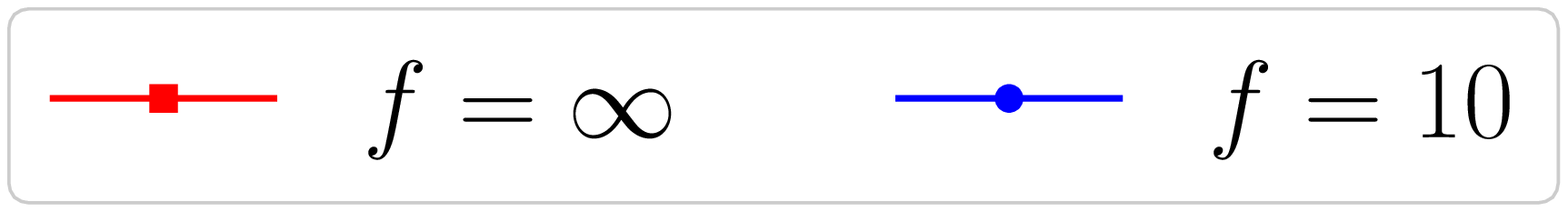}
  \includegraphics[width=0.2\textwidth]{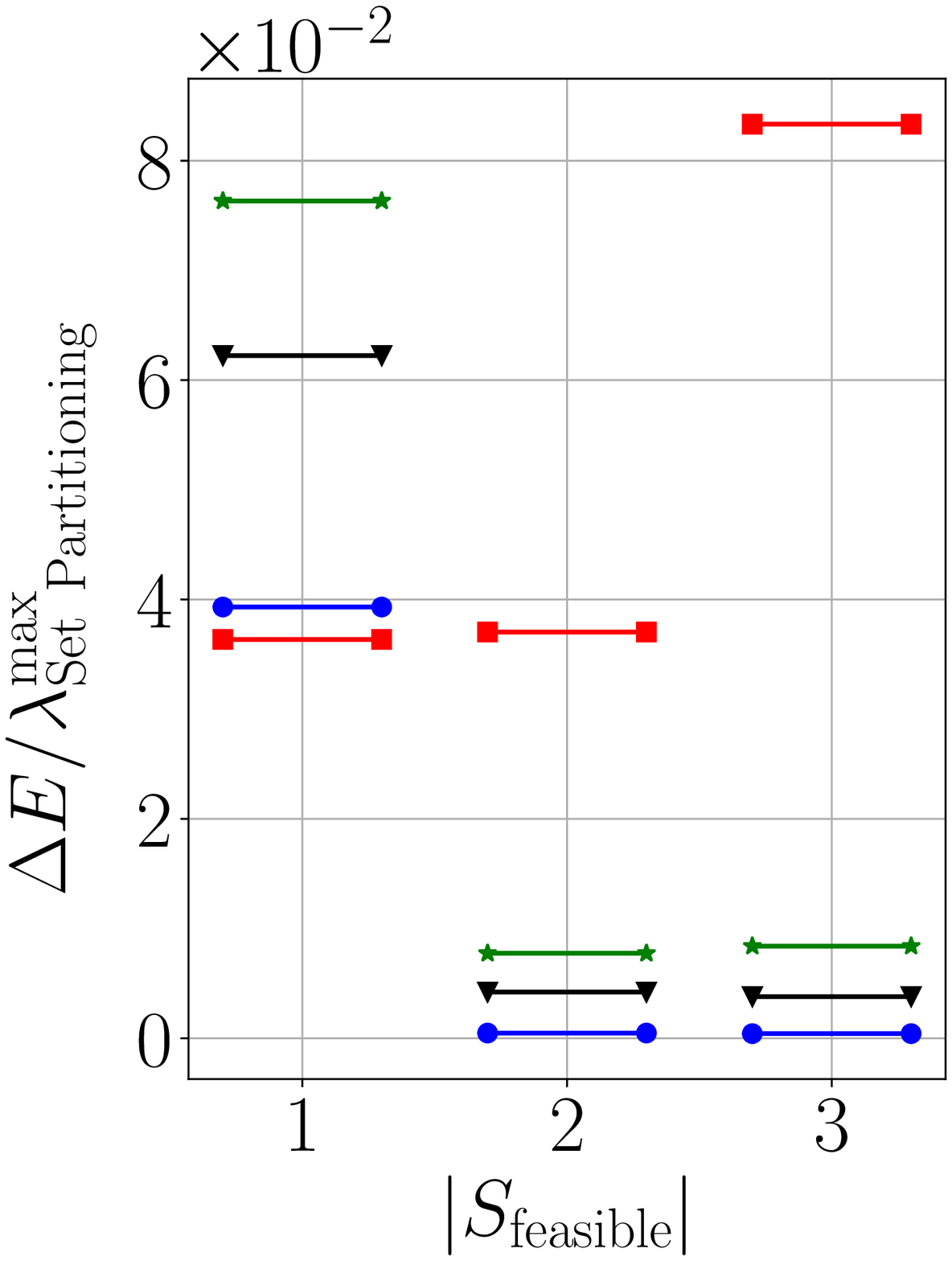}
  \includegraphics[width=0.2\textwidth]{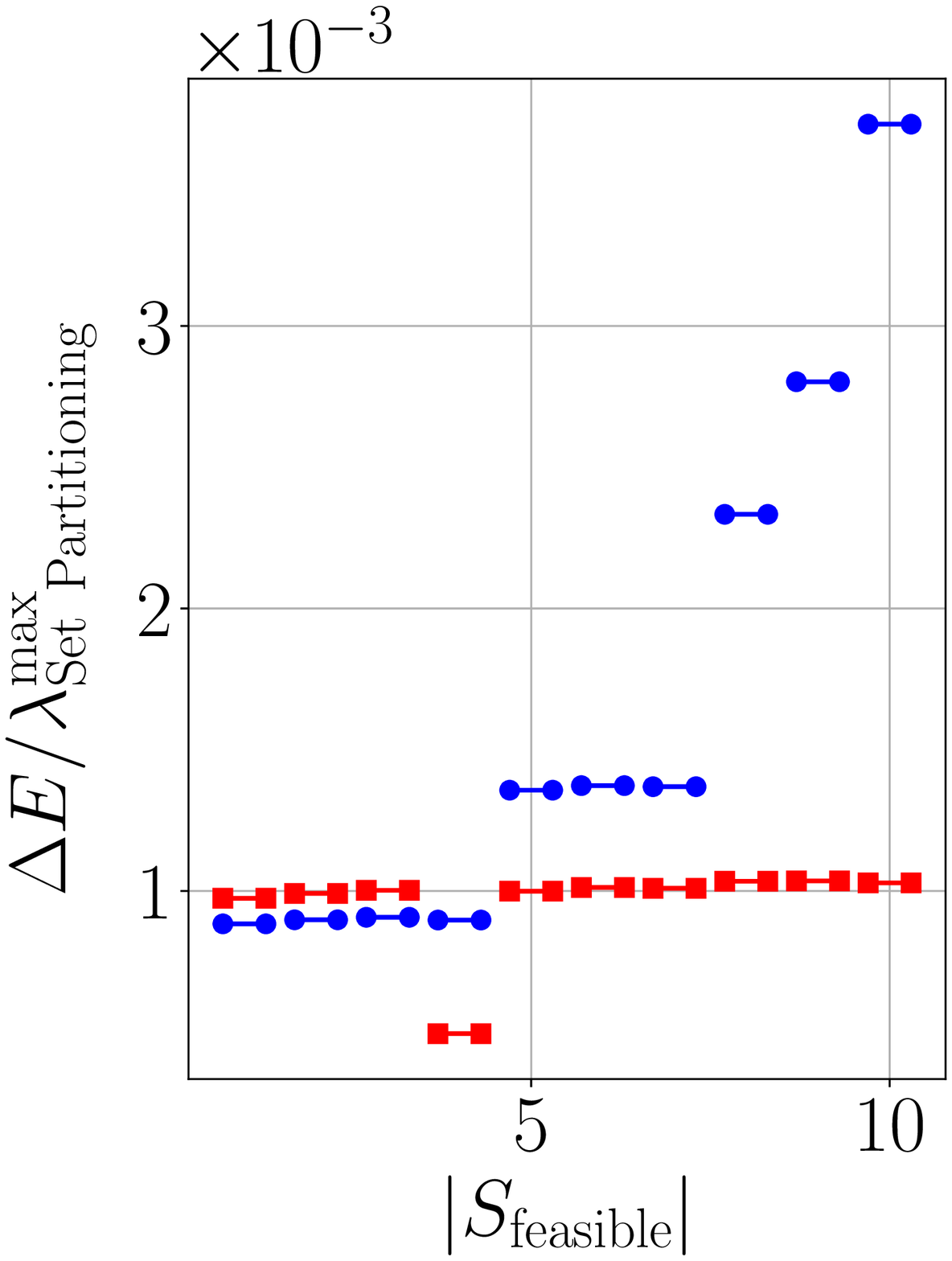}
     \caption{Minimum nonzero energy gap $\Delta E$ for cost Hamiltonian $\hat{H}^{\text{Set Partitioning}}$ depending on the factor $f$, as a ratio of the maximum eigenvalue}
   \label{fig:Eigenvalues}
\end{figure}

Since we observed that choosing a factor other than $\infty$ fails to increase the success probability for 7 instances with more than one feasible solution, we have extracted the smallest nonzero energy gap ratio with respect to the maximum eigenvalue. 
Fig.~\ref{fig:Eigenvalues} shows the ratio for instances with 6 and 20 routes. 
The graphs show for instances with 6 routes that the ratio can be increased for $G_6^{1}$ but not for $G_6^{2,3}$ by choosing a factor that considers the cost function. The lack of increased ratio corresponds to the increased required algorithm depth to obtain near unity success probability for $G_6^{2,3}$ compared to $G_6^{1}$.
Furthermore, the choice $f=10$ compared to $f=\infty$ results for instances $G_{20}^{1-3}$ in decreased ratios. Whereas the ratio is increased for $G_{20}^{4-7}$ and more distinctly for $G_{20}^{8, 9, 10}$. 
We note that as the ratio increases for the choice of factor $f$, the required algorithm depth is decreased for a given success probability here as well.
We conclude from these results that a suitable choice for weights is such that the nonzero energy gap is as large as possible as a ratio of the maximum eigenvalue of the cost Hamiltonian. 
Moreover, when we attempt to balance the objective and constraint parts of the cost Hamiltonian the smallest eigenvalues are not guaranteed to correspond to feasible solutions. This means that in the pursuit of finding the optimal solution, we can decrease the probability of finding a feasible solution. However, if we find weights such that the smallest eigenvalues correspond to feasible solutions, we do not sacrifice the probability of finding good feasible solutions for finding the optimal solution.

These results indicate that if NISQ devices are limited in algorithm depth, finding suitable weights will be crucial, requiring more computational effort. 
The task of finding suitable weights for Set Partitioning via the quadratic penalty method typically requires that several subproblems are solved, where each subproblem corresponds to a choice of weights. Typically, with the quadratic penalty method, the weight for the objective part is set to 1 and the weight for the quadratic penalty is set to be small initially. The weight of the quadratic penalty is then increased for a number of iterations or until convergence is reached. 
%
The quadratic penalty method could be executed with QAOA. We could also consider solving the problem with a classical computer, where the integer requirement could be relaxed to provide a good guess for the weights.
An initial starting point for the weights can also be chosen as $\mu_1=1$ and $\mu_2> \text{max}_{(i, j)\in E}\{ |c_i - c_j|\}$ if we assume that the smallest penalty is 1 for exchanging variables $x_i$ and $x_j$ or   
\color{black}
$\mu_2> \sum_{r=1}^{|R|} |c_r|$~\cite{FormulatingRouting} where $\mu_2$ is bounded from above. 
\color{black}
%
An alternative method to obtain suitable weights can be to 
initially attempt to solve Exact Cover where the weight is zero for the objective part and one for the constraint part. 
For the second iteration, QAOA with equal penalties set to one for the objective and constraint part of the Hamiltonian is then executed. If the solution degrades to be infeasible compared to the first solution we can assume that the objective part of the Hamiltonian dominates the constraint part. In that case, we need to increase the penalty for the constraint part for a number of iterations or until we reach a convergence.
If we, on the other hand, find that we obtain a solution of similar cost as when we attempted to solve Exact Cover, we can increase the penalty for the objective part of the Hamiltonian for a number of iterations until we observe convergence for the solutions or until the solution degrades again such that it is infeasible. 
Since each choice of weights corresponds to a subproblem to be solved with QAOA it implies a computational overhead. However, if QAOA itself is executed in polynomial time the overhead should not change the overall complexity of the algorithm. 

Finally, we conclude that the required algorithm depth of QAOA can be expected to grow with the problem size and increase as the number of feasible solutions decreases (assuming that we have identified suitable weights). Fig.~\ref{fig:SP_summary} shows that we can expect to require at least $|R|$ in algorithm depth to achieve success probability above 50\%.

\section{Conclusions \label{sec:Conclusion}}
We have proposed a method that can leverage quantum algorithms for large-scale ILPs and investigated the method by considering the quantum algorithm QAOA and the problem Tail Assignment. 
The method is useful for problems that are typically solved via Column Generation techniques, where a direct application to the problem (typically in a path-based formulation) requires in the worst case exponentially many qubits. 
%
%
%
The method can also be useful for NISQ devices as our method require less quantum resources compared to the arc-based formulations for problems as vehicle routing and Tail Assignment (defined in~\cite{Gronkvist} as model TAS in Eq. (4.1)-(4.7)). 
For Crew Pairing and Crew Rostering, in particular, some constraints are not suited to be expressed in mathematical terms as noted in~\cite{PAQS1}, utilizing a quantum algorithm in the Branch-and-Price framework for solving RMP instances can thus be the only viable option. 
Furthermore, for Tail Assignment, some constraints are recursive and non-trivial to express as an Ising model, limiting the potential to apply a quantum algorithm to the arc-based formulation directly.

The numerical results expand on the results in~\cite{vikstl2019applying} by considering more diverse and realistic, albeit small instances. 
The results indicate that the required algorithm depth decreases for a given success probability as the number of feasible solutions increases for Exact Cover, where we find the opposite results for Set Partitioning if the cost Hamiltonian is weighted poorly. 
Moreover, the reduction in success probability for Set Partitioning can be significant as the number of feasible solutions can be very large.
However, we also found that it is possible for most instances to find a suitable choice of weights such that the algorithm depth is significantly reduced to obtain a success probability above 50\%, in particular for instances where the number of solutions is larger. Even with suitable weights, we expect that instances can require an algorithm depth that grows with the problem size and node degree, where harder instances are those with few feasible solutions for QAOA with respect to both Set Partitioning and Exact Cover.
Especially hard Set Partitioning instances for QAOA are expected to be those where the minimum nonzero energy gap is small with respect to the largest eigenvalue for any weights we choose and where the minimum eigenvalues no longer correspond to the feasible solutions (whilst the ground state is still the optimal solution). These instances are more difficult because the probability of finding a  feasible solution degrades in these cases whilst favoring the optimal solution. 

%


Moreover, we have chosen to follow the mapping for both problems as presented in~\cite{Lucas2014}. Since there exists no evidence that suggests that this particular mapping, although obvious, is optimal there can exist some other more suitable mapping. Since it was observed that the node degree of the graphs affects the required algorithm depth, there might exist some more suitable mapping to be explored where the average node degree of the problem graphs can be reduced.  However, exploring alternative mappings for Exact Cover and Set Partitioning has been omitted in this work and left as a potential future challenge to consider.
 

It can further be observed that common sizes of RMP instances of Tail Assignment require approximately $10^3$-$10^4$ for practical problems. As NISQ computers were suggested to typically have 50-100 qubits initially, we would like to address this discrepancy. 
We remark that the quantum hardware is improving and new promises of NISQ devises with 1000 qubits by companies as IBM in 2023~\cite{Science} implies that the method will become applicable on NISQ devices in the near future. 
For future work, it would therefore be interesting to run QAOA on such devices for larger instances. Instances of interest to consider are generated RMP instances in Branch-and-Price frameworks for real-world problems and other hard ILP instances publicly available in operational research and mathematical optimization libraries. A remaining challenge for NISQ devices will be to realize QAOA circuits with the desired number of qubits for polynomial algorithm depths.

For future work, it could also be interesting to study if it is possible to reduce the RMP instance to be better suited for NISQ devices. For example, one could attempt to choose a subset of decision variables in RMP instances to construct smaller RMPs. However, such a reduction corresponds to options with a combinatorial behavior. Reducing the size of RMP instances can therefore require more advanced preprocessing techniques. Further techniques as those explored in~\cite{harrow2020small} can also be valuable to consider.


We note that whilst our method provides a possibility to leverage quantum algorithms to an advantage for large-scale ILPs, any quantum algorithm under consideration must be capable of either providing significant speed-up in finding solutions of similar quality as the best classical solvers or capable of finding solutions of improved quality compared to classical solvers during the same execution time. 
The numerical experiments we have considered in this paper for QAOA can not answer these open questions fully. However, it should be observed that as the average node degree of the generated instances are large, we can therefore consider that the results in Sec.~\ref{sec:resultsEC}-\ref{sec:resultsSP} to correspond to hard instances for QAOA with respect to problem size. Larger instances that are sparse can therefore have a reduced requirement on the algorithm depth, which further motivates studying instances with lower node degrees by both numerical simulations and executions on quantum devices.
%


Finally, we conclude that it is possible to integrate QAOA with a Branch-and-Price algorithm, where we achieve reasonably high success probabilities for RMP instances with a polynomial algorithm depth. 
In obtaining high quality integer solutions to RMP instances, the run-time of the general and heuristic Branch-and-Price algorithms can therefore be reduced and improve solution quality. 

\acknowledgments
This work was supported from the Knut and Alice Wallenberg Foundation through the Wallenberg Center for Quantum Technology (WACQT). G.
F. acknowledges financial support from the Swedish Research Council through the VR project QUACVA
\bibliography{refs_prx}
\appendix
\section{The heuristic Branch-and-Price algorithm for solving Tail Assignment\label{sec:BandBfixing}}
The Branch-and-Price algorithm is designed to solve large-scale Integer Linear Programs (ILPs) and combines the algorithms Column Generation and Branch-and-Bound. In this section, we review first Branch-and-Bound and second the Column Generation algorithm. Last, we review the Branch-and-Price algorithm and the fixing heuristic presented in~\cite{Gronkvist} subject to be integrated with a quantum algorithm. 
\subsection{Branch-and-Bound\label{sec:Branch-and-Bound}}
The Branch-and-Bound algorithm, given in~\cite{Land60anautomatic} and surveyed in~\cite{MORRISON201679} more recently, provides a framework for finding the optimal solution to ILPs. As the feasible region is restricted to integer points and not convex, algorithms applicable for Linear Programs (LPs) can not solve ILPs generally. The distinction here is that LPs can be solved efficiently, whereas ILPs are NP-hard problems. 

The algorithm, given in pseudo code in Alg.~\ref{alg:main}, decomposes the original ILP into subproblems recursively that can be visualized with a tree structure. Exhaustive search is avoided by pruning nodes of the tree giving  more acceptable running times in practice. Each node in the tree represents a subproblem which is the original ILP with a reduced feasible space. Each subproblem can be relaxed, i.e., the decision variables are not discrete but continuous, yielding either a lower bound (if a minimization problem), an integer solution, or that the subproblem is infeasible.

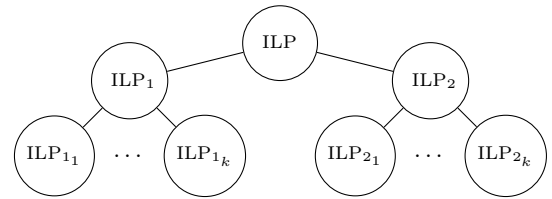
\begin{figure}[ht!]
  \centering
  \begin{tikzpicture}
  \node[shape=circle,draw=black] (p) at (0,1) {~\scriptsize ILP$_{~}$};
  \node[shape=circle,draw=black] (c1) at (-2,0.5) {~\scriptsize ILP$_{1}$};
  \node[shape=circle,draw=black] (c11) at (-3,-0.5) {\scriptsize ILP$_{1_1}$};
  \node[shape=circle,draw=black!0!white] (c1d) at (-2,-0.5) {$\dots$};
  \node[shape=circle,draw=black] (c1k) at (-1,-0.5) {\scriptsize ILP$_{1_k}$};
  
  \node[shape=circle,draw=black] (c2) at (2,0.5) {~\scriptsize ILP$_2$};
  \node[shape=circle,draw=black] (c21) at (1,-0.5) {\scriptsize ILP$_{2_1}$};
  \node[shape=circle,draw=black!0!white] (c2d) at (2,-0.5) {$\dots$};
  \node[shape=circle,draw=black] (c2k) at (3,-0.5) {\scriptsize ILP$_{2_k}$};
  
  \path [-](p) edge node[left] {} (c1);
  \path [-](c1) edge node[left] {} (c11);
  \path [-](c1) edge node[left] {} (c1k);
k);
  \path [-](p) edge node[left] {} (c2);
  
  \path [-](c2) edge node[left] {} (c21);
  \path [-](c2) edge node[left] {} (c2k);
k);
\end{tikzpicture}
  \caption{Conceptual search tree of Branch-and-Bound}
  \label{fig:BaB}
\end{figure}

Consider here that we are applying Branch-and-Bound to an integer linear program 
\begin{equation*}
  \text{ILP = min}\left\{\sum_{i=1}^n c_ix_i: \vec{x} \in S\right\} 
\end{equation*}
where $S=\left\{\vec{x} \in \mathds{Z}_+^n : \sum_{i=1}^n a_{ji} x_i \geq b_j\ \forall j =1,\dots,m\right\}$. 
The Linear Programming (LP) relaxation of the ILP is 
\begin{equation*}
  \text{LP = min}\left\{\sum_{i=1}^n c_ix_i: \vec{x} \in P\right\}
\end{equation*}
 where $P=\{\vec{x}\in \mathds{R}_+^{n}: \sum_{i=1}^{n}a_{ji}x_i \geq b_j \ \forall j =1,\dots,m\}$. 
 We know from linear programming theory that the LP relaxation of an ILP gives the relation $\text{LP}\leq\text{ILP}$.
A partition of the ILPs feasible space $S$ yields two subproblems 
\begin{align*}
  &\text{ILP}_1 = \text{min} \left\{ \sum_{i=1}^n c_ix_i : \vec{x} \in S_1 \right\},
  \\
  &\text{ILP}_2 = \text{min} \left\{ \sum_{i=1}^n c_ix_i: \vec{x} \in S_2 \right\}
\end{align*}
where $S_1$ and $S_2$ are disjoint sets that partition $S$ by a constraint on variable $x_j$ such that $S_1 = \{\vec{x}\in S: x_j \leq \lfloor x_j^0 \rfloor \}$ and 
$S_2 = \{\vec{x}\in S: x_j \geq \lceil x_j^0 \rceil \}$. The variable $x_j^0 \in \vec{x}^0$ has some fractional value and $\vec{x}^0$ is an optimal solution to LP. 
 We further know from linear programming theory that either ILP$_1$ or ILP$_2$ has the optimal solution to ILP. 
Similarly, the two subproblems can be related to new problems that correspond to the LP relaxation of ILP$_1$ and ILP$_2$ which provides lower bounds, can show that there exists no feasible integer point or can find an optimal integer solution. The three problems ILP, ILP$_1$ and ILP$_2$ can be visualized as a tree with a parent node and two child nodes, see Fig.~\ref{fig:BaB}. Clearly, ILP$_1$ and ILP$_2$ can be partitioned further into subproblems giving the tree structure rooted in a node representing the original ILP. 
If an LP relaxed subproblem is found to be infeasible, the node is pruned, i.e., the branch is not explored further and we say that the node is \emph{pruned by infeasibility}.

When the algorithm recursively explores subproblems, an incumbent, $z^*$, is maintained which is the current best feasible solution found to the ILP. Whenever a subproblem yields a solution greater or equal to the incumbent, this region cannot contain any integer solutions that would improve upon the one we already have and this particular node is pruned. We say that the node is \emph{pruned by bound}. 

If we find that a solution to a subproblem is integral, we also prune this node as we have found an optimal partial solution or candidate incumbent $z_i$ for this specific region. We say that the node is \emph{pruned by integrality.} If $z_i < z^* $ the incumbent is updated. 

Finally, if a subproblem can not be pruned by infeasibility, bound or integrality the subproblem is partitioned into $k\geq 2$ nodes representing $k$ subproblems, which are children to the current subproblem we are exploring in the tree. The $k$ subproblems are then added to a list of unexplored subproblems and a new subproblem is chosen to be explored. When there are no unexplored subproblems left the algorithm terminates and returns the incumbent solution and the corresponding assignment. 
\begin{algorithm}[H]
  \caption{Branch-and-Bound($S$)}
  \label{alg:main}
  \begin{algorithmic}[1]
    \STATE{$\vec{x}^*\gets \emptyset$}
    \STATE{$z^* \gets \infty$}
    \STATE{$z_U^* \gets \infty$}
    \STATE{$\mathcal{L}\gets \{S\}$}
    \WHILE{$|\mathcal{L}|>0$}
      \STATE {$S_i\gets chooseSubProblem(\mathcal{L}) $ }
      \STATE {$\mathcal{L} \gets \mathcal{L}\backslash\{ S_i\} $ }
      \IF{$S_i$ has feasible solution to LP relaxation}
        \STATE{$(z_L, \vec{x}_L)\gets solveLP relaxation(S_i)$}
        \IF{$z_L < z_U^*$}
          \IF{$\vec{x}_L$ feasible to ILP}
            \STATE{\hspace*{0pt}\hfill $\triangleright$ Prune by integrality}
            \IF{$z_L < z^*$}
              \STATE {$\vec{x}^*\gets \vec{x}_L$ }
              \STATE{$z^* \gets z_L$}
              \STATE{$z_U^* \gets z^*$}
            \ENDIF
          \ELSE 
             \STATE{$(z_U, \vec{x}_U)\gets getFeasibleSolution(S_i)$}
             \STATE{$z_U^* \gets$min$(z_U^*, z_U)$}
             \STATE{$\{S_{i_1}, \dots, S_{i_k}\}\gets partition(S_i)$ }
             \STATE{$\mathcal{L}\gets \mathcal{L}\cup \{ S_{i_1}, \dots, S_{i_k}\}$}
          \ENDIF
        \ELSE
          \STATE {\hspace*{0pt}\hfill $\triangleright$ Prune by bound}
        \ENDIF
      \ELSE
        \STATE {\hspace*{0pt}\hfill$\triangleright$ Prune by infeasibility}
      \ENDIF
    \ENDWHILE
    \STATE \textbf{return }$(z^*, \vec{x}^*)$
  \end{algorithmic}
\end{algorithm}
\subsection{Column Generation \label{sec:colgen}}
In the previous section we mentioned that the LP relaxation of an ILP could be efficiently solved. However, consider the case where the number of variables is exponentially large so that even generating the LP would take exponential time and space. This is exactly the case for large-scale ILPs as the Tail Assignment formulation in~\cite{Gronkvist}, which has an exponential number of possible routes in the worst case. 


The Column Generation algorithm~\cite{LubbeckeDesrosiers}, depicted with green colored boxes with dotted borders in Fig.~\ref{HeuristicBranchAndBound} and presented in pseudo code in Alg.~\ref{alg:golgen}, is based on well known duality concepts from linear programming theory. It has been proved successful for both linear programs and ILPs, particularly when the number of decision variables is very large.
Instead of attempting to construct and solve the complete problem
it is decomposed into a Master Problem (MP) 
\begin{align*}
  z_{\text{MP}}^*=& \text{ minimize }  \sum_{j \in J} c_j x_j, \\
          & \text{ subject to } \sum_{j \in J} a_{ij}x_j \geq b_i \ \forall i \in I\\
   & ~~~~~~~~~~~~~~~~ x_j \geq 0 \ \forall j \in J
\end{align*}
and a Pricing Problem (PP)
\begin{equation*}
  \text{argmin} \left\{\bar{c}_j = c_j - \sum_{i\in I} a_{ij}\pi_i : j \in J \right\},
\end{equation*}
here $\pi_i$ are the dual variables that correspond to the primal variables, $x_j$, found by solving the MP. 
 The PP often encapsulates most of the problem specific details and difficult constraints and generates new columns, also referred to as entering variables.

Since the number of decision variables is very large, the MP is further reduced to a restricted version, denoted the Restricted Master Problem (RMP), meaning that the number of decision variables is smaller, often much smaller, than the original problem. The reduced size of the RMP is tractable to solve with some LP solver such as the dual simplex~\cite{DualSimplex} or primal simplex~\cite{PrimalSimplex} algorithm, compared to the MP. 
 
The decomposition results in an iterative algorithm where the RMP and the PP are solved for a number of iterations or until optimal conditions hold. For each iteration, we attempt to find entering and exiting variables where the exiting variables are removed from the RMP and the entering variables are added to the RMP, resulting in new RMP and PP instances. 
 
The PP is thus some problem that when solved can generate improving columns and decision variables to the RMP, based on given input of the dual variables from the RMP, such that the cost of the new solution, which at this point is not guaranteed to be integral, is improved. 
%
%
Improving columns are identified by having a negative reduced cost $\bar{c}_j$ and optimal conditions hold when no variables with negative reduced cost can be found, which is the same condition as in the simplex algorithm. 

If the original problem is an ILP, the MP is the LP relaxation of the ILP. In the case of Tail Assignment the RMP corresponds to a restricted and LP relaxed Set Partitioning or Exact Cover problem, see Sec.~\ref{sec:SPEC}, where the decision variables are continuous real variables.
%
%
The PP  can thus be defined as 
\begin{equation}
  \text{argmin} \left\{\bar{c}_r = c_r - \sum_{f\in F} a_{fr}\pi_f : r \in R \right\}.\label{eq:RCSPPobj}
\end{equation}
for Tail Assignment, where $\pi_f$ is the dual variable of flight $f$ obtained when solving the RMP.

 To be noted, the first step of Column Generation is to construct an initial RMP, which for Tail Assignment can be $A=\mathds{1}_{|F|\times|F|}$ where the costs $c_r$ are set to some large number and thus unlikely to be part of a solution. Variables can be chosen as exiting variables when the value of the reduced cost is above a given threshold, however, removing variables from the RMP does not necessarily as improve convergence as removing variables also removes dual information. Further investigations in deleting columns can be found in~\cite{Gronkvist}, in Sec. 6.4.

\begin{figure}[ht!]
  \centering
  \begin{tikzpicture}
  \node[shape=rectangle,draw=black] (f1) at (-4,2) {~\scriptsize $f_1, c_1, \pi_1$};
  \node[shape=rectangle,draw=black] (f10) at (-4,0.5) {~\scriptsize $f_{10}, c_{10}, \pi_{10}$};
  \node[shape=rectangle,draw=black] (f2) at (-2,1.5) {~\scriptsize $f_2, c_2, \pi_2$};
  \node[shape=rectangle,draw=black] (f9) at (-2,0) {~\scriptsize $f_9, c_9, \pi_9$};
  
  \node[shape=rectangle,draw=black] (f6) at (0,1) {\scriptsize $f_6, c_6, \pi_6$};
  \node[shape=rectangle,draw=black] (f7) at (0,2) {\scriptsize $f_7, c_7, \pi_7$};
  \node[shape=rectangle,draw=black] (f3) at (0,0) {\scriptsize $f_3, c_3, \pi_3$};
  \node[shape=rectangle,draw=black] (f4) at (+2,+1) {\scriptsize $f_4, c_4, \pi_4$};
  \node[shape=rectangle,draw=black] (f8) at (+2,0) {\scriptsize $f_8, c_8, \pi_8$};
  \node[shape=rectangle,draw=black] (f5) at (+3,+2) {\scriptsize $f_5, c_5, \pi_5$};
  
  \node[shape=circle,draw=black] (s) at (3.4,0) {\scriptsize sink};
  
  \path [->](f1) edge node[left] {}(f2) ;
  \path [->](f1) edge node[left] {}(f7) ;
  
  \path [->](f10) edge node[left] {}(f2) ;
  \path [->](f10) edge node[left] {}(f6) ;
  
  \path [->](f2) edge node[left] {}(f6) ;
  \path [->](f2) edge node[left] {}(f7) ;
  \path [->](f2) edge node[left] {}(f3) ;
  
  \path [->](f3) edge node[left] {}(f8) ;
  \path [->](f3) edge node[left] {}(f4) ;
  
  \path [->](f9) edge node[left] {}(f3) ;
  \path [->](f9) edge node[left] {}(f6) ;
  
  \path [->](f6) edge node[left] {}(f4) ;
  \path [->](f6) edge node[left] {}(f5) ;
  \path [->](f6) edge node[left] {}(f8) ;
  
  \path [->](f7) edge node[left] {}(f4) ;
  \path [->](f7) edge node[left] {}(f5) ;
  
  \path [->](f4) edge node[left] {}(s) ;
  \path [->](f5) edge node[left] {}(s) ;
  \path [->](f8) edge node[left] {}(s) ;

\end{tikzpicture}
  \caption{Pricing problem}
  \label{fig:PP}
\end{figure}
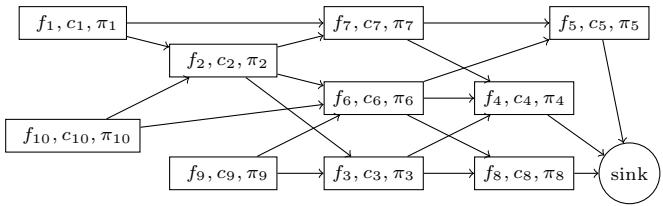

%
Furthermore, solving the PP at first glance appears intractable as the number of reduced costs can be exponentially large. By formulating the problem as a Resource Constrained Shortest Path Problem (RCSPP) we avoid to explicitly construct all routes. The RCSPP is described by a connection network, depicted in Fig.~\ref{fig:PP} with a unique sink vertex and other vertices representing flights with edges that represent legal connections where the nodes are associated with a flight cost $c_f$ and a dual variable $\pi_f$ found by solving the RMP.
 The problem depicted in Fig.~\ref{fig:PP} is a shortest path problem, where the objective is given by Eq.~\eqref{eq:RCSPPobj} and additional costs for each edge, i.e., flight connection. The problem becomes an RCSPP problem when we introduce cumulative constraints as resources, where a resource is a value accumulated throughout  the route and is required to not go above some limit, hence giving a resource constraint. Resources and subsequently resource constraints are introduced for each maintenance type where a requirement can be given by the maximum flying hours that are allowed prior to a check or the maximum number of landings an aircraft is allowed to make prior to a check. 

The task is then to find the shortest path with respect to the reduced costs in the network and the resource constraints. The PP is NP-hard where for example a label-setting algorithm~\cite{labelSettingAlg} can be applied to solve instances.

\begin{algorithm}[H]
  \caption{ColumnGeneration ($F,\ T$)}
  \label{alg:golgen}
  \begin{algorithmic}[1]
    \STATE {$RMP \gets FindInitialSolution(F, T)$}
    \STATE {$\vec{\pi}, \vec{x}\gets SolveRestrictedMasterProblem(RMP)$}
    \STATE {$negativeReducedCosts \gets solvePricingProblem(\vec{\pi}) $}
    \STATE {$exitingVariables \gets findExitingVariables(RMP)$}
    \WHILE {$negativeReducedCosts \neq \emptyset $}
      \STATE {$RMP \gets RMP \cup \{negativeReducedCosts\}$}
      \STATE {$RMP \gets RMP\backslash \{exitingVariables\}$}
      \STATE {$\vec{\pi}, \vec{x} \gets SolveRestrictedMasterProblem(RMP)$}
      \STATE {$negativeReducedCosts \gets solvePricingProblem(\vec{\pi}) $}
      \STATE {$exitingVariables \gets findExitingVariables(RMP)$}
    \ENDWHILE
    
    \STATE \textbf{return }$z=\vec{c}^T\vec{x}, \vec{x}$
  \end{algorithmic}
\end{algorithm}
\subsection{Branch-and-Price and fixing heuristics \label{sec:Branch-and-Price}}
Since only integral solutions are accepted for ILPs (and the original formulation of Tail Assignment), the Column Generation algorithm is typically augmented to Branch-and-Price~\cite{Branch-and-Price}, by combining Column Generation and Branch-and-Bound. 
In Branch-and-Price, we add an additional branching step, which occurs when no columns with a negative reduced cost can be found via solving the PP and the optimal solution is not integral. The fractional solution from the Column Generation provides a lower bound, if we are considering a minimization problem, as the algorithm solves the LP relaxed subproblem in Branch-and-Bound.
In the branching step the search space is partitioned, where the Column Generation algorithm is executed for each subproblem created. 
Therefore, we point out that Branch-and-Price can be thought of as Branch-and-Bound where Column Generation is utilized as a subroutine to compute bounds, show infeasibility or find an integer solution.

Moreover, Grönkvist~\cite{Gronkvist} noticed that  Branch-and-Price might be unnecessarily slow when applied to Tail Assignment 
and introduced a fixing heuristic where the branching step is replaced. 
The fixing heuristic finds the variable $x_i$ closest to 1 and fixes it to 1, which forces the corresponding route to be part of the solution.
It can be noted that the difference between the fixing heuristic and the typical branching is that the search space is restricted and not partitioned, meaning that the fixing heuristic is a dive into a specific branch of the search tree. Additional backtracking methods are utilized but are beyond this section's scope where such further information can be found in~\cite{Gronkvist}. 
We denote the modified Branch-and-Price algorithm as the heuristic Branch-and-Price and depict the algorithm with the blue and green colored boxes with dotted and dashed borders in Fig.~\ref{HeuristicBranchAndBound} subject to be integrated with a quantum algorithm in Sec.~\ref{sec:integQAOA}.
\section{Mapping problems to the Ising spin glass model\label{sec:mapping}}
If we consider the  Set Partitioning problem in Eq.~\eqref{eq:sp1}-\eqref{eq:sp3}   and apply a quadratic penalty  on the constraints we obtain a nonlinear integer optimization problem. If we further assume  constants $\mu_1\in \{ \mathds{Z}^+ \cup \{0\}\},\ \mu_2 \in \mathds{Z}^+$ that balance the objective function and the constraints we obtain a new optimization problem
\begin{alignat}{2}
   & \text{min. }  & \mu_1\sum_{r \in R}c_r x_r  +  \mu_2\sum_{f \in F}\left( \left[\sum_{r \in R} a_{fr}x_r\right]- 1\right)^2, & \label{eq:MinCostQuad} 
   \\
    & \text{s.t. } & x_r \in \{0, 1\} \  \forall r \in R &  \label{eq:Integrality}.
\end{alignat}
The new optimization problem in Eq.~\eqref{eq:MinCostQuad}-\eqref{eq:Integrality} can subsequently be modified to have variables $s_r \in \{-1, 1\}$ by 
replacing the variables $x_r = \frac{1 + s_r}{2}$, as presented by Lucas for several combinatorial optimization problems~\cite{Lucas2014}. The variable change results in the following classical Hamiltonian
\begin{align*}
H(s_1,\dots,s_{|R|})
& =  
\mu_1\cdot\sum_{r \in R}c_r \frac{1 + s_r}{2}  \\
 & +
\mu_2\cdot\sum_{f \in F}\left( \left[\sum_{r \in R} a_{fr}\frac{1 + s_r}{2}\right]- 1\right)^2   \\
&=  \mu_1 H^{\text{Objective}}(s_1,\dots,s_{|R|}) \\
&+   \mu_2 H^{\text{Exact Cover}}(s_1,\dots,s_{|R|})
\end{align*}
 which we expand separately for the objective Hamiltonian and the Exact Cover Hamiltonian, where the Exact Cover Hamiltonian can be referred to as the constraint Hamiltonian.
For the objective part we obtain
\begin{align*}
    H^{\text{Objective}}(s_1,&\dots,s_{|R|}) = \\
    & \sum_{r \in R} h^{\text{Objective}}_r s_r + \sum_{r'>r} J^{\text{Objective}}_{rr'} s_r s_{r'}  \\
    =& \sum_{r \in R} \frac{c_r}{2} s_r+ \sum_{r \in R} \frac{c_r}{2} = \sum_{r \in R} \frac{c_r}{2} s_r,
\end{align*}
by ignoring the constant energy shift. Thus  
\begin{align*}
   & h^{\text{Objective}}_r =  \frac{c_r}{2}, \\
   & J^{\text{Objective}}_{rr'} = 0.
\end{align*}
For the constraints, i.e., the Exact Cover Hamiltonian, it was showed in~\cite{vikstl2019applying}  that the classical Hamiltonian takes the form
%
\begin{align*}
 H^{\text{Exact Cover}}(s_1, \dots, s_{|R|}) &= \sum_{r \in R} h^{\text{Exact Cover}}_r s_r  \\
 &+\sum_{r'>r} J^{\text{Exact Cover}}_{rr'} s_r s_{r'}
\end{align*}
where 
\begin{align*}
    & h^{\text{Exact Cover}}_r =  \sum_{f \in F} a_{fr}\left( \sum_{r' \in R}\frac{a_{fr'}}{2} -1 \right), \\
    & J^{\text{Exact Cover}}_{rr'} = \sum_{f \in F}\frac{a_{fr} a_{fr'}}{2}.
\end{align*}
For the Set Partitioning problem we then obtain the following Hamiltonian 
\begin{align*}
     H^{\text{Set Partitioning}}&(s_1, \dots, s_{|R|})= \\    
     & \sum_{r\in R} \left[\mu_1\cdot h_r^{\text{Objective}} + \mu_2\cdot h_r^{\text{Exact Cover}}\right]s_r + \\
     & \mu_2\cdot\sum_{r'>r} J_{rr'}^{\text{Exact Cover}}s_r s_{r'}.
\end{align*}
Finally, the quantum Hamiltonian is obtained by promoting $s_r$ to $\hat{\sigma}_r^z$
\begin{align*}
 \hat{H}^{\text{Set Partitioning}}&(\hat{\sigma}_1^z, \dots, \hat{\sigma}_{|R|}^z)= \\    
     & \sum_{r\in R} [\mu_1\cdot h_r^{\text{Objective}} + \mu_2\cdot h_r^{\text{Exact Cover}}]\hat{\sigma}_r^z  \\
     +& \mu_2\cdot\sum_{r'>r} J_{rr'}^{\text{Exact Cover}}\hat{\sigma}_r^z \hat{\sigma}_{r'}^z.
\end{align*}

It can be noted  
that the mapping holds for any ILP of the form 
\begin{align*}
\text{minimize}  \    & \sum_{r \in R} c_r x_r,  
\\
\text{subject to} \ & \sum_{r \in R} a_{fr}x_r = b_f \ \forall f \in F, 
\\
             & x_r \in \{0, 1\}\ \forall r\in R , 
\end{align*}
if $h^{\text{Exact Cover}}_r$ is modified to
\begin{equation*}
    h^{\text{Exact Cover}}_r =  \sum_{f \in F} a_{fr}\left(\sum_{r' \in R}\frac{a_{fr'}}{2} -b_f\right).
\end{equation*}

\section{Expectation value for algorithm depth one\label{sec:expvalproof}}
In this section we derive the expression of the expectation value  in Eq.~\eqref{eq:expvalend} 
 for algorithm depth $p=1$ of QAOA. 
The expectation value of a general Ising spin glass  Hamiltonian $\hat{H}=\sum_{i=1}^n h_i \hat{\sigma}_i^z + \sum_{(i,j)\in E}J_{ij}\hat{\sigma}_i^z\hat{\sigma}_j^z$  associated to an undirected  graph $G= (V, E)$ with $n=|V|$ nodes and $|E|$ edges  can be computed accordingly
\begin{align*}
     \langle E \rangle &= \text{Tr}[\rho \hat{H}] =\sum_{i=1}^n h_i \text{Tr}[\rho \hat{\sigma}_i^z ] + \sum_{(i,j)\in E} J_{ij}\text{Tr}[\rho \hat{\sigma}_i^z\hat{\sigma}_j^z]. 
\end{align*}
The undirected graph $G$ has  no self loops, which  means that  no edge $(i, i)$ is present in the graph. We furthermore consider  the  edge  $(i,j)$ as identical to edge $(j,i)$ and the sum over edges thus include the edge between node $i$ and  $j$ exactly once. In other words, the edges are unordered pairs  that connect the two nodes  without a particular direction, hence in graph $G$ that we consider $(j, i)$ is simply another way of referring to edge $(i, j)$ which means that $J_{ij}=J_{ji}$.
The density matrix in the expression for the expectation value is $\rho = U_M(\beta)U_c(\gamma)\ket{+}\bra{+} U_c^\dag(\gamma)U_M^\dag(\beta)$,
where QAOA operators are defined as  
\begin{align*}
     & U_M(\beta) = \prod_{i=1}^n e^{-i \beta \hat{\sigma}_i^x}
    \\
     & U_c(\gamma) = \prod_{i=1}^n e^{-i \gamma h_i \hat{\sigma}_i^z} \prod_{(i,j)\in E}e^{-i \gamma J_{ij}\hat{\sigma}_i^z\hat{\sigma}_j^z} =  U_c^1(\gamma)U_c^2(\gamma).
\end{align*}
We can rewrite the expectation value as
\begin{align*}
    \langle E \rangle &= \sum_{i=1}^n h_i \text{Tr}\left[\ket{+}\bra{+} U_c^\dag(\gamma) U_M^\dag(\beta)\hat{\sigma}_i^zU_M(\beta)U_c(\gamma) \right]
   \\
   & + \smashoperator{\sum_{(i,j)\in E}} J_{ij}\text{Tr}\left[\ket{+}\bra{+}U_c^\dag(\gamma) U_M^\dag(\beta) \hat{\sigma}_i^z\hat{\sigma}_j^zU_M(\beta)U_c(\gamma)\right]  
   \\
   & = \sum_{i=1}^n h_i \langle E_{i} \rangle + \sum_{(i,j) \in E} J_{ij}\langle E_{ij} \rangle
\end{align*}
by the cyclic property of the trace.
We remark that partial terms Tr$\left[\ket{+}\bra{+}\hat{a}\right]$ of the expectation value contribute if $\hat{a}$ is a combination of $\hat{\sigma}^x$  and/or $\mathds{1}$. 
The resulting value for terms $\langle E_{i} \rangle$ and $\langle E_{ij} \rangle$ have been derived for  triangle free graphs in~\cite{willsch2019}. However, the resulting value for a graph with  triangles was shown  via Mathematica in~\cite{ozaeta2020expectation}. In this section we show the same  general form of $\langle E_{ij} \rangle$ by analytical means.

We begin by considering some edge $(i,j)$, clearly all terms in  $U_M(\beta)$ commute with $\hat{\sigma}_i^z\hat{\sigma}_j^z$ except 
 for $e^{-i \beta \hat{\sigma}_i^x}$ and $e^{-i \beta \hat{\sigma}_j^x}$.  We use the following relation
\begin{align}
     F( \hat{a}, \eta \hat{b}) & =   e^{i \eta \hat{b}}
     \hat{a} e^{-i \eta \hat{b}} \nonumber\\
     & =  c_\eta^2\hat{a} + s_\eta^2\hat{b}\hat{a}\hat{b} + i \frac{s_{2\eta}}{2}[\hat{b},\hat{a}] \label{eq:F}
\end{align}
where $c_x^y = \text{cos}^y(x)$ and  $s_x^y = \text{sin}^y(x)$ for convenience. The terms resulting from the mixing operator $U_M(\gamma)$ are therefore
\begin{align*}
U_M^\dag(\beta) \hat{\sigma}_i^z\hat{\sigma}_j^zU_M(\beta)
&= F(\hat{\sigma}_i^z, \beta \hat{\sigma}_i^x)
F( \hat{\sigma}_j^z, \beta \hat{\sigma}_j^x) \\
&= 
     c_{2\beta}^2\hat{\sigma}_i^z\hat{\sigma}_j^z 
    \\
    &+  c_{2\beta}s_{2\beta}[\hat{\sigma}_i^z\hat{\sigma}_j^y 
    +  \hat{\sigma}_i^y\hat{\sigma}_j^z] \\
    &+ s_{2\beta}^2\hat{\sigma}_i^y\hat{\sigma}_j^y
\end{align*}
by evaluating Eq.~\eqref{eq:F}.
For the ease of future derivitions, we separate these parts as 
$ \langle E_{ij} \rangle =  \langle E_{ij}^{zz}\rangle   + \langle E_{ij}^{zy}\rangle +  \langle E_{ij}^{yz}\rangle + \langle E_{ij}^{yy}\rangle$ where the sinus and cosinus terms are temporarily ignored. To clarify, here we defined  
$\langle E_{ij}^{ab} \rangle = \text{Tr}\left[\ket{+}\bra{+} U_c^\dag(\gamma) \hat{\sigma}_i^a \hat{\sigma}_j^b U_c(\gamma)\right]$.

We note that 
$[U_c(\gamma), \hat{\sigma}_i^z\hat{\sigma}_j^z] = 0$ and hence  $\langle E_{ij}^{zz} \rangle$ does not contribute to the overall expectation value $\langle E\rangle$.
However, for $\hat{\sigma}_i^y\hat{\sigma}_j^z$ all terms in $U_c^2(\gamma)$ with operators corresponding to edges with  node $i$ contribute, i.e.,
$[e^{-i \gamma J_{kp}\hat{\sigma}_k^z\hat{\sigma}_p^z}, \hat{\sigma}_i^y\hat{\sigma}_j^z] \neq 0 \ \forall p:(k=i, p) \in E $ and only $[e^{-i \gamma h_{i}\hat{\sigma}_i^z}, \hat{\sigma}_i^y\hat{\sigma}_j^z] \neq 0$ of all operator terms corresponding to nodes in $U_c^1(\gamma)$. 
Similarly  for $\hat{\sigma}_i^z\hat{\sigma}_j^y$, all operator terms in $U_c^2(\gamma)$ corresponding to edges that include node $j$  and  the term $e^{-i \gamma h_{j}\hat{\sigma}_j^z}$ in $U_c^1(\gamma)$ contribute. 
For $\hat{\sigma}_i^y\hat{\sigma}_j^y$ we note that  
all operator terms in $U_c^2(\gamma)$ for edges that include node $i$ or $j$ contributes as 
$[e^{-i \gamma J_{kp}\hat{\sigma}_k^z\hat{\sigma}_p^z}, \hat{\sigma}_i^y\hat{\sigma}_j^y] \neq 0$ $\forall p \neq i:(k=j, p) \in E$ and  $\forall p\neq j:(k=i, p) \in E $. Furthermore, both terms in $U_c^1(\gamma)$ that correspond to node $i$ and $j$ contribute to the expectation value as well.

We now wish to evaluate the terms  $\langle E_{ij}^{yy}\rangle, \langle E_{ij}^{zy}\rangle$ and  $\langle E_{ij}^{yz}\rangle$. We begin with the most complex case, 
 $\langle E_{ij}^{yy}\rangle$, which is the only term that changes if triangles are present in the graph compared to the expression given in~\cite{willsch2019}. 
 Since the only  terms in $U_c^1(\gamma)$ that do not commute with $\hat{\sigma}_i^y\hat{\sigma}_j^y$ are $e^{-i \gamma h_i \hat{\sigma}_i^z}$  and $e^{-i \gamma h_j \hat{\sigma}_j^z}$, the terms that we obtain from $U_c^1(\gamma)$ are thus 
\begin{align*}
U_c^{1\dag}(\gamma) \hat{\sigma}_i^y \hat{\sigma}_j^y U_c^1(\gamma) &=  
F(\hat{\sigma}_i^y, \gamma h_i  \hat{\sigma}_i^z) F(\hat{\sigma}_j^y, \gamma h_j  \hat{\sigma}_j^z)  \\&= 
     c_{2h_i \gamma}c_{2h_j \gamma}\hat{\sigma}_i^y\hat{\sigma}_j^y +  s_{2h_i \gamma}c_{2h_j \gamma}\hat{\sigma}_i^x\hat{\sigma}_j^y 
    \\
    &+  
    c_{2h_i \gamma}
    s_{2h_j \gamma} 
    \hat{\sigma}_i^y 
    \hat{\sigma}_j^x +  
    s_{2h_i \gamma} 
    s_{2h_j \gamma}
    \hat{\sigma}_i^x 
    \hat{\sigma}_j^x  
\end{align*}
 which gives us four terms to consider. 
We will now use the following relation 
\begin{align}
G(\hat{c}, \eta_a &\hat{a}, \eta_b \hat{b})  = 
     e^{i \eta_a \hat{a}} e^{i \eta_b \hat{b}}  \hat{c} e^{-i \eta_b \hat{b}} e^{-i \eta_a \hat{a}} \nonumber
     \\
     & = e^{i \eta_a \hat{a}}(c_{\eta_b}^2\hat{c} + s_{\eta_b}^2\hat{b}\hat{c}\hat{b} + i \frac{s_{2\eta_b}}{2}[\hat{b},\hat{c}])e^{-i \eta_a \hat{a}}  \nonumber  
    \\
    & = c_{\eta_b}^2[c_{\eta_a}^2\hat{c} + s_{\eta_a}^2\hat{a}\hat{c}\hat{a} + i \frac{s_{2\eta_a}}{2}[\hat{a},\hat{c}]]  \nonumber
    \\
    & +s_{\eta_b}^2[c_{\eta_a}^2\hat{b}\hat{c}\hat{b} + s_{\eta_a}^2\hat{a}(\hat{b}\hat{c}\hat{b})\hat{a} + i \frac{s_{2\eta_a}}{2}[\hat{a},\hat{b}\hat{c}\hat{b}]]  \nonumber
    \\
     +i \frac{s_{2\eta_b}}{2}&[c_{\eta_a}^2[\hat{b},\hat{c}] + s_{\eta_a}^2\hat{a}[\hat{b},\hat{c}]\hat{a} + i \frac{s_{2\eta_a}}{2}[\hat{a},[\hat{b},\hat{c}]]]\label{eq:G}
\end{align}
to evaluate the contributing terms when $U_c^2(\gamma)$ is applied.
It then becomes clear that the operators $\hat{\sigma}_i^x\hat{\sigma}_j^y$ and $\hat{\sigma}_i^y\hat{\sigma}_j^x$ do not contribute to  $\langle E_{ij}^{yy} \rangle$ as there are no terms that can result in a  pure $\hat{\sigma}^x$ and/or $\mathds{1}$ combination when evaluating Eq.~\eqref{eq:G}. We explicitly give the expressions for all terms below for the case of $\hat{\sigma}_i^x\hat{\sigma}_j^y$ (which by symmetry also allow us to throw away $\hat{\sigma}_i^y\hat{\sigma}_j^x$):
 \begin{align*}
    \quad \hat{a} =  \hat{\sigma}_i^z \hat{\sigma}_k^z, 
    \quad \hat{b} =  \hat{\sigma}_j^z \hat{\sigma}_p^z,    
    \quad \hat{c} =  \hat{\sigma}_i^x \hat{\sigma}_j^y, 
 \end{align*}
  \begin{align*}
     &  \hat{a}\hat{c}\hat{a}  \propto  \hat{\sigma}_i^x \hat{\sigma}_j^y, 
     \quad [\hat{a},\hat{c}]  \propto  \hat{\sigma}_i^y \hat{\sigma}_j^y \hat{\sigma}_k^z,
     \quad \hat{b}\hat{c}\hat{b} \propto \hat{\sigma}_i^x \hat{\sigma}_j^y,
 \end{align*}
  \begin{align*}
    &  \hat{a}(\hat{b}\hat{c}\hat{b})\hat{a} \propto \hat{\sigma}_i^x \hat{\sigma}_j^y, 
    \quad[\hat{a},\hat{b}\hat{c}\hat{b}]  \propto \hat{\sigma}_i^y \hat{\sigma}_j^y \hat{\sigma}_k^z,
    \quad [\hat{b},\hat{c}]  \propto  \hat{\sigma}_i^x \hat{\sigma}_j^x \hat{\sigma}_p^z,
 \end{align*}
  \begin{align*}
    &  \hat{a}[\hat{b},\hat{c}]\hat{a} \propto  
    \begin{cases}
         \hat{\sigma}_i^x \hat{\sigma}_j^x \hat{\sigma}_{k}^z  & \text{ if $k=p$} \\
         \hat{\sigma}_i^x \hat{\sigma}_j^x \hat{\sigma}_{p}^z & \text{ else}
    \end{cases},
     \\
    & [\hat{a},[\hat{b},\hat{c}]] \propto 
        \begin{cases}
         \hat{\sigma}_i^y \hat{\sigma}_j^x   & \text{ if $k=p$} \\
         \hat{\sigma}_i^y \hat{\sigma}_j^x \hat{\sigma}_{k}^z\hat{\sigma}_{p}^z & \text{ else}
    \end{cases}.
 \end{align*}
 This means that we only need to consider $\hat{\sigma}_i^y\hat{\sigma}_j^y$ and $\hat{\sigma}_i^x\hat{\sigma}_j^x$. We can rewrite the contributing terms as 
\begin{align*}
 & \frac{1}{2}\left(c_{2(h_i - h_j) \gamma}[ \hat{\sigma}_i^y \hat{\sigma}_j^y + \hat{\sigma}_i^x \hat{\sigma}_j^x ] - c_{2(h_i + h_j) \gamma}[  \hat{\sigma}_i^x \hat{\sigma}_j^x -\hat{\sigma}_i^y \hat{\sigma}_j^y]\right) 
\end{align*}
by using the relations 
\begin{align}
    & s_{2h_i \gamma} s_{2h_j \gamma} = \frac{1}{2}\left[c_{2(h_i - h_j) \gamma} -c_{2(h_i + h_j) \gamma}\right], \label{eq:ss}\\
    & c_{2h_i \gamma} c_{2h_j \gamma} = \frac{1}{2}\left[c_{2(h_i - h_j) \gamma} + c_{2(h_i + h_j) \gamma}\right].\label{eq:cc}
\end{align}
For the contributing terms we now  consider how terms in $U_c^2(\gamma)$ corresponding to a triangle $(i,j,p)$ act on $\hat{\sigma}_i^y \hat{\sigma}_j^y$ and $\hat{\sigma}_i^x \hat{\sigma}_j^x$.
By evaluating the function in Eq.~\eqref{eq:G} for   $\hat{\sigma}_i^x\hat{\sigma}_j^x$  we get the following contributing terms
\begin{align*}
G(\hat{\sigma}_i^x\hat{\sigma}_j^x, 
  \gamma J_{ip} \hat{\sigma}_i^z\hat{\sigma}_p^z, \gamma J_{jp} \hat{\sigma}_j^z \hat{\sigma}_p^z) & = 
       c_{2\gamma J_{jp}} c_{2\gamma J_{ip}}\hat{\sigma}_i^x\hat{\sigma}_j^x     \\
    &+s_{2\gamma J_{jp}}s_{2\gamma J_{ip}}\hat{\sigma}_i^y\hat{\sigma}_j^y
\end{align*}
since
 \begin{align*}
    \quad \hat{a} =  \hat{\sigma}_i^z \hat{\sigma}_p^z, 
    \quad \hat{b} =  \hat{\sigma}_j^z \hat{\sigma}_p^z,  
    \quad \hat{c} =  \hat{\sigma}_i^x \hat{\sigma}_j^x, 
 \end{align*}
  \begin{align*}
 &  \hat{a}\hat{c}\hat{a}  =  -\hat{\sigma}_i^x \hat{\sigma}_j^x, 
 \quad [\hat{a},\hat{c}]  =   2i\hat{\sigma}_i^y \hat{\sigma}_j^x \hat{\sigma}_p^z,
 \quad \hat{b}\hat{c}\hat{b} = -\hat{\sigma}_i^x \hat{\sigma}_j^x, 
 \end{align*}
  \begin{align*}
&  \hat{a}(\hat{b}\hat{c}\hat{b})\hat{a} = \hat{\sigma}_i^x \hat{\sigma}_j^x, 
\ [\hat{a},\hat{b}\hat{c}\hat{b}]  = -2i\hat{\sigma}_i^y \hat{\sigma}_j^x \hat{\sigma}_p^z,
\  [\hat{b},\hat{c}]  = 2i\hat{\sigma}_i^x \hat{\sigma}_j^y \hat{\sigma}_p^z,
 \end{align*}
\begin{align*}
&  \hat{a}[\hat{b},\hat{c}]\hat{a} 
= -2i\hat{\sigma}_i^x \hat{\sigma}_j^y \hat{\sigma}_p^z,
\quad [\hat{a},[\hat{b},\hat{c}]] = -4 \hat{\sigma}_i^y \hat{\sigma}_j^y.
 \end{align*}
The final expression for $\hat{\sigma}_i^x\hat{\sigma}_i^x$ is found by the relation $c_{2a} = (c_{a}^2 - s_{a}^2)$.
For $\hat{\sigma}_i^y\hat{\sigma}_j^y$ we get the following contributing terms 
\begin{align*}
G( \hat{\sigma}_i^y\hat{\sigma}_j^y,\gamma J_{ip} \hat{\sigma}_i^z \hat{\sigma}_p^z, \gamma J_{jp} \hat{\sigma}_j^z \hat{\sigma}_p^z ) & = 
    c_{2\gamma J_{jp}} c_{2\gamma J_{ip}}\hat{\sigma}_i^y\hat{\sigma}_j^y \\
   & +s_{2\gamma J_{jp}}s_{2\gamma J_{ip}}\hat{\sigma}_i^x\hat{\sigma}_j^x.
\end{align*}
by inspecting the partial terms we obtain by considering Eq.~\eqref{eq:G}:
 \begin{align*}
    \quad \hat{a} =  \hat{\sigma}_i^z \hat{\sigma}_p^z, 
    \quad \hat{b} =  \hat{\sigma}_j^z \hat{\sigma}_p^z,  
    \quad \hat{c} =  \hat{\sigma}_i^y \hat{\sigma}_j^y, 
 \end{align*}
  \begin{align*}
 &  \hat{a}\hat{c}\hat{a}  =  -\hat{\sigma}_i^y \hat{\sigma}_j^y, 
 \quad [\hat{a},\hat{c}]  =  -2i\hat{\sigma}_i^x \hat{\sigma}_j^y \hat{\sigma}_p^z,
 \quad \hat{b}\hat{c}\hat{b} = -\hat{\sigma}_i^y \hat{\sigma}_j^y, 
 \end{align*}
  \begin{align*}
&  \hat{a}(\hat{b}\hat{c}\hat{b})\hat{a} = \hat{\sigma}_i^y \hat{\sigma}_j^y, 
\ [\hat{a},\hat{b}\hat{c}\hat{b}]  = 2i\hat{\sigma}_i^x \hat{\sigma}_j^y \hat{\sigma}_p^z,
\  [\hat{b},\hat{c}]  = -2i\hat{\sigma}_i^y \hat{\sigma}_j^x \hat{\sigma}_p^z,
 \end{align*}
\begin{align*}
&  \hat{a}[\hat{b},\hat{c}]\hat{a} 
= 2i\hat{\sigma}_i^y \hat{\sigma}_j^x \hat{\sigma}_p^z,
\quad [\hat{a},[\hat{b},\hat{c}]] = -4 \hat{\sigma}_i^x \hat{\sigma}_j^x.
 \end{align*}
%
By  using  trigonometric relations
\begin{align}
    & c_{a+b} = c_ac_b - s_as_b, \label{eq:ss2}\\
    & c_{a-b} = c_ac_b + s_as_b.\label{eq:cc2}
\end{align}
it is clear that the operators corresponding to a triangle $(i,j,p)$ act on $\hat{\sigma}_i^y\hat{\sigma}_j^y + \hat{\sigma}_i^x\hat{\sigma}_j^x$ and $ \hat{\sigma}_i^x\hat{\sigma}_j^x -\hat{\sigma}_i^y\hat{\sigma}_j^y $ as
\begin{align*}
 G( \hat{\sigma}_i^y\hat{\sigma}_j^y + \hat{\sigma}_i^x\hat{\sigma}_j^x,\gamma J_{ip} \hat{\sigma}_i^z \hat{\sigma}_p^z,& \gamma J_{jp} \hat{\sigma}_j^z \hat{\sigma}_p^z ) = \\
     & c_{2\gamma({J_{jp}} - J_{ip} ) } [\hat{\sigma}_i^y\hat{\sigma}_j^y + \hat{\sigma}_i^x\hat{\sigma}_j^x],
\\
 G(\hat{\sigma}_i^x\hat{\sigma}_j^x-\hat{\sigma}_i^y\hat{\sigma}_j^y,\gamma J_{ip} \hat{\sigma}_i^z \hat{\sigma}_p^z,& \gamma J_{jp} \hat{\sigma}_j^z \hat{\sigma}_p^z ) = \\
     &c_{2\gamma({J_{jp}} + J_{ip} ) } [\hat{\sigma}_i^x\hat{\sigma}_j^x-\hat{\sigma}_i^y\hat{\sigma}_j^y ].  
\end{align*} 
Furthermore, other terms in $U_c^2(\gamma)$ corresponding to edges $(i,k)$ and $(j,l)$ that are not part of a triangle will give rise to the following expressions for $\hat{\sigma}_i^x\hat{\sigma}_j^x$    
\begin{align*}
 F( \hat{\sigma}_i^x\hat{\sigma}_j^x,\gamma J_{ik} \hat{\sigma}_i^z\hat{\sigma}_k^z )&  = 
    c_{2\gamma J_{ik}}\hat{\sigma}_i^x\hat{\sigma}_j^x -s_{2\gamma J_{ik}}\hat{\sigma}_i^y\hat{\sigma}_j^x\hat{\sigma}_k^z,
    \\
 F( \hat{\sigma}_i^x\hat{\sigma}_j^x,\gamma J_{jl} \hat{\sigma}_j^z \hat{\sigma}_l^z)& =  c_{2\gamma J_{jl}}\hat{\sigma}_i^x\hat{\sigma}_j^x -s_{2\gamma J_{jl}}\hat{\sigma}_i^x\hat{\sigma}_j^y\hat{\sigma}_l^z
\end{align*}
and for $\hat{\sigma}_i^y\hat{\sigma}_j^y$ we get 
\begin{align*}
F( \hat{\sigma}_i^y\hat{\sigma}_j^y,\gamma J_{ik} \hat{\sigma}_i^z \hat{\sigma}_k^z)&  = c_{2\gamma J_{ik}}\hat{\sigma}_i^y\hat{\sigma}_j^y +s_{2\gamma J_{ik}}\hat{\sigma}_i^x\hat{\sigma}_j^y\hat{\sigma}_k^z,
    \\
F( \hat{\sigma}_i^y\hat{\sigma}_j^y,\gamma J_{jl} \hat{\sigma}_j^z \hat{\sigma}_l^z) & = c_{2\gamma J_{jl}}\hat{\sigma}_i^y\hat{\sigma}_j^y +s_{2\gamma J_{jl}}\hat{\sigma}_i^y\hat{\sigma}_j^x\hat{\sigma}_l^z.
\end{align*}
Thus, for  $\langle E_{ij}^{yy}\rangle $ we get the following contributing parts to the expectation value  
\begin{align}
     \langle E_{ij}^{yy} \rangle = &\ \frac{1}{2}s_{2\beta}^2
    \prod_{\substack{(i,k)\in E \\ (j,k) \notin E}}  c_{2\gamma J_{ik}} \ 
    \prod_{\substack{(j,l)\in E \\ (i,l) \notin E}}  c_{2\gamma J_{jl}} \times 
    \nonumber
    \\
    & \ \text{\Huge[}c_{2(h_i - h_j) \gamma}
    \smashoperator{\prod_{\substack{(i,p)\in E \\(j,p) \in E}}}
    c_{2(J_{ip} - J_{jp}) \gamma} 
    \nonumber
    \\
    &-c_{2(h_i + h_j) \gamma}\smashoperator{\prod_{\substack{(j,p)\in E \\(i,p) \in E}}}
    c_{2(J_{ip} + J_{jp}) \gamma}  \text{\Huge]}
    \label{eyy}
\end{align}
since only the $\hat{\sigma}_i^x\hat{\sigma}_j^x$ terms are nonzero when the trace is taken. 

For  $\langle E_{ij}^{yz} \rangle$ on the other hand, the only non commuting term of $U_c^1(\gamma)$ with $\hat{\sigma}_i^y\hat{\sigma}_j^z$  is $e^{-i \gamma h_i \hat{\sigma}_i^z}$, which results in
\begin{equation*}
F( \hat{\sigma}_i^y\hat{\sigma}_j^z,\gamma h_i \hat{\sigma}_i^z) = 
    c_{2\gamma h_i}\hat{\sigma}_i^y\hat{\sigma}_j^z  + s_{2\gamma h_i} \hat{\sigma}_i^x \hat{\sigma}_j^z.
\end{equation*}
For the operator, $e^{-iJ_{ij} \gamma \hat{\sigma}_i^z\hat{\sigma}_j^z}$, corresponding to edge $(i,j)$, we have that
\begin{align*}
    & F(\hat{\sigma}_i^y\hat{\sigma}_j^z, \gamma J_{ij}\hat{\sigma}_i^z\hat{\sigma}_j^z) =   c_{2\gamma J_{ij}}\hat{\sigma}_i^y\hat{\sigma}_j^z + s_{2\gamma J_{ij}}\hat{\sigma}_i^x,
    \\
    & 
    F(\hat{\sigma}_i^x\hat{\sigma}_j^z, \gamma J_{ij}\hat{\sigma}_i^z\hat{\sigma}_j^z) =  c_{2\gamma J_{ij}}\hat{\sigma}_i^x\hat{\sigma}_j^z - s_{2\gamma J_{ij}}\hat{\sigma}_i^y.
\end{align*}
Other operators of $U_c^2(\gamma)$ corresponding to edges that include node $i$ further gives the expressions
\begin{align*}
    & F(\hat{\sigma}_i^y\hat{\sigma}_j^z, \gamma J_{ip} \hat{\sigma}_i^z\hat{\sigma}_p^z)  = c_{2\gamma J_{ip}}\hat{\sigma}_i^y\hat{\sigma}_j^z + s_{2\gamma J_{ip}}\hat{\sigma}_i^x\hat{\sigma}_j^z\hat{\sigma}_p^z,
    \\
    & 
    F(\hat{\sigma}_i^x, \gamma J_{ip} \hat{\sigma}_i^z\hat{\sigma}_p^z)
       =  c_{2\gamma J_{ip}}\hat{\sigma}_i^x - s_{2\gamma J_{ip}}\hat{\sigma}_i^y\hat{\sigma}_p^z,
    \\
    & 
    F(\hat{\sigma}_i^x\hat{\sigma}_j^z,\gamma J_{ip} \hat{\sigma}_i^z\hat{\sigma}_p^z)
     =   c_{2\gamma J_{ip}}\hat{\sigma}_i^x\hat{\sigma}_j^z  - s_{2\gamma J_{ip}}\hat{\sigma}_i^y\hat{\sigma}_j^z\hat{\sigma}_p^z,
    \\
    & 
    F(\hat{\sigma}_i^y,  \gamma J_{ip} \hat{\sigma}_i^z\hat{\sigma}_p^z)
        =   c_{2\gamma J_{ip}}\hat{\sigma}_i^y + s_{2\gamma J_{ip}}\hat{\sigma}_i^x\hat{\sigma}_p^z.
\end{align*}
We can again conclude that since only the $\hat{\sigma}_i^x$ term contributes here,  the final contribution  to the expectation value is
\begin{equation}
    \langle E_{ij}^{yz} \rangle = \frac{ s_{4\beta}}{2}c_{2\gamma h_{i}}s_{2\gamma J_{ij}} 
    \smashoperator{\prod_{p\neq j : (i,p)\in E}}  c_{2\gamma J_{ip}}
    \label{eyz}
\end{equation}
and   
\begin{equation}
    \langle E_{ij}^{zy} \rangle =  \frac{ s_{4\beta}}{2}c_{2\gamma h_{j}}s_{2\gamma J_{ij}} 
    \smashoperator{\prod_{p\neq i : (j,p)\in E}}  c_{2\gamma J_{jp}}
    \label{ezy}
\end{equation}
by exchanging index $i$ and $j$.

Finally, for  $\langle E_i \rangle$ the only non commuting term of $U_M(\beta)$ is $e^{-i \beta \hat{\sigma}_i^x}$, which results in
\begin{align*}
     F(\hat{\sigma}_i^z,e^{-i \beta \hat{\sigma}_i^x} ) = 
     c_{2\beta}\hat{\sigma}_i^z + s_{2\beta}\hat{\sigma}_i^y.
\end{align*}
We can exclude $\hat{\sigma}_i^z$ as it commutes with $U_c(\gamma)$. On the other hand $\hat{\sigma}_i^y$ does not commute with  $e^{-i \gamma h_i \hat{\sigma}_i^z}$. The operator $U_c^1(\gamma)$ therefore gives rise to the following expression
\begin{align*}
    F(\hat{\sigma}_i^y, \gamma h_i \hat{\sigma}_i^z ) =  c_{2\gamma h_i}\hat{\sigma}_i^y + s_{2\gamma h_i}\hat{\sigma}_i^x.
\end{align*}
When we act with $e^{-i \gamma J_{ip} \hat{\sigma}_i^z\hat{\sigma}_p^z}$  for an edge $p:(i,p) \in E$ we get 
\begin{align*}
    F(\hat{\sigma}_i^y, \gamma J_{ip} \hat{\sigma}_i^z\hat{\sigma}_p^z) =   c_{2\gamma J_{ip}}\hat{\sigma}_i^y + s_{2\gamma J_{ip}}\hat{\sigma}_i^x\hat{\sigma}_p^z,
    \\
    F(\hat{\sigma}_i^x, \gamma J_{ip} \hat{\sigma}_i^z\hat{\sigma}_p^z) =   c_{2\gamma J_{ip}}\hat{\sigma}_i^x - s_{2\gamma J_{ip}}\hat{\sigma}_i^y\hat{\sigma}_p^z
\end{align*}
which only contributes with $c_{2\gamma J_{ip}}\hat{\sigma}_i^x$.
The resulting contribution of  $\langle E_i \rangle$  to the overall expectation value is therefore
\begin{align}
    \langle E_i \rangle =  s_{2\beta} s_{2\gamma h_i} \prod_{p:(i,p)\in E}  c_{2\gamma J_{ip}}.
    \label{ez}
\end{align}
We conclude by noting that if we add all the terms in Eq.~\eqref{ez},~\eqref{ezy},~\eqref{eyz} and~\eqref{eyy} with their coefficients $h_i$ and $J_{ij}$ we get the expression in Eq.~\eqref{eq:expvalend}.
~\\
~\\
~\\
~\\
\section{Success probabilities of Set Partitioning\label{sec:SP_success_prob}}
In this section,  a summary is given of the results of ideal simulations of QAOA circuits for all instances applied to the Set Partitioning problem. Table~\ref{tab:success_prob} 
shows success probabilities for  Hamiltonians constructed for factors $f=\infty$, $f^*$ and intermediate choices. As a shorthand, $P^{f}$ denotes $P_{\text{success}}^{\text{Set Partitioning}}$ given a Set Partitioning Hamiltonian with weights $\mu_1$ and $\mu_2$ for a factor $f$.
~\\
~\\
~\\
~\\
\begin{table}[h!]
\caption{Success probabilities for QAOA applied to Set Partitioning for problem sizes 6-20, given algorithm depth $p$ for multiple choices of factor $f$}
\begin{subtable}[h]{0.45\textwidth}
    \centering 
    \begin{tabular}{cccccccc} \hline \hline
    $|R|$ & $p$ & $|S_{\text{feasible}}|$ & $P^{f=\infty}$ & $P^{f=100}$ &  $P^{f=10}$  & $P^{f=1}$ &
    \\ \hline 
6 & 40 & 1 & 99.55 & 99.52 & 99.96 & 99.98 & \\
  &   & 2 & 50.   & 51.36 & 59.93 & 72.67 & \\
  &   & 3 & 32.39 & 36.   & 67.43 & 99.71 & \\
\hline 
    $|R|$ & $p$ & $|S_{\text{feasible}}|$ & $P^{f=\infty}$ & $P^{f=100}$ &  $P^{f=10}$  & $P^{f=1}$
    & \\ \hline 
8 & 40 & 1 &  99.68 & 99.54 & 99.83 & 99.83 & \\
    &  & 2 & 52.47 & 57.29 & 95.46 & 99.33 & \\
    &  & 3 & 27.62 & 38.48 & 99.22 & 99.99 & \\
    &  & 4 & 26.69 & 96.68 & 99.75 & 99.4 & \\
\hline 
    $|R|$ & $p$ & $|S_{\text{feasible}}|$ & $P^{f=\infty}$ & $P^{f=100}$ &  $P^{f=10}$ &  $P^{f=33.33}$
    & \\ \hline 
10 & 40 & 1 & 96.75 & 5.19 &   &   & \\
    &  & 2 & 51.15 &  0.28 &   & 47.18 & \\
    &  & 3 & 33.37 & 38.33 & 99.03 &   & \\
    &  & 4 & 25.38 & 39.52 & 99.95 &   & \\
    &  & 5 & 18.89 & 89.3  & 99.91 &   & \\
     \hline 
$|R|$ & $p$ & $|S_{\text{feasible}}|$ & $P^{f=\infty}$ & $P^{f=100}$ &  $P^{f=10}$ & $P^{f=20}$ & $P^{f=25}$
\\ \hline 
12 & 40 & 1 & 82.72 & 45.85 &   &   &  \\
 &  & 2 & 60.41 & 64.52 & 87.83   &   & \\
 &  & 3 & 23.78 & 27.19 &   & 40.21   & \\
 &  & 4 & 31.35 & 37.62 &   &   & 70.46 \\
 &  & 5 & 13.06 & 20.15 & 99.5 &   &   \\
 &  & 6 & 16.54 & 24.31 & 99.53 &   &   \\ \hline 
    \end{tabular}
%
            \begin{tabular}{cccccccc} 
$|R|$ & $p$ & $|S_{\text{feasible}}|$ & $P^{f=\infty}$ & $P^{f=100}$ &  $P^{f=10}$ &  &
            \\ \hline  
        14 & 40 & 1 & 67.07 & 38.76   &  & $~~\quad \quad$ & $\qquad \qquad$\\
             &  & 2 & 47.59 & 63.63 & 21.06 & &\\
             &  & 3 & 34.73 & 43.78 & 52.09 & &\\
             &  & 4 & 19.94 & 42.91 & 99.92 & &\\
             &  & 5 & 20.99 & 31.98 & 97.46 & &\\
             &  & 6 & 20.06 & 36.09 & 99.63 & &\\
             &  & 7 & 12.25 & 25.68 & 99.22 & &\\
             \hline 
            $|R|$ & $p$ & $|S_{\text{feasible}}|$ & $P^{f=\infty}$ & $P^{f=10} $ & & &
            \\ \hline 
        20 & 20 & 1 & 12.71 & 12.79 & & &\\
             &  & 2 & 14.42 & 12.8 & & &\\
             &  & 3 & 10.27 & 12.64 & & &\\
             &  & 4 & 15.84 & 12.68 & & &\\
             &  & 5 & 16.34 & 17.26 & & &\\
             &  & 6 & 13.23 & 17.53 & & &\\
             &  & 7 & 12.44 & 19.09 & & &\\
             &  & 8 & 11.32 & 86.52 & & &\\
             &  & 9 &  8.19 & 77.11 & & &\\
             &  & 10 & 7.75 & 86.39 & & &\\ \hline \hline
            \end{tabular}    
\end{subtable}
\label{tab:success_prob}
\end{table}
~\\
~\\
~\\
~\\
~\\
~\\
~\\
\end{document}